\documentclass[aps,prx,twocolumn,superscriptaddress, longbibliography,nofootinbib]{revtex4-2}
\usepackage{graphicx}
\usepackage{latexsym}
\usepackage{amssymb}
\usepackage{amsmath}
\usepackage{amsfonts}
\usepackage{upgreek}
\usepackage{bm}
\usepackage{multirow}
\usepackage{enumitem}
\usepackage{color}
\usepackage{hyperref}
\usepackage{physics}
\usepackage{tcolorbox}
\usepackage{manfnt}
\usepackage{lipsum}
\usepackage[ruled,linesnumbered]{algorithm2e}
\usepackage{tabularx}

\hypersetup{
colorlinks = true,
linkcolor = [rgb]{0.70,0.13,0.13},
citecolor = [rgb]{0.13,0.55,0.13},
urlcolor = [rgb]{0.25, 0.41, 0.88}}
\newcommand{\U}{\mathrm{U}}

\newcommand{\bsigma}{{\bm{\sigma}}}

\newcommand{\bs}{{\bm{s}}}
\newcommand{\bt}{{\bm{t}}}
\newcommand{\bmm}{{\bm{m}}}

\newcommand{\secref}[1]{Sec.\,\ref{#1}}
\newcommand{\appref}[1]{Appendix.\,\ref{#1}}
\newcommand{\eqnref}[1]{Eq.\,\eqref{#1}}
\newcommand{\figref}[1]{Fig.\,\ref{#1}}

\newcommand{\rd}{\partial}

\newcommand{\cP}{{{\cal P}}}

\newcommand{\cC}{{{\cal C}}}
\newcommand{\Xfactor}{{{\cos \theta}}}
\newcommand{\Zfactor}{{{\sin \theta}}}

\newcommand{\ii}{\hspace{1pt}\mathrm{i}\hspace{1pt}}
\newcommand{\fh}{\frac{1}{2}}

\SetAlgorithmName{Protocol}{protocol}{List of Protocols}

\begin{document}

\title{Decoding Measurement-Prepared Quantum Phases and Transitions: \\
from Ising model to gauge theory, and beyond}

\author{Jong Yeon Lee}\email{jongyeon@kitp.ucsb.edu}
\affiliation{Kavli Institute for Theoretical Physics, University of California, Santa Barbara, CA, USA}

\author{Wenjie Ji}
\affiliation{Department of Physics, University of California, Santa Barbara, CA 93106, USA}

\author{Zhen Bi}
\affiliation{Department of Physics, The Pennsylvania State University, University Park, Pennsylvania 16802, USA}

\author{Matthew P. A. Fisher}
\affiliation{Department of Physics, University of California, Santa Barbara, CA 93106, USA}

\date{\today}
\begin{abstract}  
Measurements allow efficient preparation of interesting quantum many-body states with long-range entanglement, conditioned on additional transformations based on measurement outcomes. Here, we demonstrate that the so-called conformal quantum critical points (CQCP) can be obtained by performing general single-site measurements in an appropriate basis on the cluster states in $d\geq2$. 
The equal-time correlators of the said states are described by correlation functions of certain $d$-dimensional classical model at finite temperatures, and feature spatial conformal invariance. This establishes an exact correspondence between the measurement-prepared critical states and conformal field theories of a range of critical spin models, including familiar Ising models and gauge theories. 
Furthermore, by mapping the long-range entanglement structure of measured quantum states into the correlations of the corresponding thermal spin model, we rigorously establish the stability condition of the long-range entanglement in the measurement-prepared quantum states deviating from the ideal setting. 
Most importantly, we describe protocols to decode the resulting quantum phases and transitions without post-selection, thus transferring the exponential measurement complexity to a polynomial classical computation. Therefore, our findings suggest a novel mechanism in which a quantum critical wavefunction emerges, providing new practical ways to study quantum phases and conformal quantum critical points.

\end{abstract}

\maketitle

\section{Introduction}

Recently, the perplexing and exciting effects of measurements on the evolution of quantum many-body states are attracting growing interest from both the condensed matter and quantum information communities.
There are two branches of studies. In the first branch, one focuses on how quantum entanglement propagates and builds up under random measurements and unitary dynamics. Initiated by the discovery of the transition in the entanglement structure under random measurements and circuit evolution~\cite{YaodongFisher2018,SkinnerNahum2018, Chan2019}, there has been extensive work in this direction~\cite{YaodongFisher2019, Vasseur2019, XiangyuLuca2019, Gullans2020, Soonwon2020, TangZhu2020, JianLudwig2020, LopezVasseur2020, Bao2020, Rossini2020, Fan2021, Yaodong2021, BenZion2020, BarrattDecoding2022}.   
In the second branch, one focuses on systematic measurements in a static situation, preparing quantum states by performing measurements on a subsystem of a so-called resource state, which is related to the measurement-based quantum computation (MBQC). Remarkably, various families of quantum states with long range entanglement, such as Greenberger-Horne-Zeilinger (GHZ)  state or those with certain topological and fracton orders among others~\cite{1Dcluster_GHZ, 2Dcluster, 2Dcluster_toric, Piroli2021, 3dCluster_fracton1, 3dCluster_fracton2, NatRydberg,CSScode, CSScode2, ClusterCSS, Lu2022} can be prepared through measurements on cluster states. 

\begin{figure}[!t]
    \centering
    \includegraphics[width = 0.48 \textwidth]{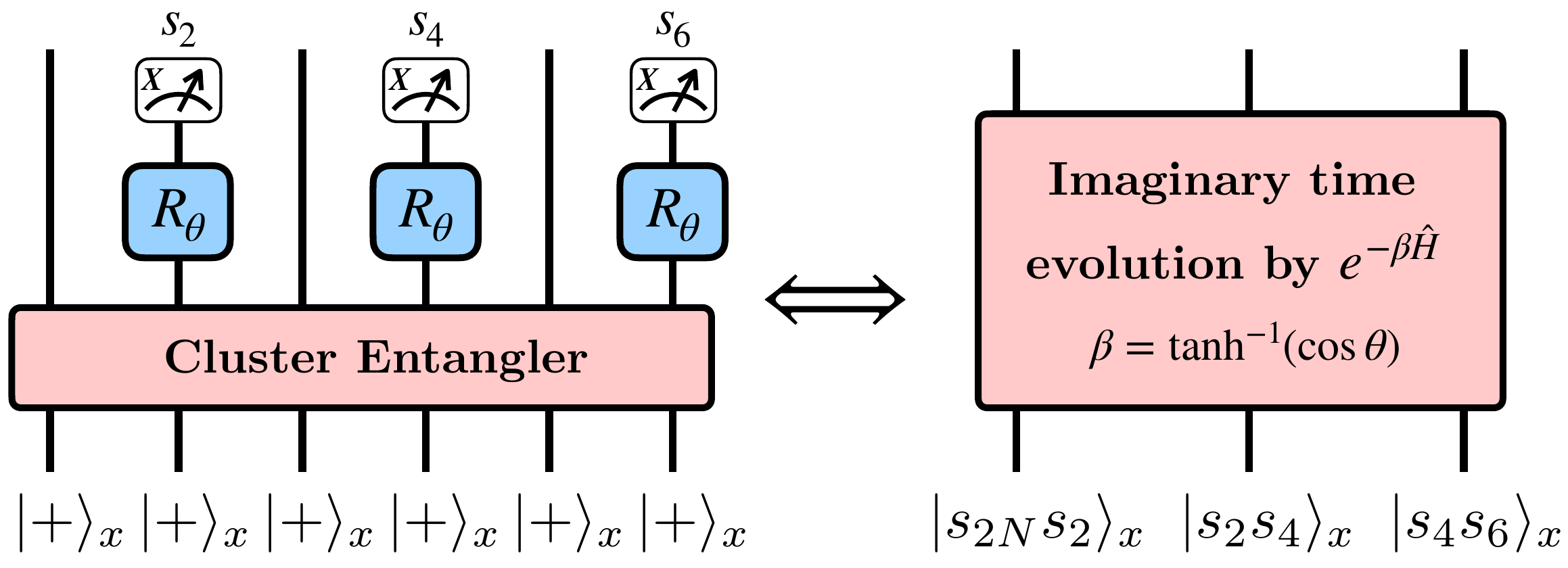}
    \caption{\label{fig:setting} Measurement-based quantum state preparation with the measurement angle $\theta$ away from the $X$-basis. The cluster entangler is a product of controlled-$Z$ gates between all neighboring qubits in a given geometry. Here, we measure the operator $O_\theta = X \cos \theta + Z \sin \theta$ on a subset of sites, whose outcome is denoted by $s_i$. Our state preparation is equivalent to the imaginary time evolution of a specific product state in $X$-basis by some Hamiltonian $\hat{H}$ for $\beta = \tanh^{-1}(\cos \theta)$. With this mapping, at $\theta = 0$ ($\beta \rightarrow \infty$), it is straightforward that we will prepare the ground state of $\hat{H}$. At intermediate $\beta$ ($\theta > 0$), we may find an interesting critical point.   }
\end{figure}

\begin{figure*}[!t]
    \centering
    \includegraphics[width = 0.97 \textwidth]{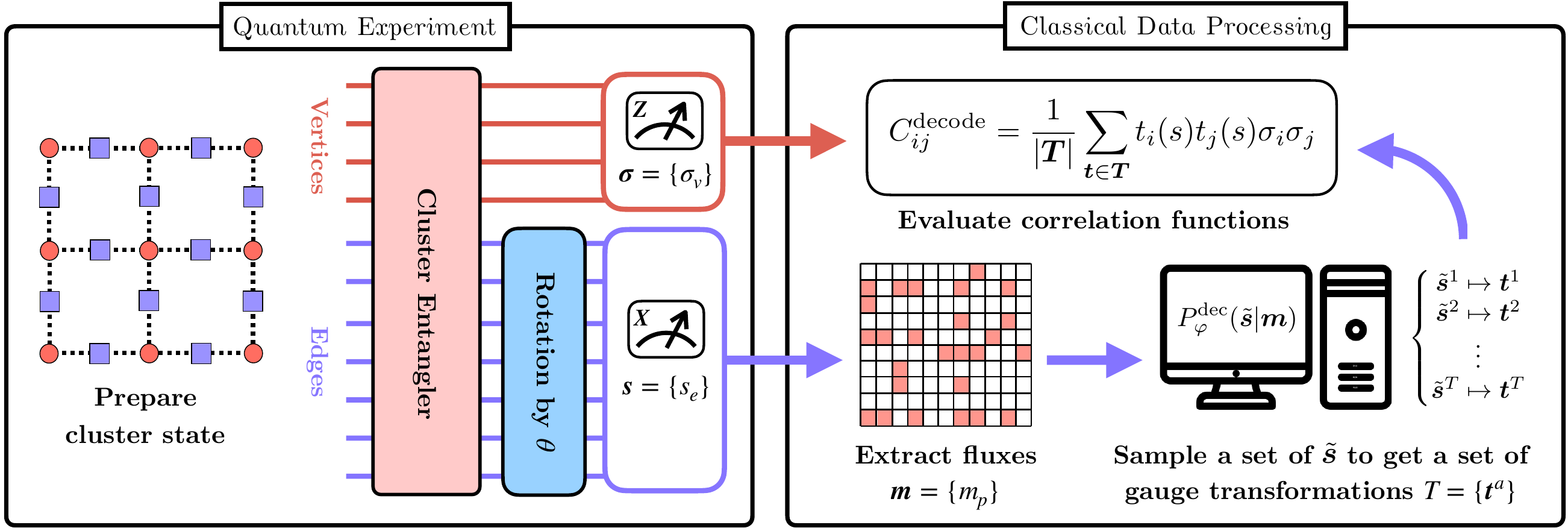}
    \caption{\label{fig:cao} Schematic diagram to prepare and decode a 2D GHZ state by measurements, where basis rotation by $\theta$ is applied for the edges. The diagram divides in two parts: quantum experiment and classical data processing. In the first part, a quantum experiment is performed to obtain measurement outcomes on edges ($X$-basis) and vertices ($Z$-basis) denoted by $(\bs, \bsigma)$. In the second part, a classical data processing is performed to \emph{decode} a given experimental outcome $(\bs, \bsigma)$. Here, the set of gauge transformations $\bm{T} = \{ \bt^a \}$ is generated based on the decoder denoted by $P^\textrm{dec}_\varphi(\tilde{\bs}|\bm)$ (See \secref{sec:Random} for details). Without classical data processing of $\bsigma$ based on the outcomes on edges $\bs$, the GHZ-ness of the quantum state on vertices (after measuring edges) is not identifiable.  }
\end{figure*}

Practically, many of the measurement-based state preparation require a precise control of the measured operators. In addition, classical data processing and quantum feedback based on the measurement outcomes are necessary to transform the resulting quantum state into the desired one. For instance, to generate the GHZ and toric code states from the cluster states in $1d$ and $2d$, single-spin measurements of the Pauli $X$ operators on the sites are needed in the paradigmatic example~\cite{1Dcluster_GHZ, 2Dcluster, 2Dcluster_toric, Piroli2021}, followed by a series of spin-flips based on the measurement outcomes. However, to adopt the measurement scheme to prepare these quantum states in experiments, it is important to understand whether measurements deviating from the $X$-axis can still produce a quantum state with the same entanglement properties, since noise in controlling the measurement angle could be present in experiments.
Thereby motivated, in this work, we explore the effects of general single-site measurements on the cluster states, namely measurements of single spins along an arbitrary direction, which will shed light on the stability of many measurement-based state preparation schemes.

With the general single-qubit measurements, a natural and perhaps more exciting question to ask is whether we can tune the resulting state through a phase transition, i.e. access certain quantum critical states, by tuning the measurement directions.  
The primary result of this paper is that a family of quantum critical states at so-called conformal quantum critical points (CQCP)~\cite{z2CFT1, z2CFT2} can be realized by measuring a subset of spins on a cluster state in a direction rotated away from the $X$-axis by a special angle $\theta_c$. We prove that the series of quantum states labeled by the measurement angle $\theta$ in $d$ space dimension has wavefunction amplitudes given by the Boltzmann weights of the corresponding classical spin model in $d$-dimension at inverse temperature $\beta(\theta) = \tanh^{-1}(\cos \theta)$. In $d \geq 2$, certain quantum states can undergo a phase transition at a critical measurement angle $\theta_c$. At criticality, the spatial correlation functions of the state exhibit conformal invariance~\cite{z2CFT1}.

Strikingly, we show that such a critical quantum state can be obtained even without post-selection, by employing a classical decoding protocol. 
This contrasts to two recent progresses: $(i)$ if the single-qubit measurement is performed on a subsystem of a generic highly entangled quantum state, the remaining pure wavefunction exhibits no feature (random), characterized by completely uniform distribution in the Hilbert space~\cite{CotlerChoi, ChoiShaw2021}. ($ii$) for the random circuit dynamics with measurements, the resulting quantum state becomes maximally mixed state after averaging over measurement outcomes, although post-selection can reveal interesting features~\cite{YaodongFisher2018,SkinnerNahum2018, Chan2019,YaodongFisher2019, Vasseur2019, XiangyuLuca2019, Gullans2020, Soonwon2020, TangZhu2020, JianLudwig2020, LopezVasseur2020, Bao2020, Rossini2020, Fan2021, Yaodong2021, BenZion2020}.
On the other hand, although our setup (\figref{fig:setting}) involves extensive number of measurements, we do not require post-selection to identify nontrivial correlation structure in the resulting states.  
This facilitates the experimental preparation and detection without exponential scaling of the measurement complexity. 
Thus, our discovery provides a complementary viewpoint in essential ways to the critical steady states in monitored quantum dynamics either with both unitary evolution and measurements or with measurements only. The preparation of critical states in our study can be thought of as a shallow depth unitary circuit followed by  single-site measurements as in \figref{fig:setting}.

The rest of the paper is organized as follows. 

In \secref{sec:1D}, we provide a detailed analysis of the 1D cluster state under measurements in a rotated basis. By calculating correlation functions of the post-measurement state, we provide a rigorous understanding of the stability of long-range entanglement there. Furthermore, we establish that the post-measurement wavefunction amplitude is proportional to the Boltzmann weight of a classical 1D Ising model at finite temperature. 
More generally, the post-measurement state can be expressed as the product state in $X$-basis, further evolved by a certain quantum Hamiltonian in imaginary time. 
With this observation, we find a parent Hamiltonian for the measured state, parametrized by the measurement angle $\theta$. 

In \secref{sec:2D}, we generalize our idea and construction to the 2D cluster state defined on vertices and edges of a square lattice. By measuring edges in a rotated basis, we obtain the classical 2D Ising model as the classical Hamiltonian describing the post-measurement wavefunction amplitudes, while by measuring vertices, we obtain the 2D Ising gauge theory as the corresponding classical model. Interestingly, we find that the long-range entanglement structure of the GHZ state in the case of measuring edges is robust against a finite angle deviation from the $X$-axes. 
The transition of the post-measurement state from a symmetry breaking phase to a disordered phase is found at a finite measurement angle $\theta_c$ with associated critical behavior. 
On the other hand, the long-range entanglement structure of the toric code ground state is unstable as soon as we deviate from the $X$-axes. We show that the stability of these long-range entanglement structures can be understood from the phase transitions of the corresponding classical models.

In \secref{sec:3D}, we turn to three dimensions, where we construct two different cluster states, namely symmetry protected topological (SPT) phases associated with $\mathbb{Z}_2^{(0)} \times \mathbb{Z}_2^{(2)}$ and $\mathbb{Z}_2^{(1)} \times \mathbb{Z}_2^{(1)}$ symmetries. We show that the post-measurement state of the first case corresponds to the ordinary 3D Ising model and the 3D 2-form Ising model\footnote{Ising model and Ising gauge theory can be thought of as a model with  0-form and 1-form $\mathbb{Z}_2$ symmetries. A model where six spins at faces interact has a 2-form $\mathbb{Z}_2$ symmetry, which we call 2-form Ising model.} when measured on the vertices and edges respectively, while the state of the second case corresponds to the 3D Ising gauge theory. By mapping to the classical partition functions, we show that preparation of the 3D toric code topological order with 1-form symmetry breaking is stable under measurements away from the $X$-axes, while that of the 3D toric code with 2-form symmetry breaking is unstable. 

In \secref{sec:Random}, we discuss the critical issue of post-selection, focusing on  the 2D cluster state measured on edges as a representative example. We show that measurement outcomes are under a correlated probability distribution, which is a gauge-symmetrized version of the uncorrelated random bond Ising model (RBIM). Based on this, we establish the correspondence between the ensemble of quantum states prepared by measurements and the ensemble of the ferromagnetic RBIM along the special manifold called Nishimori line~\cite{Nishimori,Nishimori2}. Furthermore, although the density matrix after measurement on the edges has no order on vertices if edge outcomes are averaged over, we show that there is a classical \emph{decoding} protocol which allows us to efficiently obtain a long-range correlation (ferromagnetic ordering) in the vertices for every measurement outcome on the edges as illustrated in \figref{fig:cao}.
In other words, we show that the resulting quantum state exhibits a \emph{hidden} ordering that can be efficiently revealed through the proposed decoding protocol. 
Thereby, we establish the stability of detectable long-range entanglement and correlations for generic post-measurement cluster states away from the $X$-basis, avoiding post selection. Based on the proposed decoding protocol, we outline an experimental blueprint to detect the long-range ordered phases and their transitions,
without exponential overhead. We also briefly discuss the generalization of our state-preparation setup, where one is not confined to the Nishimori line.

In \secref{sec:generalization}, we discuss the general idea of interactive quantum phases. Furthermore, we address the issue of post-selection of other examples, demonstrating the presence of long-range entanglement structures which can be classically decoded in other measurement-prepared quantum states.  In particular, these examples correspond to the random bond Ising model or random plaquette Ising gauge theory along the Nishimori line~\cite{Nishimori,Nishimori2,NishimoriPoint}, a special manifold in the parameter space of these random interaction models.

In \secref{sec:moreSPT}, we provide more examples of measuring cluster states with subsystem symmetries and fracton physics. In \secref{sec:summary}, we conclude with a discussion on conformal quantum critical states and outlook.

\begin{figure}[!t]
    \centering
    \includegraphics[width = 0.43 \textwidth]{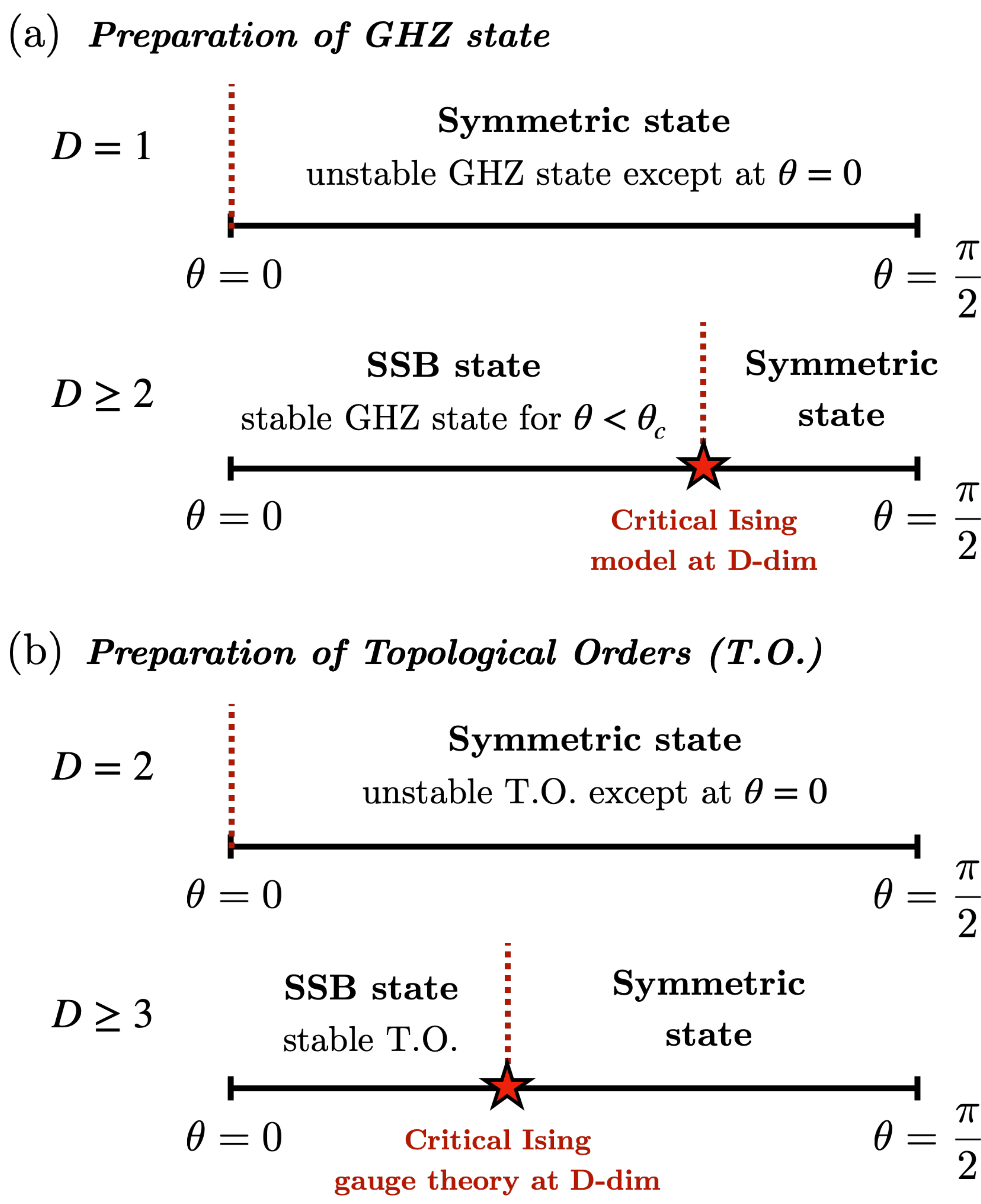}
    \caption{\label{fig:summary} Schematic phase diagrams for the post-measurement state with varying measurement angle $\theta$ away from the $X$-basis with post-selections for (a) GHZ state and (b) topological order (T.O.) preparations at $D$ spatial dimensions. The red dotted lines represent critical points. For the preparation of topological orders in $D = 3$, we consider a cluster state at $\mathbb{Z}_2^{(1)} \times \mathbb{Z}_2^{(1)}$ SPT phase. We note that preparing 3D toric code from $\mathbb{Z}_2^{(0)} \times \mathbb{Z}_2^{(2)}$ SPT is unstable if $\theta > 0$. The presence of 0-form or 1-form symmetry in the unmeasured lattice translates into the global spin-flip or local gauge symmetries in the corresponding classical model. Without post-selections, the stable regions get smaller with random bond universality classes for critical theories. }
\end{figure}

\section{Warm-up: measurements on 1D cluster states} \label{sec:1D}

As a warm-up, we start our discussion from measurements on 1D cluster states. Consider a chain of qubits which are in a 1D cluster state $\ket{\psi}$ stabilized by the following Hamiltonian: $H_\textrm{cluster}^\textrm{1D} = - \sum_n Z_{n-1} X_n Z_{n+1}$. 
The cluster state has interesting properties upon measurements, which can be understood from its nature as a decorated domain wall SPT~\cite{Chen2014}.
The Hamiltonian has $G = \mathbb{Z}_2 \times \mathbb{Z}_2$ symmetry where $g_1 = \prod_n X_{2n+1}$ and $g_2 = \prod_n X_{2n}$ are the two generators. Note that $Z_{2n-1} Z_{2n+1}$ measures the $g_1$ domain wall, while $X_{2n}$ measures the $g_2$ charge. Therefore, the ground state can be understood as a superposition of all possible $g_1$ domain wall configurations where a $g_2$ charge is attached to a $g_1$ domain wall. 
This special structure of the wavefunction indicates that measuring the $\mathbb{Z}_2$ charges on the even sites will specify the domain wall structure of the $g_1$ symmetry. If the measurement outcomes are $X_{2n}=1$, the resulting wavefunction will have no $g_1$ domain walls. Since the measurements on the even sites commute with the $\mathbb{Z}_2$ ($g_1$) symmetry defined on the odd sites, this measurement generates a GHZ state on the odd sites, i.e. $\ket{\uparrow \uparrow...}+\ket{\downarrow\downarrow...}$, a symmetric superposition of spontaneously symmetry-breaking (SSB) states of the $g_1$ symmetry. On the other hand, if we measure qubits on the even sites in the $Z$-basis and get all $+1$ outcomes, this proliferates the domain walls of the $g_1$ symmetry resulting in a disordered wavefunction on the odd sites, namely $\ket{++...}$.

\subsection{General single-site measurements}
Seeing that measuring the subset of qubits along the $X$-direction and $Z$-direction give us states with completely different characteristics, one may wonder what would happen if a general measurement is conducted along the axis rotated away from $X$-axis by an angle $\theta$. Will the resulting state have the same universal properties as the GHZ state for small $\theta$? Is there a transition from the GHZ state to the paramagnetic state at a certain angle? To answer these questions, let us consider the following projective measurement operator
of the $n^{th}$  qubit $P_n$ along the spin axis $\hat{n} \equiv (\Xfactor, \Zfactor \sin \phi, \Zfactor \cos \phi)$,
\begin{equation} \label{eq:projector}
    P_{s_n} = \frac{1}{2} \qty[I + s_n(X_n n_x  + Y_n n_y + Z_n n_z))],\quad P_{s_n}^2 = P^{\phantom{2}}_{s_n}
\end{equation}
where $s_n = \pm 1$ is the measurement outcome. 
For now, we assume $\phi=0$, which would not change the physics of primary interest. 
Let ${\cal P}_{\{s_{2n}\}} = \prod_{n} P_{{s_{2n}}}$ denote the projection operator for measurements outcomes $\{s\}$ on all the even sites. For simplicity, we denote it by $\cP_{\bm{s}}$. To characterize the resulting state $\cP_{\bm{s}} \hspace{-2pt} \ket{\psi} \equiv \ket{\cP_{\bm{s}} \psi}$, we will calculate correlation functions, in particular, $\langle Z_1Z_{2n+1}\rangle_{\cal P\psi}$. To that end, the following lemma is useful throughout the work: 

\smallskip
{\bf Lemma} Consider a $d$-dimensional stabilizer SPT state protected by symmetry groups  $G^{(n)}_1 \times G^{(d-n-1)}_2$, which have a mixed anomaly. Here the superscripts denote $n$ and $(d-n-1)$ form symmetries, while $G_1$ and $G_2$ act on qubits on two different sublattices. Then, the expectation value of an operator defined on a given sublattice is non-vanishing only if the operator is a symmetry action on that sublattice. 
\smallskip

The lemma can be understood quite intuitively. As the ground state is stabilized by local stabilizer terms, if a certain operator does not commute with all the stabilizers, its expectation value should vanish. However, if it commutes with all the stabilizers, it simply means that the operator is nothing but a symmetry of the given stabilizer Hamiltonian. 
For example, for the 1D cluster state $\ket{\psi}$, for any operator $O$ defined on the even sublattice, $\bra{\psi} O \ket{\psi}$ vanish unless $O$ is the identity or $\prod_n X_{2n}$ (See \appref{app:lemma}). 

With this lemma, we can show that the correlation function in the measured state with outcomes $\{s_{2n} \}$ has the following form
\begin{align} \label{eq:1dcorr}
    & \expval{Z_1 Z_{2n+1}}_{\cP_{\bm{s}} \psi} = \frac{  \bra{\cP_{\bm{s}} \psi}  Z_{1} Z_{2n+1} \ket{\cP_{\bm{s}} \psi}  }{ \bra{\cP_{\bm{s}} \psi}  \ket{\cP_{\bm{s}} \psi} } \nonumber \\
    &=  \frac{\qty[ \prod_{m=1}^n s_{2m}(\Xfactor)^n + \prod_{m=n+1}^N s_{2m} (\Xfactor)^{N-n} ]}{\qty[ 1 + \prod_{m=1}^N s_{2m} (\Xfactor)^{N} ]} \nonumber \\
    & \xrightarrow{N\rightarrow \infty} \qty(\prod_{m=1}^n s_{2m}) e^{-n/\xi},\ \ \text{with}\ \xi=|\ln \cos \theta|^{-1}.
\end{align}
Here $N$ (even) is the number of unmeasured sites and we have assumed periodic boundary conditions. In this derivation, we employed both the lemma and the equality; $Z_1 Z_{2n+1}| \psi \rangle = \prod_{m=1}^n X_{2m} | \psi \rangle$. The correlation length only depends on $\theta$ characterizing the deviation of the measurement angle from the $\hat{x}$-axis. This means that the long-range correlation of the GHZ state disappears at any finite $\theta$ in the thermodynamic limit. Practically, one would get an approximate GHZ state for system size smaller than the length scale $\xi(\theta)$. 

\subsection{Connection to Classical Partition Function} 

A keen reader may have noticed that \eqnref{eq:1dcorr} closely resembles the low temperature series expansion for the 1D classical Ising model. 
Based on this observation, one can show that the norm of the post-measurement wavefunction is proportional to the partition function of a classical Ising model at the inverse temperature $\beta \equiv \tanh^{-1}(\cos \theta)$, where the actual amplitude can have an additional complex phase factor (See \appref{app:1dIsing} for more details):
\begin{align} \label{eq:wavefunction}
    &\ket{\cP_{\bm{s}} \psi} =  \sum_{\{ \cC \}} w(\cC) \ket{ \cC  } \otimes \ket{M},\quad \left|w(\cC)\right|^2 \propto  \frac{1}{Z} e^{- \beta H(\cC)}
\end{align}
where $\cC = \{ \sigma \}$ denotes the spin configuration (in the computational basis) on the odd sites,  $H[\{ s \}]\{\sigma \}) = -  \sum_n s_{2n} \sigma_{2n-1} \sigma_{2n+1}$ is the Ising Hamiltonian with bond signs determined by the measurement outcomes $\{ s_{2n} \}$, $\ket{M}$ is the measured state on the even sites, and $Z \equiv \sum_{\cC}e^{- \beta H(\cC)}$ is the thermal partition function of the corresponding spin model $H[\{s\}]$. 
Therefore, the stability of the long-range entanglement of the post-measurement 1D cluster state can be understood in terms of the 1D classical Ising model at the inverse temperature $\beta(\theta)$. In 1D, one can always make a variable change on the spins ($\sigma_i \rightarrow t_i \sigma_i$ and $s_{ij} \rightarrow t_i t_j s_{ij}$ with $t_i,t_j = \pm1$), which we refer to as a gauge transformation in the rest of the paper, to make the 1D Ising model ferromagnetic, except possibly for the last bond if $\prod_{m=1}^{N} s_{2m} = -1$. 
It is well-known that the 1D Ising model is disordered at any finite $T>0$. Therefore, we conclude that the long-range entanglement is unstable at $\theta \neq 0$, which is the same conclusion as the one made from the explicit calculation above.

\subsection{The post-measurement wavefunction}

In this section, we directly derive the wavefunction after measurements from the decorated domain-wall construction~\cite{Chen2014}. The result shows an intriguing structure for the wavefunction on the unmeasured odd sites: 
\begin{align} \label{eq:1d_true_wavefunction}
    \ket{\cP_{\bm{s}} \psi}_\textrm{odd} &= \frac{1}{\sqrt{Z_0}} e^{-\frac{\beta}{2} \hat{H}} \Big[\otimes_{n=1}^N \ket{ X_{2n+1} = s_{2n} s_{2n+2} } \Big] \nonumber \\
    &= \frac{1}{\sqrt{2^N \cdot Z_0}} \sum_{ \{ \sigma \} }  e^{-\frac{\beta}{2} \hat{H}[\bs](\{ \sigma \})} \ket{ \{ \sigma \} } \nonumber \\
    \hat{H}[\bs] &\equiv -\sum_n s_{2n} Z_{2n-1}Z_{2n+1}, \,\,\, \tanh \beta = \cos \theta \,\,
\end{align} 
where $Z_0 = (2 \cosh \beta)^{N_d}$ and $N_d$ is the number of measured qubits (domain walls). This is the imaginary time evolution of the product state $\otimes_{n=1}^N \ket{ X_{2n+1} = s_{2n} s_{2n+2} }$ by the Hamiltonian $\hat{H}$. This form of the wavefunction immediately implies that in the limit $\beta\rightarrow\infty$ as $\theta\rightarrow 0$, the system should relax into the ground state of $\hat{H}$.

Let us describe the derivation. The pre-measurement cluster state wavefunction is written as the equal weight superposition of all domain wall configurations with charges attached accordingly:
\begin{equation}
    \ket{\psi} = \frac{1}{\sqrt{2^{N-1}} } \sum_{\{ d_{2n} \}}  \ket{\{ d_{2n} \} }_\textrm{ddw}
\end{equation}
where $\{ d_{2n}=\pm 1 \}$, and $d_{2m}=-1$ denotes a domain wall between sites $2m-1$ and $2m+1$.  Here the subscript ddw stands for the state that is a decorated domain wall basis, where domains (charges) are defined on odd (even) sites. The summation is over $2^{N-1}$ configurations since with periodic boundary conditions the domain walls are under the constraint $\prod_{n=1}^{N} d_{2n} = 1$. Here,
\begin{align} \label{eq:odd_basis}
    \ket{\{ d  \} }_\textrm{ddw} & \equiv  \ket{ \{\sigma_{\{d \}}\} }_\textrm{odd} \otimes  \ket{ \{ d\} }_\textrm{even} \nonumber \\
    \ket{ \{\sigma_{\{d \}}\} }_\textrm{odd} &\equiv \frac{1}{\sqrt{2}} \Bigg[\sum_{\xi = \pm 1} \otimes_{n=1}^N \Big| Z_{2n-1} = \xi \prod_{m=1}^n  d_{2m} \Big\rangle \Bigg] \nonumber \\
    \ket{ \{ d\} }_\textrm{even} &\equiv  \otimes_{n=1}^N \Big| X_{2n} = d_{2n} \Big\rangle
\end{align}
In this definition, the state labeled by $\ket{\{d_{2n}\}}_\textrm{ddw}$ is the cat state of two different spin configurations giving the same domain-wall configuration.
For example, the state with no domain wall, namely $\ket{\{d_{2n} = 1\}}_\textrm{ddw}$, would be the GHZ state on odd sites, and accordingly, all $\ket{+}$ states on even sites:
\begin{equation}
    \frac{1}{\sqrt{2}} \Big[ \ket{\uparrow \uparrow \cdots} + \ket{\downarrow \downarrow \cdots} \Big]_\textrm{odd} \otimes \Big[ \ket{+}^{\otimes N} \Big]_\textrm{even}.
\end{equation}

With an explicit representation in hand for $\ket{\psi}$, we now   want to obtain the amplitude of the post measurement wavefunction, $\cP_{\bm{s}} \ket{\psi}$, which can be written as  
\begin{equation}
    \ket{\cP_{\bm{s}} \psi} =  \sum_{\{ d_{2n} \}} C(\{ d_{2n} \}) \ket{ \{\sigma_{\{d \}}\} } \otimes \ket{M}   
\end{equation}
where now $\ket{M} = \otimes_{n=1}^N \ket{M_{s_{2n} }}$ stands for the measured component on even sites. 
To obtain $C(\{ d\})$, we first decompose the kets $\ket{\pm}  = [1,\pm 1]^T/\sqrt{2}$ into the measurement basis: 
\begin{equation}
    \ket{\pm} = a_\pm \ket{M_+} +  b_\pm \ket{M_-},
\end{equation}
with 
\begin{equation}
    \ket{M_{\pm}} = \frac{1}{\sqrt{2(1 \pm  \sin \theta)}}\mqty( \sin (\theta) \pm 1 \\ \cos(\theta) )
\end{equation}
satisfying 
$O_\theta |M_\pm \rangle = \pm | M_{\pm} \rangle$ with measurement operator $O_\theta = X\cos \theta + Z \sin \theta$ (assuming $\phi=0$).  
The coefficients follow from
$a_\pm = \braket{M_+}{\pm}$ and $b_\pm = \braket{M_-}{\pm}$. 

Then, note that for a measurement outcome $\{ s_{2n} \}$, the projection is defined as 
\begin{equation}
    \cP_{\bm{s}} \mapsto \otimes_{n=1}^N \ket{M_{s_{2n}}} \bra{M_{s_{2n}}}.
\end{equation}
The coefficient can then be obtained by
\begin{align}
        C(\{ d_{2n} \})   = \frac{\bra{M} \otimes_{n=1}^N \ket{d_{2n}}}{\sqrt{2^{N-1}}}  = \frac{\prod_{n=1}^N \braket{M_{s_{2n}}}{d_{2n}} }{\sqrt{2^{N-1}}},
\end{align}
where (see \appref{app:wavefunction})
\begin{align} \label{eq:amplitude}
    \braket{M_{s_{2n}}}{d_{2n}}  &= \varphi_{2n} \sqrt{ \frac{ (1 + s_{2n} d_{2n} \cos \theta ) }{2 } } \nonumber \\
    & = \varphi_{2n}  e^{\frac{\beta}{2} s_{2n} d_{2n}} / \sqrt{2 \cosh \beta}.
\end{align}
with $\varphi_{2n} \equiv (-1)^{(1-s_{2n})(1-d_{2n})/4} = d_{2n}^{(1-s_{2n})/2}$. Using $d_{2n} = \sigma_{2n-1} \sigma_{2n+1}$, we have (at $\phi=0$)
\begin{equation}
    \ket{\cP_{\bm{s}} \psi}_\textrm{odd} = \sum_{\{d \} } \frac{e^{\frac{\beta}{2} \sum_n s_{2n} Z_{2n-1} Z_{2n+1}}}{Z_0^{1/2}}  \frac{\prod_n \varphi_{2n} }{\sqrt{2^{N-1}}}\ket{ \{\sigma_{\{d \}}\} },
\end{equation}
where $Z_0 = (2 \cosh \beta)^{N}$. The last factor of the above wavefunction can be simplified as
\begin{align} \label{eq:time_evolve_begining}
    &\sum_{\{d \} } \frac{1}{\sqrt{2^{N-1}} }\Big( \prod_n (\sigma_{2n-1}\sigma_{2n+1})_{2n}^{(1-s_{2n})/2} \Big) \ket{ \{\sigma_{\{d \}}\} } \nonumber \\
    & = \prod_n (Z_{2n-1}Z_{2n+1})^{(1-s_{2n})/2} \Big[ \otimes_{n=1}^N \ket{+}_{2n-1} \Big] \nonumber \\
    & = \prod_n Z_{2n-1}^{[(1-s_{2n-2})+(1-s_{2n})]/2} \Big[ \otimes_{n=1}^N \ket{+}_{2n-1} \Big] \nonumber \\
    & = \otimes_{n=1}^N \ket{ X_{2n-1} = s_{2n-2} s_{2n} }
\end{align}

Putting everything together, we finally obtain the wavefunction in \eqnref{eq:1d_true_wavefunction}. This expression gives consistent results for the norm $\braket{\cP_{\bm{s}} \psi}$. For $\phi \neq 0$, we can obtain non-trivial complex phase factors as detailed in the \appref{app:wavefunction}. Although these phase factors can affect the expectation values when we measure correlations of $Y$ or $X$ operators, they do not change any physics in the $Z$ correlations.

There are two solvable limits: $\theta = 0$ and $\theta = \pi/2$. At $\theta=0$, the result simply implies that unless $s_{2n} = d_{2n}$, the wavefunction component is zero. Therefore, the only surviving component would be $\ket{\{ \sigma_{\{s\}} \} }$, whose explicit form is defined in \eqnref{eq:odd_basis}. 
At $\theta=\pi/2$, the wavefunction amplitudes become uniform with phase factors $\varphi_{2n}$ depending on the measurement comes. For $s_{2n} = 1$, it gives $\ket{+}^{\otimes N}$, and for $s_{2n} = (-1)^n$, it gives $\ket{-}^{\otimes N}$. 
This aligns with the expectation from the stabilizer correlation $Z_{2n}X_{2n+1}Z_{2n+2} | \psi 
\rangle   = | \psi \rangle $.

We remark that the derivation here is completely general for any wavefunction constructed by a decorated domain wall method. This is because decorated domain wall method completely specifies the relation between measured and unmeasured sites in a simple manner. In a higher dimensional case, we can show that the measurement-projected amplitude in \eqnref{eq:amplitude} can be calculated in terms of domain wall variable, which can be converted into the operator action on unmeasured sites. The expression for the product state to be time-evolved in \eqnref{eq:time_evolve_begining} can also be derived in a similar manner, where the state in $X$-basis would be given by the product of measurement outcomes neighboring an unmeasured site.


\subsection{Parent Hamiltonian for the measured states}\label{subsec:1dparent}

Interestingly, we find that the family of states $\ket{\cP_{\bm{s}} \psi}$ on the odd sites with post-selection $s_{2n} = 1$ is the ground state of the following Hamiltonian $H= \sum_n H_n$:
\begin{align} \label{eq:1dparent}
    H_n &=  - \Big[ X_{2n-1} -  \cos^2 \theta  Z_{2n-3} X_{2n-1} Z_{2n+1} \nonumber \\
    & + \cos \theta  \qty( Z_{2n-3} Z_{2n-1} +  Z_{2n-1} Z_{2n+1}) \Big].
\end{align}
We derive the Hamiltonian following the Witten conjugation method~\cite{WITTEN1982, Wouters2021, pivot}. It is a simple procedure that later allows us to generate the parent Hamiltonians for our post-measurement states in various settings. The crucial premise is that $\ket{\cP_{\bm{s}} \psi}$ is given by the imaginary time evolution of a certain product state~\eqnref{eq:1d_true_wavefunction} for the inverse temperature $\beta = \beta(\theta)$. 
This is to say, at $\beta=0$, the state is the ground state of $\hat{H}_0 = - \sum_{n} s_{2n} s_{2n+1} X_{2n+1}$, and at $\beta>0$, the state can be thought of as the evolution of the ground state under the non-unitary operator $M_\beta = e^{ \frac{\beta}{2} \sum_{n} s_{2n} Z_{2n-1} Z_{2n+1} }$. 

To proceed further, let us perform a Kramers-Wannier duality, where we can write $X_{2n-1} \rightarrow X_{2n-2}' X_{2n}'$ and $Z_{2n-1} Z_{2n+1} \rightarrow Z_{2n}'$. The state becomes 
\begin{align}
    &|\mathcal{P}\psi\rangle\propto M'_\beta |\Psi_0\rangle', \qquad M'_\beta = \prod_n e^{\frac{\beta}{2}s_{2n}Z_{2n}'}\nonumber \\
    &|\Psi_0\rangle'=|\{ X'_{2n}X'_{2n+2}=s_{2n}s_{2n+2},\text{ for all }n\}\rangle'.
\end{align}
The Hamiltonian for $|\Psi_0\rangle'$ is
\begin{align}
    H_0'&=\sum_n \frac{1}{2}\left(1-s_{2n}s_{2n+2}X_{2n}'X_{2n+2}'\right) \nonumber \\
    &=\sum_{n}\Gamma'^\dagger (n) \Gamma' (n) \nonumber \\
    &\Gamma' (n)=\frac{1}{2}\left(s_{2n}X_{2n}'-s_{2n+2}X_{2n+2}'\right).
\end{align}
where we intentionally write $H_0'$ in positive semi-definite structure. Then it follows that one choice of the Hamiltonian for $|\mathcal{P}\psi\rangle'$ is 
\begin{align}
    H'&=\sum_n H_n' = \sum_{n}{\Gamma'_\beta}^\dagger (n){\Gamma_\beta}'(n)   \nonumber \\
    \Gamma_\beta'(n)&=M_\beta'\Gamma'(n)M_\beta'^{-1}
\end{align}
As $e^{-\beta Z} = \cosh \beta(1 - Z \tanh \beta) = \cosh \beta( 1 - Z \cos \theta)$, up to an overall constant prefactor of $\cosh \beta(\theta)$, 
\begin{align}
    \Gamma_\beta'(n)&\propto(s_{2n}X_{2n}'-s_{2n+2}X_{2n+2}') \nonumber \\
    & \quad - i \cos\theta (Y_{2n}'-Y_{2n+2}')
\end{align}
Thus, $H_n'$ reads
\begin{align}\label{eq:1dparentKW}
    H'_n=&-s_{2n}s_{2n+2}X_{2n}'X_{2n+2}'-\cos^2\theta Y_{2n}'Y_{2n+2}'\nonumber \\
    &- \cos\theta (s_{2n}Z_{2n}'+s_{2n+2}Z_{2n+2}')+\textrm{const}.
\end{align}
As $H_n'$ is a positive semi-definite Hamiltonian that annihilates $\ket{\cP_{\bm{s}} \psi}$, $\sum_n H_n'$ is a valid parent Hamiltonian whose ground state is $\ket{\cP_{\bm{s}} \psi}$.
Once we reverse the Kramers-Wannier duality, we obtain  
\begin{align}
    H_n &= -  s_{2n} s_{2n+2} X_{2n+1} + \cos^2\theta Z_{2n-1} X_{2n+1} Z_{2n+3} \nonumber \\
    &- \cos\theta (s_{2n} Z_{2n-1} Z_{2n+1} + s_{2n+2} Z_{2n+1} Z_{2n+3}) 
\end{align}
For the measurement outcome $s_{2n}=1$ for all $n$, we obtain the Hamiltonian \eqnref{eq:1dparent}. Furthermore, note that applying $X$-gate (basis flip) to a set of sites is equivalent to flipping $\{ s \}$ that are emanating from the set of sites. In fact, if $\prod s_{2n} = 1$, we can always find a basis where the model is entirely ferromagnetic. Even when $\prod s_{2n} = -1$, we can find a basis where only a single bond is antiferromagnetic and the other bonds are ferromagnetic.

\begin{figure}[!t]
    \centering
    \includegraphics[width = 0.34 \textwidth]{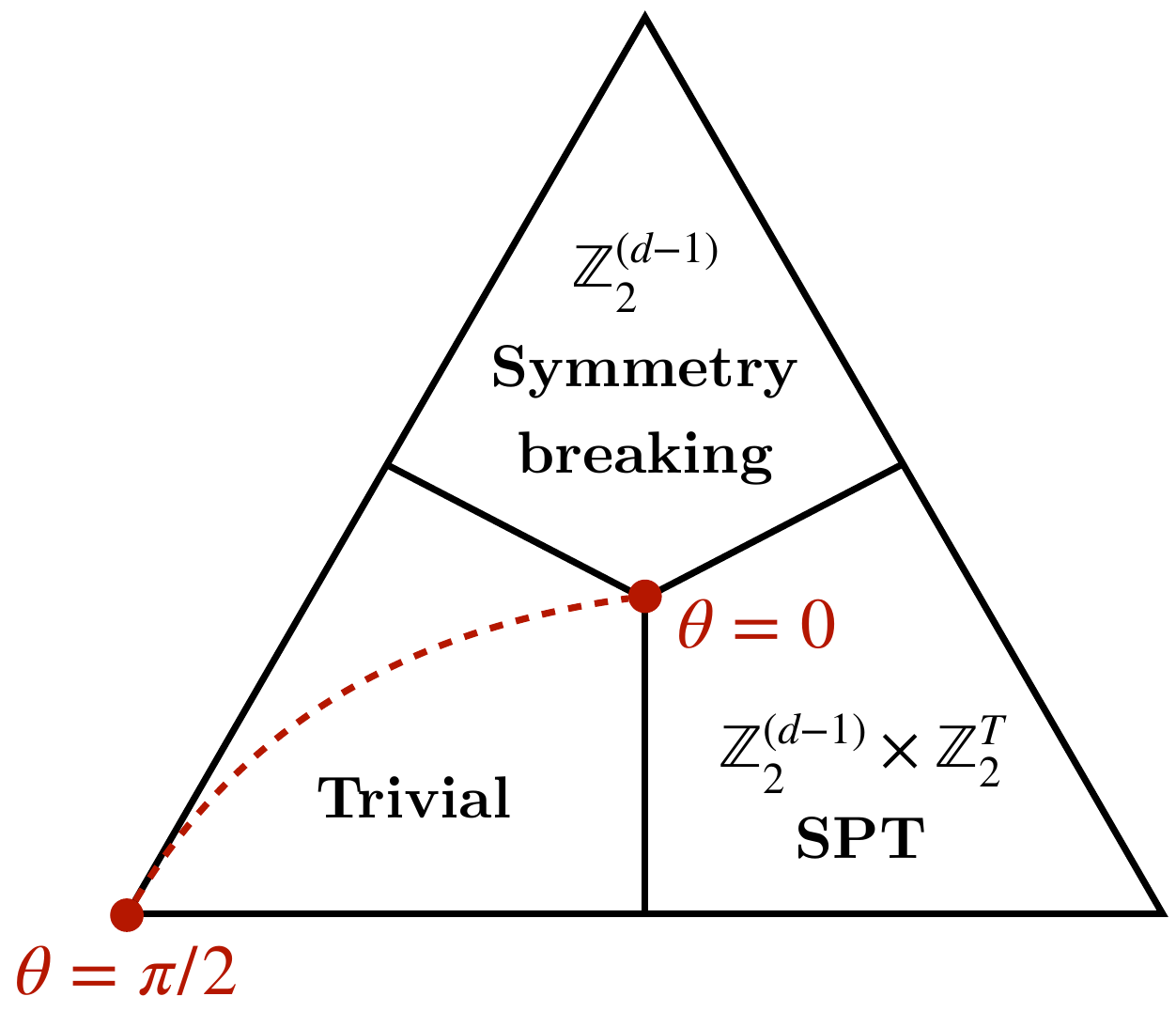}
    \caption{\label{fig:phase_diagram} Adaptation from \cite{pivot}. Schematic phase diagram for the parent Hamiltonian of the $d$-dimensional cluster state (defined on vertices and edges of the cubic lattice) measured on vertices. In $d=1$, $\mathbb{Z}_2^{(0)}$ symmetry breaking simply corresponds to the ferromagnet, while in $d>1$, $\mathbb{Z}_2^{(d-1)}$ symmetry breaking corresponds to a topological order. Red dotted line is a trajectory that can be prepared by our measurement scheme. Note that at $\theta=0$, the ground state at the multicritical point coincides with the ground state at the fixed point of $\mathbb{Z}_2^{(d-1)}$ SSB phase. In this phase, since $\mathbb{Z}_2^{(d-1)}$ symmetry defect is given by lines, the groundstate must be written as all fluctuating configurations of $Z=-1$ loops in computational basis. As a result, it already saturates the maximum area law entanglement capacity, and if we can prepare the state, it has to be located at the critical point with a special property.    }
\end{figure}

The spectrum of this Hamiltonian is known to be gapless at $\theta=0$~\cite{parentHam}, which is a multicritical point neighboring the paramagnet, ferromagnet ($\mathbb{Z}_2^{(0)}$ SSB), and $\mathbb{Z}_2^{(0)} \times \mathbb{Z}_2^{\cal T}$ SPT (i.e., the cluster state)~\cite{Chen2013} as illustrated in \figref{fig:phase_diagram}. The gaplessness at $\theta=0$ is easily seen from \eqnref{eq:1dparentKW}, where the terms are nothing but the XY model with perpendicular magnetic field~\cite{pivot} (which can be written in terms of free fermions under a Jordan-Wigner transformation).
Although the Hamiltonian is gapless, the ground state entanglement entropy does not diverge; this can be directly inferred from the fact that the initial 1D cluster state is described by a matrix product state (MPS) with bond dimension $\chi=2$, and the measurement projection cannot change the MPS structure.
This seemingly inconsistent behavior can be resolved by realizing that the criticality is not captured by a 2D conformal field theory but by a critical theory of free fermions with dynamic critical exponent $z = 2$~\cite{PEPS_cirac}.
In fact, the ground state trajectory of \eqnref{eq:1dparent} is a paradigmatic example of a 1D phase transition that can be expressed by a simple MPS with bond dimension $\chi=2$~\cite{gaplessMPS}.

\section{2D cluster states} \label{sec:2D}

\begin{figure}[!t]
    \centering
    \includegraphics[width = 0.48 \textwidth]{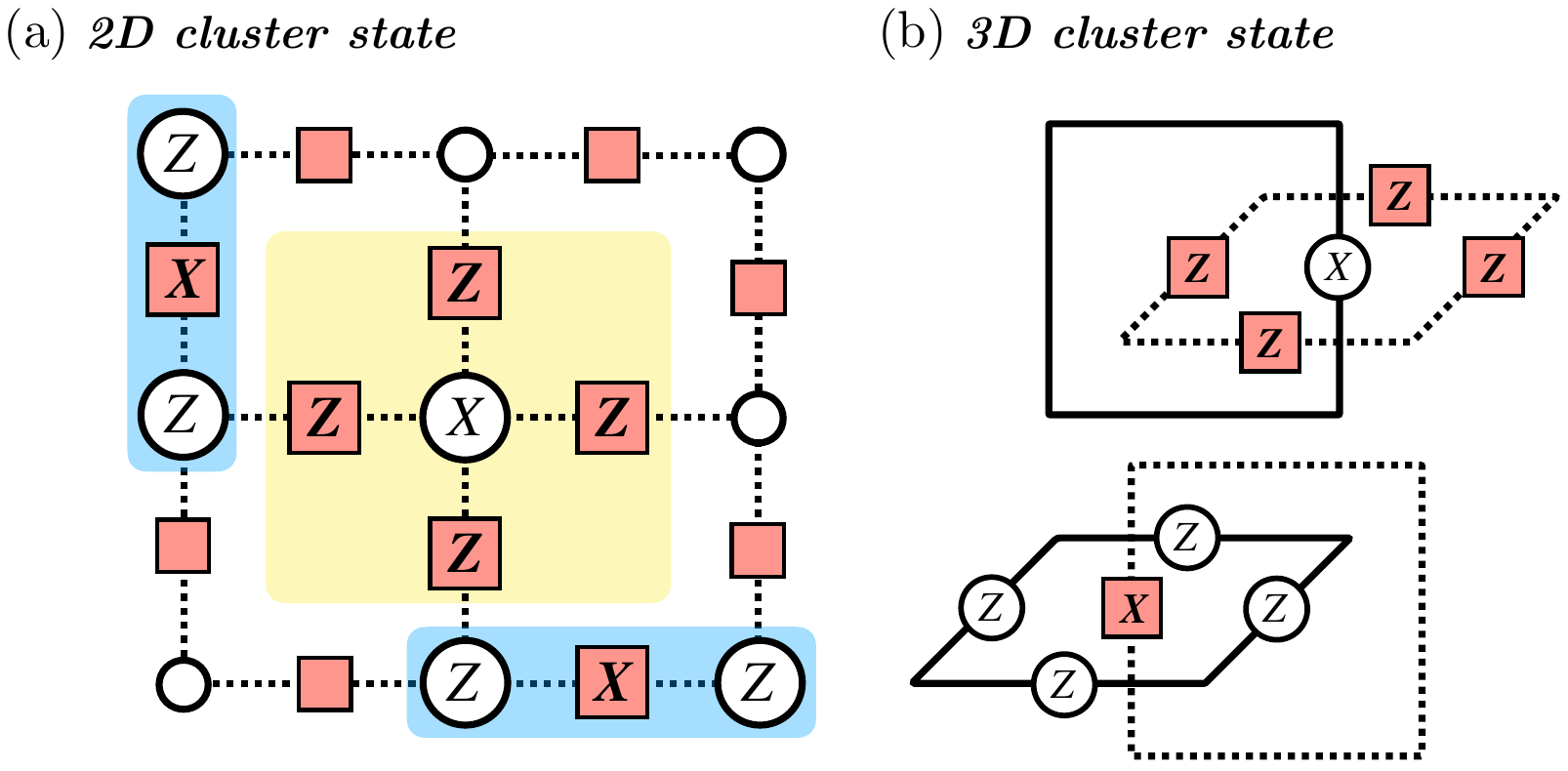}
    \caption{\label{fig:2d3d} Schematic visualization of (a) 2D and (b) 3D cluster state models. In 2D, we have qubits on edges and vertices. In 3D, we have qubits on edges of the cubic lattice and its dual lattice, where we have mixed anomaly between two 1-form symmetries. }
\end{figure}

Now that we have a thorough understanding of the post-measurement states in one dimension, we turn next to two spatial dimensions, where higher form symmetries~\cite{Kapustin2014} become important. Consider the 2D cluster state Hamiltonian where qubits reside at vertices and edges of 2d square lattice as illustrated in \figref{fig:2d3d}(a):
\begin{equation} \label{eq:2dclusterHam}
    H = - \sum_{v} \qty( X_v \prod_{e \ni v } \bm{Z}_e ) - \sum_{e} \qty( \bm{X}_e \prod_{v \in e} Z_v ).
\end{equation}
Bold symbols $\bm{Z}$ and $\bm{X}$ act on edges, and unbold symbols $Z$ and $X$ act on vertices. 
Here, all terms in the Hamiltonian commute with one another, and the groundstate satisfies each term to be 1. This implies that $B_p \equiv \prod_{e \in p} \bm{X}_e=1$ for any plaquette $p$. There are two symmetries in this Hamiltonian: 
\begin{align} \label{symm01}
    \textrm{$\mathbb{Z}_2^{(0)}$ 0-form: } & g = \prod_{v} X_v \nonumber \\
    \textrm{$\mathbb{Z}_2^{(1)}$ 1-form: } & h_\gamma =  \prod_{e \in \gamma} \bm{X}_e
\end{align}
where $\gamma$ is any closed loop along the bonds. Again, note that the ground state of \eqnref{eq:2dclusterHam} has the decorated domain wall (defect) structure: the creation of a pair of 1-form charged objects by $\prod_{e \ni v} \bm{Z}_e$ is accompanied with the creation of 0-form domain walls by $X_v$; also, the creation of 1-form domain walls by $\bm{X}_e$ is accompanied with the creation of a pair of 0-form charges by $Z_v Z_{v'}$.

For the 2D cluster states, one can choose to measure the spins either on the vertices or on the edges. The two measurement schemes exhibit qualitatively different physics as we will show below.

\subsection{Measurements on vertices: Ising gauge theory}

We apply the projective measurement in \eqnref{eq:projector} on every vertex spin. At $\theta = 0$ with all measurement outcomes being $+1$, we would expect the resulting state to satisfy the following constraint, $A_v \equiv \prod_{e \ni v} \bm{Z}_e=1$ and $B_p \equiv \prod_{e \in p} \bm{X}_e=1$, giving rise to a topological order of the 2D toric code model~\cite{kitaev2006}.
In this state, the operator $C^Z_{\gamma^*} \equiv \prod_{j \in \gamma^\perp} \bm{Z}_j$ defined on the contractible loop $\gamma^\perp = \rd S$ perpendicular to the edges has the expectation value $\langle C^Z_{\gamma^\perp} \rangle = 1$, which is a signature of the spontaneously broken 1-form symmetry as $C^Z_{\gamma^\perp}$ counts the 1-form symmetry defects enclosed by the loop. 
In a complementary point of view, we can embed the system into the torus geometry, and consider a logical qubit operator $C^Z_{\gamma^\perp}$ for $\gamma^\perp$ being a non-contractible loop along the cycle. Here, we can show that $\langle C^Z \rangle  = 0$. As $C^X \equiv \prod_{e \in \gamma} \bm{X}_e = 1$ for any non-contractible loop $\gamma$ due to the stabilizer structure and $\{ C^X, C^Z \} = 0$, the post-measurement state must have $\langle C^Z \rangle = 0$. This implies that the resulting quantum state is the symmetric superposition of four different configurations under $C^Z$ operators. However, we can show that the correlation function of two non-contractible loops separated by the distance $l$ is constant, i.e., the state develops a long-range order with a spontaneously broken 1-form symmetry.

We want to detect whether the topological order is robust when the measurements are moved away from $\theta=0$. To do so, we calculate the expectation value of $C^Z_{\rd S}$ defined on the boundary of a surface $S$. Using the same formalism, we can show that (See \appref{app:2dIsingGauge})
\begin{equation} \label{eq:2d_area}
    \langle C^Z_{\rd S}\rangle_{\cP_{\bm{s}} \psi}=\frac{\bra{\cP_{\bm{s}} \psi} C^Z_{\rd S} \ket{\cP_{\bm{s}} \psi}}{ \bra{\cP_{\bm{s}} \psi}  \ket{\cP_{\bm{s}} \psi}  } \sim (\Xfactor)^\abs{S}    
\end{equation}
where we employed both the lemma and the equality $\prod_{e \in \rd S} \bm{Z}_e = \prod_{v \in S} X_v$. The \emph{area law} of the loop expectation value, instead of the \emph{perimeter law}, indicates that 1-form symmetry is intact in the thermodynamic limit. 
Therefore, the $\ket{\cP_{\bm{s}} \psi}$ cannot be topologically ordered. A complementary fact in support of this observation is that the expectation value of $\prod_{e \in l} \bm{X}_l$, an operator that creates a pair of anyons on the two ends of a open string $l$, is non-vanishing, namely
\begin{equation}
    \big\langle \prod_{e \in l} \bm{X}_l \big\rangle_{\cP_{\bm{s}} \psi} =  \big\langle \prod_{v \in \rd l} Z_v \big\rangle_{\cP_{\bm{s}} \psi} \approx (\sin \theta)^2.
\end{equation}
Non-vanishing expectation value for any open string operator is the signature of the anyon condensation which gives a trivial symmetric phase for $\theta > 0$. However, we remark that even though the state is not the SSB of 1-form symmetry, \eqnref{eq:2d_area} gives a quantitative answer for how far it exhibits the correlation structure that can be approximated as being topologically ordered.

We observe that the above loop expectation value can be mapped to the area-law correlation function in the 2D classical Ising gauge theory at finite temperature, where the 1-form symmetry exactly maps to the local gauge symmetry. More precisely, the corresponding Ising gauge theory is defined on the edges of the dual lattice, and the local gauge transformation is defined as flipping the spins on the edges emanating from the set of dual vertices. Such a gauge transformation is equivalent to the 1-form symmetry action $\prod \bm{X}$ along the loop defined on the boundary of the set of dual sites. Similar to the 1D example in \eqnref{eq:1d_true_wavefunction}, the post-measurement wavefunction can be expressed in the following form:
\begin{align} \label{eq:2d_IGT_true_wavefunction}
    \ket{\cP_{\bm{s}} \psi} &\propto e^{-\frac{\beta}{2} \hat{H}} \Big[\otimes_{n=1}^N \big|  \bm{X}_e = \prod_{v \in e} s_v \big\rangle \Big] \nonumber \\
    \hat{H} &\equiv - \sum_{v}  s_v \prod_{e \ni v} \bm{Z}_e =  -   \sum_{\tilde{p}}  s_{\tilde{p}} \prod_{\tilde{e} \in \tilde{p} } \bm{Z}_{\tilde{e}},
\end{align}
where the tilde subscript is for the dual lattice label. Expanded in $\bm{Z}_{\tilde{e}}$ basis in the dual lattice, the above expression gives rise to the wavefunction amplitude given by the Boltzmann weight of the 2d Ising gauge theory (see \appref{app:2dIsingGauge}). 2d Ising gauge theory is exactly solvable and known to enter a trivial phase with area law loop expectation value at any finite temperature, which aligns with \eqnref{eq:2d_area}.

We remark that in \eqnref{eq:2d_IGT_true_wavefunction}, $\ket{\cP_{\bm{s}} \psi}$ remains the same if one replaces the imaginary time evolution $e^{-\beta \hat{H}/2}$ by $e^{-\beta \hat{H}_\textrm{toric}^\textrm{2D}}$ since $B_p = \prod_{e \in p} \bm{X}_e$ commutes with all other terms in $\hat{H}$ and acts trivially on the state. Therefore, at $\beta\rightarrow \infty$ ($\theta \rightarrow 0$), we expect the projected wavefunction to be the toric code ground state. Based on this observation, we can find a series of parent Hamiltonian which stabilizes the wavefunction in \eqnref{eq:2d_IGT_true_wavefunction} parametrized by $\beta(\theta)$. The parent Hamiltonian on the dual lattice reads~\cite{pivot} 
\begin{align}     \label{eq:2dIsgauge_parent}
    H &=  H_0 + \cos^2 \theta H_\textrm{SPT} +   4 \cos \theta   H_\textrm{toric}, \nonumber \\
    H_0 &= - \sum_{\tilde{e}} \Big[\prod_{\tilde{p} \ni \tilde{e}} s_{\tilde{p}} \Big] \bm{X}_{\tilde{e}} \nonumber \\
    H_\textrm{SPT} &=  \sum_{\tilde{e}} \bm{X}_{\tilde{e}} \prod_{ \tilde{e}' \in \tilde{n}(\tilde{e})} \bm{Z}_{\tilde{e}'} \nonumber \\
    H_\textrm{toric} &= - \sum_{\tilde{p}}  s_{\tilde{p}} \prod_{\tilde{e} \in \tilde{p} } \bm{Z}_{\tilde{e}}
\end{align}
Here $\tilde{n}(\tilde{e})$ is the set of neighboring edges that are boundaries of two dual plaquettes sandwiching $\tilde{e}$. The derivation is a direct generalization of the procedure described in \secref{subsec:1dparent}.

This parent Hamiltonian is gapless at $\theta = 0$. Similar to the 1D case, the gapless point is multicritical, neighboring a $\mathbb{Z}_2$ topological order, a $\mathbb{Z}^{(1)}_2 \times \mathbb{Z}_2^{\cal T}$ SPT, and a trivial paramagnetic phase as illustrated in \figref{fig:phase_diagram}. Interestingly, the ground state at $\theta=0$ gapless point is exactly the toric code ground state as our construction demonstrates. However, for any $\theta>0$, the Hamiltonian enters the trivial confined phase. At the multicritical point, the model has $\U(1)$ pivot symmetry~\cite{pivot} and the low energy effective theory can be shown to be dual to a dilute interacting Bose gas at zero density in $2d$~\cite{uzunov1981, BEC, fisher1989boson}. This critical theory is known to have exact dynamical exponent $z=2$ as the self energy correction from the boson self-interaction vanishes at every order of perturbation theory due to the absence of particles in the vacuum state~\cite{fisher1989boson}.

\subsection{Measurements on edges: Ising model} \label{sec:2dIsing}

Next, we apply the projective measurement in \eqnref{eq:projector} on every edge spin. At $\theta = 0$ with all measurement outcomes $s_e = 1$, we would expect the resulting state to have $Z^{\phantom{'}}_v Z_v' = 1$ for any edge $e=(v,v')$. Therefore, the post-measurement state would be the GHZ state. At $\theta > 0$, we can show that the two-point correlation $\expval{Z_i Z_j}_{\cP_{\bm{s}} \psi}$ is given by the two-point correlation function of a classical 2D Ising model at temperature $\beta(\theta)$ (See \appref{app:2dIsingModel}). Furthermore, the wavefunction on the unmeasured sites is expressed as:
\begin{align} \label{eq:2d_IT_true_wavefunction}
    \ket{\cP_{\bm{s}} \psi} &\propto e^{-\frac{\beta}{2} \hat{H}} \Big[\otimes_{n=1}^N \big| {X}_v = \prod_{e \in n(v)} s_e \big\rangle \Big] \nonumber \\
    \hat{H} &\equiv - \sum_{e = (v,v')} s_e Z_v Z_{v'}.
\end{align}

If $s_e = 1$ for all measurements, the physical properties of this wavefunction can be understood from the 2D classical ferromagnetic Ising model. 
Unlike the 1D case, the 2D Ising model has a finite temperature ordering transition which implies that long-range entanglement is robust for $\theta \neq 0$ and that we can prepare a quantum critical state at a specific measurement angle $\theta_c = \cos^{-1} \tanh(\beta_c)$. Since the 2D Ising model has an exact self-duality, we obtain that the transition happens when $\beta_c = \sqrt{2} - 1$, i.e. $\theta_c \sim 65^\circ$. Therefore, by measuring the 2D cluster state, we can  prepare the wavefunction which goes through the phase transition across this measurement angle. In particular, the ordered state will appear as the GHZ state of two SSB configurations, which is long-range entangled and robust up to a finite $\theta$ as illustrated in \figref{fig:summary}(a).

At $\theta = \theta_c$, we have a quantum phase transition, where the critical state has an area law entanglement since the pre-measurement cluster state is parameterized by 2D tensor-network state, called projected entangled pair states (PEPS),  with bond dimension $\chi = 2$~\cite{PEPS_cirac}. As its correlation functions in $Z$-basis are determined by the critical 2D Ising model, the wavefunction has a spatial conformal structure with power-law decaying correlation functions. This turns out to be a specific example of a conformal quantum critical point (CQCP)\cite{z2CFT1, z2CFT2}. For a CQCP with a known statistical weight, one can always construct a parent Hamiltonian for the critical state, which is the generalization of the RK Hamiltonian~\cite{Henley2004}. Such a parent (quantum) Hamiltonian will have a dynamic critical exponent, $z$, to be equal to the dynamic exponent for relaxational critical dynamics for the corresponding classical statistical model. Thereby, we can compute the dynamic critical exponent for a given CQCP Hamiltonian. For CQCPs with a $\U(1)$ symmetry, the critical theory is analytically known to be a quantum Lifschitz theory with a dynamic critical exponent $z=2$. However, the present model does not have any $\U(1)$ symmetry. Indeed, a numerical analysis reveals  that its dynamic critical exponent is $z\approx 2.2$~\cite{IsingCQCP, IsakovCQCP}.

\begin{figure}[!t]
    \centering
    \includegraphics[width=0.4\textwidth]{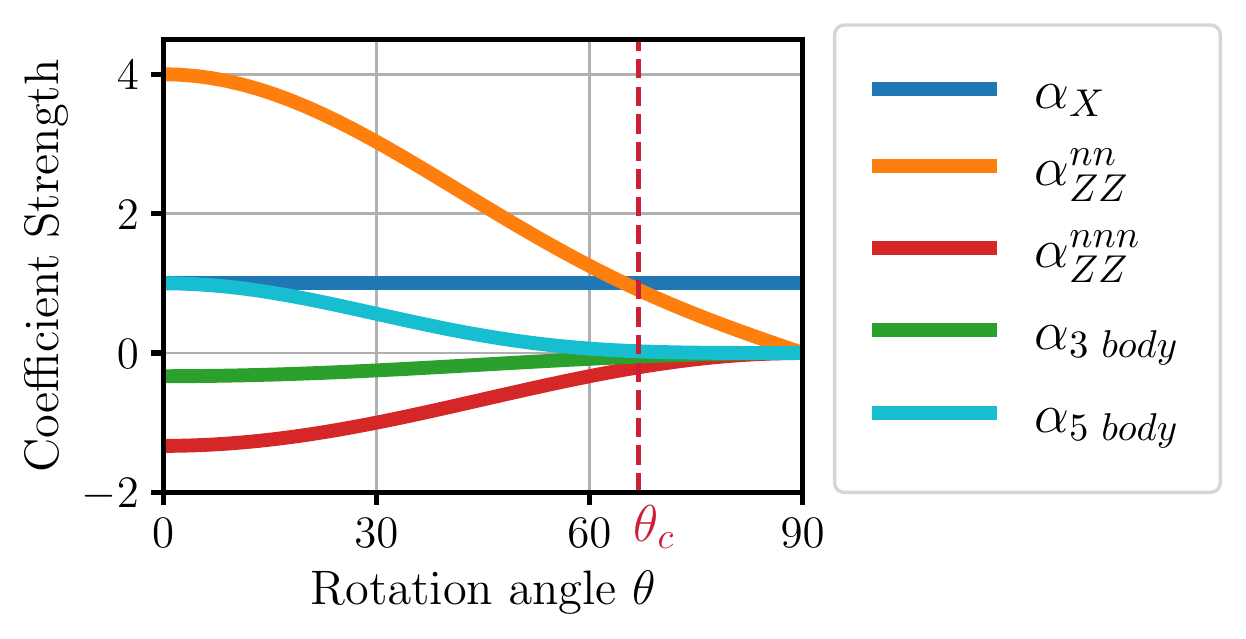}
    \caption{The coefficients of the parent Hamiltonian in \eqnref{eq:2dparent_postselect} as a function of measurement angle $\theta\in [0,\frac{\pi}{2}]$. The red dashed line marks the position of $\theta_c$. Coefficients are normalized in such a way that the strength of the transverse field term $\alpha_X$ is 1 throughout the range. }
    \label{fig:2dparent_postselect_coeff}
\end{figure}

Similar to previous sections, we can obtain the parent Hamiltonian for the post-measurement state using the general method outlined in \secref{sec:1D}. For measurement outcomes $s_e = 1$, $H$ is given by
\begin{align}
    H =& -\alpha_X \sum_i X_i -\alpha_{ZZ}^\textrm{nn} \sum_{\langle ij\rangle}Z_iZ_j -\alpha_{ZZ}^\textrm{nnn} \sum_{\langle \langle ij \rangle \rangle} Z_i Z_j \nonumber\\
    &-\alpha_\textrm{3-body} \sum_i X_i B_i -\alpha_\textrm{5-body} \sum_i X_i\prod_{j\in n(i)}Z_j
    \label{eq:2dparent_postselect}
\end{align}
where $B_i=\sum_{j,k\in n(i), j\neq k} Z_jZ_k$ and $n(i)$ is the set of sites neighboring the site $i$. The coefficients are graphically shown in \figref{fig:2dparent_postselect_coeff}. At the critical point $\theta_c$, $\frac{\alpha_3}{\alpha_1}\approx 1.1$, while the other coefficients $\alpha_2,\alpha_4,\alpha_5$ are relatively small. For a comparison, the critical point of the 2d transverse field Ising model is given by $(\frac{\alpha_3}{\alpha_1})_{\text{TFIS}}\approx 1.5$\cite{2dIsing}.

In the case without post selection, the state after measurements is still the ground state of a local Hamiltonian, whose structure is similar to \eqnref{eq:2dparent_postselect} but with the signs of the coefficients depending on the measurement outcomes (See \appref{app:2dlinks_parent}). With randomly signed interaction coefficients, one may wonder whether there still exists a phase transition in this Hamiltonian as a function of $\theta$. If it exists, what is the nature of the transition? We will address this in \secref{sec:Random}.

\section{3D Cluster States} \label{sec:3D}

\subsection{3D SPT with \texorpdfstring{$\mathbb{Z}_2^{(1)} \times \mathbb{Z}_2^{(1)}$}{Z2 x Z2} symmetry}

The generalization of the previous results into three spatial dimensions is more diverse as there can be different types of cluster states. In 3D, we can consider a new type of cluster state, illustrated in \figref{fig:2d3d}(b), which has a generalization of decorated domain wall (defect) construction for two 1-form symmetries. In this model, qubits are defined on the edges and faces of a cubic lattice. Note that qubits on the faces live on the edges of the dual cubic lattice. 
This particular 3D cluster stabilizer Hamiltonian is written as the following:
\begin{equation}
    H_{\text{3D SPT}} = - \sum_{e}  X_e \prod_{f \ni e}  \bm{Z}_{f}   - \sum_{f} \bm{X}_{f} \prod_{e \in f}  {Z}_{e}  \label{eq:1form1form}
\end{equation}
where $f$ runs over all faces of the cubic lattice. Bolded symbols act on faces, and unbolded symbols act on edges. By multiplying stabilizers, we obtain that $\prod_{f \in c} \bm{X}_f = 1$ for any cube $c$ and $\prod_{e \ni v} X_e = 1$ for any vertex $v$.
Here, generators of two 1-form symmetries are defined on two-dimensional surfaces: 
\begin{align} 
    \textrm{$\mathbb{Z}_2^{(1)}$ 1-form: } & h_{\rd V} \equiv \prod_{f \in \rd V} \bm{X}_{f} \nonumber \\
    \textrm{$\mathbb{Z}_2^{(1)}$ 1-form: } & g_{\rd V} \equiv \prod_{e \perp \rd \tilde{V}} {X}_{e}
\end{align}
where $V$ is a  three-dimensional volume enclosed by cubic faces, and $\tilde{V}$ is an infinitesimally inflated version of $V$ which intersects with edges emanating from $V$. Therefore, $\rd V$ is a set of  faces, while $\rd \tilde{V}$ is a set of edges. Without loss of generality, if we measure all faces to be $\bm{X}_f = 1$, then we obtain that the resulting state has $\prod_{e \in f} Z_e = 1$ and $\prod_{e \ni v} X_e = 1$ for all $f$ and $v$, which gives the 3D toric code ground state.

Now, let us measure the qubits on the faces at angle $\theta$ (due to the dual nature of the system, we get the same physics by measuring edges). In this case, the post-measurement state in the $Z$-basis has its amplitudes given by the Boltzmann weight of the 3d Ising gauge theory (c.f. \eqnref{eq:wavefunction}). One can show that the 1-form symmetry of the unmeasured model maps into the local gauge symmetry in the 3d Ising gauge theory. With measurement outcomes $\{s_f\}$, the post-measurement state is expressed by 
\begin{align} \label{eq:3d_IGT_true_wavefunction}
    \ket{\cP_{\bm{s}} \psi} &\propto e^{-\frac{\beta}{2} \hat{H}} \Big[\otimes_{n=1}^N \big| {X}_e = \prod_{f \ni e} s_f \big\rangle \Big] \nonumber \\
    \hat{H} &\equiv - \sum_{f} s_f \prod_{e \in f} Z_e.
\end{align}
Assume we shift $\hat{H}$ by the sum of local terms $A_v \equiv \prod_{e \ni v} X_e$ to convert it into $\hat{H}_\textrm{toric}^\textrm{3D}$. Still, if we define $\ket{\cP_{\bm{s}} \psi}$ as above with this new Hamiltonian $\hat{H}_\textrm{toric}^\textrm{3D}$, the state would be the same because $A_v$ commutes with $\hat{H}$ and $A_v$ acts trivially on the initial state\footnote{This is because $\prod_{e \ni v} (\prod_{f \ni e} s_f ) = 1$. In fact, for any $\theta$, these $\mathbb{Z}_2$ gauge charge configurations remain frozen. }. Therefore, the above equation is nothing but an imaginary time evolution by 3D toric code Hamiltonian $\hat{H}_\textrm{toric}^{\textrm{3D}}$.
On the other hand, note that $\hat{H}$ itself corresponds to the 3D Ising gauge theory with a random interaction sign $s_f$ at $\beta = \tanh^{-1}(\cos \theta)$. Since the 3D Ising gauge theory has a finite temperature transition at $\beta_c^{\textrm{3D gauge}} = 0.76$, the correspondence implies that the confinement transition would happen at $\theta_c \approx 50^\circ$ with post-selection $s_f = 1$.

For $\theta < \theta_c$,  the expectation of the Wilson loop operator $W_{\Gamma}=\prod_{e\in \Gamma}Z_e$ over a loop $\Gamma$ decays exponentially with a perimeter law, which can be predicted based on the correspondence to the 3d Ising gauge theory (See \appref{app:3D_IGT}). As a result, the preparation of three-dimensional deconfined phase is robust. 
For $\theta > \theta_c$, $\expval{W_{\Gamma}}$ decays exponentially with an area law, which implies that the phase belongs to the trivial confined phase. At $\theta = \theta_c$, the state becomes critical. Its dynamical critical exponent can be obtained from the dynamics of 3d classical Ising model~\cite{CASTELNOVO2005, IsakovCQCP}, which can be calculated by various methods. 
The parent Hamiltonian for the case without post-selection is in \appref{app:3dparent11}.

In comparison, the transition between the same two gapped phases is more commonly described by a $3d$ toric code model in a single transverse field, where the transverse field term generates flux loop excitations. In that model, the direct transition between the two phases are mapped to Wegner's 4-dimensional lattice gauge theory \cite{wegner1971duality}, and it is known to be of first order~\cite{balian1975gauge,creutz1979experiments}.

\subsection{3D SPT with \texorpdfstring{$\mathbb{Z}_2^{(0)} \times \mathbb{Z}_2^{(2)}$}{Z2 x Z2}     symmetry}

Note that one can also consider a different geometry for a 3D cluster state, where qubits reside at vertices and edges of a cubic lattice. This cluster state has a decorated domain wall construction of a $Z_2$ 0-form and a $Z_2$ 2-form symmetries. By measuring vertices in $X$-basis with $s_v = 1$, one can get the 3D toric code state. However, upon measuring vertices at angle $\theta$, the resulting topological order becomes unstable, which get mapped into a 3D 2-form Ising model (See \appref{app:3dIsing2form}).  
The post-measurement wavefunction is given by
\begin{align}
    \ket{\cP_{\bm{s}} \psi} \propto e^{\frac{\beta}{2} \sum_vs_v \prod_{e\ni v}Z_e} \Big[ \otimes_e |X_e=\prod_{v\in e}s_v \Big\rangle.
\end{align} 

At finite $\theta$, the expectation value of Wilson surface operator $M_{\rd V} \equiv \prod_{e \in \rd V} \bm{Z}_e$ that measures 2-form symmetry defects decays with the volume enclosed by the surface, \begin{equation} \label{eq:3d_twoform}
    \frac{\bra{\cP_{\bm{s}} \psi} \prod_{e \in \rd V} \bm{Z}_e \ket{\cP_{\bm{s}} \psi}}{\bra{\cP_{\bm{s}} \psi}  \ket{\cP_{\bm{s}} \psi}} \sim  (\Xfactor)^\abs{V}.  
\end{equation}
For the state to be a topologically ordered, the above quantity should decay at most exponentially with the surface area. Therefore, there is no SSB of the 2-form symmetry nor long-range entanglement. In a complimentary point of view, we can show that $\expval{\prod_{e \in \mathcal{L}} \bm{X}_e}_{\cP_{\bm{s}} \psi} \sim (\sin \theta)^2$ ($\mathcal L$ is an open string), which indicates the anyon condensation occurs for any $\theta > 0$. 

In fact, $\theta=0$ is again a multicriticality point with the 3D toric code state as the ground state. As one may guess from its similarity to the 2D toric code preparation case, this multi-critical point is captured by the Hamiltonian with $\U(1)$ pivot symmetry generated by $H_\textrm{toric}^\textrm{3D}$~\cite{pivot}:
\begin{align}     \label{eq:3dIsgauge_parent}
    H &=  H_0 + \cos^2 \theta H_\textrm{SPT} +   6 \cos \theta   H^\textrm{3D}_\textrm{toric}, 
\end{align}
whose detailed structure is illustrated in \appref{app:3d02SPT}. Similar to the 1D and 2D examples, the state has parent Hamiltonian which is the interpolation of disordered, $\mathbb{Z}_2^{(2)} \times \mathbb{Z}_2^{\cal T}$ SPT, and topologically ordered states as illustrated in \figref{fig:phase_diagram}.

On the other hand, if we measure edges of this model, we obtain the wavefunction with amplitude given by the 3D Ising model (See \appref{app:3d_IT}). In this case, the long-range entanglement, i.e., GHZ-ness of the state, is robust as long as $\theta < \theta_c \approx 78^\circ$ if we post-select on $s_e = 1$.


\section{Post Selection Issues} \label{sec:Random}

In the previous sections, we primarily discussed the properties of the post-measurement wavefunctions assuming post-selection on the measurement outcome $s_i = 1$. 
However, this is highly restrictive since every time we run the experiment we will likely get a different measurement outcome. Indeed, the probability of getting the same measurement outcome decreases exponentially with the system size, so that correlations of the post-measurement wavefunction cannot be experimentally verified in a reasonable time, unless the system size is small.

In general, the measurement procedure generates an ensemble of quantum states with different measurement outcomes $\bs$, whose structure is given by the mixed state density matrix in the Kraus representation\footnote{$\sum_s \cP_{\bm{s}} = \mathbb{I}$, $\cP^2_s = \cP_{\bm{s}}$, and $\cP_{\bs} \cP_{\bs'} = \delta_{\bs,\bs'} \cP_{\bs}$.}:
\begin{equation} \label{eq:density_matrix}
    | \psi \rangle \langle  \psi| \rightarrow \rho_\theta \equiv \sum_\bs \cP_{\bm{s}} | \psi \rangle \langle  \psi| \cP_{\bm{s}} =  \sum_\bs P_s(\bm{s}) \frac{| \cP_{\bm{s}}  \psi \rangle \langle   \cP_{\bm{s}} \psi| }{\braket{\cP_{\bm{s}} \psi}}
\end{equation}
where $P_s(\bm{s}) = \braket{\cP_{\bm{s}} \psi}$ is the probability for the measurement outcome to be $\bs = \{ s\}$. Without any post selection, one should consider the mixed state density matrix $\rho_\theta$, instead of pure states $\ket{\cP_\bs \psi}$. As we repeat the experiment, we will prepare a series of states $\{ \ket{\cP_{\bs_1} \psi}, \ket{\cP_{\bs_2} \psi}, ... \}$, where the unmeasured degrees of freedom on each state can be measured only once. Then, one might well ask: Is it possible to extract any non-trivial  information about the general properties of post-measurement wavefunctions? 
In the following, we will demonstrate that this is possible, provided one classically ``decodes", using the measurement results $\{\bs_1,\bs_2,..\}$.

\subsection{Probability distribution: measurement outcomes}

To demonstrate the simplest yet interesting case, throughout this section, we consider measuring the qubits on the edges of the 2D cluster state where we have $N$ vertices, see \secref{sec:2dIsing}. The probability distribution of the measurement outcome $\bs$ directly follows from \eqnref{eq:1d_true_wavefunction}:
\begin{equation} \label{eq:Ps}
    P_s(\bm{s}) = \braket{\cP_{\bm{s}} \psi} = \frac{1}{2^N Z_0} \sum_{\sigma=\pm 1} e^{\beta \sum_{\expval{ij}} s_{ij} \sigma_i \sigma_j } ,
\end{equation}
where $Z_0 = (2 \cosh \beta)^{2N}$ since there are $N$ vertices and $2N$ edges. Here, the sites are labelled by $i,j$ and $s_{ij}$ is the measurement outcome on the edge interconnecting the two nearest-neighbor sites.  Using this probability distribution, one can show that the measurement outcomes satisfy the following correlations:
\begin{align}\label{eq:correlation_2d}
    \mathbb{E}[\hspace{-5pt}\prod_{e \in \gamma_\textrm{loop}} \hspace{-5pt}  s_e ] &= \sum_{\bs} P_s(\bs) \prod_{e \in \gamma_\textrm{loop}} \hspace{-5pt}  s_e =  (\cos \theta)^\abs{\gamma_\textrm{loop}} 
\end{align} 
where $|\gamma_\textrm{loop}|$ is the length of an arbitrary closed loop along the edges. The calculation of the above expectation value is equivalent to the application of the Lemma described in \appref{sec:1D}.  Or alternatively, can be obtained directly from \eqnref{eq:Ps}, upon using $\tanh \beta = \cos \theta$.

\subsection{Random Bond Ising Model}

The correlation of the measurement outcomes characterized by \eqnref{eq:correlation_2d} is in a very particular form, and one may ask whether there is any simpler probability distribution that reproduces the same correlation structure. Indeed, consider the distribution of $\bs$ where each edge is chosen independently with probability $p(s_{e}) = (1 + s_{e} \cos \theta)/2$. This is the distribution of bonds in the canonical random bond Ising model (RBIM)~\cite{BinderYoung1986},  
\begin{equation}
    P^\textrm{RBIM}_s(\bs) \equiv \prod_{{e}} p^\textrm{RBIM}(s_{e}) = \prod_{e} \qty(\frac{1 + s_e \cos \theta}{2}).
\end{equation}
This distribution function can  be conveniently re-expressed
as $p^\textrm{RBIM}(s_e) = \frac{\sin \theta}{2} e^{b s_e}$ where $e^{b} = \frac{1 + \cos \theta}{\sin \theta}$.  Equivalently, $b = \tanh^{-1}(\cos \theta)\equiv \beta$, as in \eqnref{eq:1d_true_wavefunction}. Under this probability distribution, one can show that
\begin{align} \label{eq:loop_corr}
    \mathbb{E}\Big[\prod_{e \in \gamma} s_e \Big] &= \sum_{n=0}^l \binom{l}{n} (-1)^{l-n} p_+^{n} (1-p_+)^{l-n} \nonumber \\
    &= (p_+ - (1-p_+))^l = (\cos \theta)^l ,
\end{align}
where $l$ is the length of the loop $\gamma$. This is exactly the same as in \eqnref{eq:correlation_2d}, and one may suspect a close connection between $P_s(\bs)$ and $P^\textrm{RBIM}(\bs)$. Indeed, $P_s(\bs)$ can be obtained by a \emph{gauge symmetrization} procedure on $P^\textrm{RBIM}_s(\bs)$, defined as the following:
\begin{align} \label{eq:gauge_sym}
        \tilde{P}^\textrm{RBIM}_s(\bs) &\equiv \frac{1}{2^N} \sum_{t_i = \pm 1} \prod_{\expval{ij}} p^\textrm{RBIM}( t_i s_{ij} t_j ) \nonumber \\
        & = \frac{1}{2^N} \sum_{t_i = \pm 1} \qty( \frac{\sin \theta}{2} )^{2N} e^{\sum_{\expval{ij}} \beta t_i s_{ij} t_j } \nonumber \\
        & = \frac{1}{2^N Z_0 } \sum_{t_i = \pm 1}   e^{\sum_{\expval{ij}} \beta t_i s_{ij} t_j } = P_s(\bs)
\end{align}
where the set of variables $\bt = \{ t_i \}$ are defined on the vertices.

\subsection{Gauge-Invariant Structure: Frustration}

We have demonstrated that there is a connection between $P_s(\bs)$ and $P^\textrm{RBIM}_s(\bs)$. In order to gain more insight, we can think about correlation structures of the post-measurement wavefunction. According to \eqnref{eq:2d_IT_true_wavefunction}, correlations of the post-measurement wavefunction $\ket{\cP_\bs \psi}$ are determined by the partition function $Z_\beta[\bs] = \sum_{\{\sigma\}} e^{\beta \sum_{\expval{ij}} s_{ij} \sigma_i \sigma_j }$,
since 
\begin{equation} \label{eq:Correlator}
   \frac{ \langle \cP_\bs \psi |Z_i Z_j | \cP_\bs \psi \rangle } {\langle \cP_\bs \psi |  \cP_\bs \psi \rangle} = \frac{1}{Z_\beta[\bs]}\sum_{\{\sigma\}} \sigma_i \sigma_j e^{\beta \sum_{\expval{ij}} s_{ij} \sigma_i \sigma_j }.
\end{equation} Viewing $\bs = \{ s_{ij} \}$ as a set of gauge fields and $\bsigma = \{ \sigma_i \}$ as a set of matter fields, one can think about the gauge transformation defined by $\bt = \{ t_i \}$, 
\begin{equation}\label{eq:gauge_transformation}
    \tilde{\sigma}_i = t_i \sigma_i, \quad \tilde{s}_{ij} = t_i s_{ij} t_j, \quad t_i = \pm 1.
\end{equation}
Since the partition function is invariant under the change of variables to be summed over, $Z_\beta[\bs] = Z_\beta[\tilde{\bs}]$.  Moreover, the probability distribution function, $P_s(\bs) \propto Z_\beta[\bs]$ is also invariant under any gauge transformation by $\bt$.

This \emph{gauge transformation} defines  equivalence classes: $\bs$ and $\bs'$ belong to the same equivalence class if there exists $\bt$ such that $\bs$ can be transformed into $\bs'$ under \eqnref{eq:gauge_transformation}, denoted by $\bs \sim \bs'$. Then, correlation functions of two gauge-equivalent post-measurement wavefunctions $\ket{\cP_\bs \psi}$ and $\ket{\cP_{\bs'} \psi}$ are related as
\begin{equation} \label{eq:gauge_relation}
    \langle \cP_\bs \psi | Z_i Z_j | \cP_\bs \psi \rangle  = t_i t_j \langle \cP_{\bs'} \psi | Z_i Z_j  |  \cP_{\bs'} \psi \rangle   .
\end{equation}
Therefore, up to an overall sign  $t_i t_j$ (which is important in an actual experimental detection), one concludes that the correlation functions only depend on the gauge-invariant structure.

\begin{figure}[!t]
    \centering
    \includegraphics[width = 0.48 \textwidth]{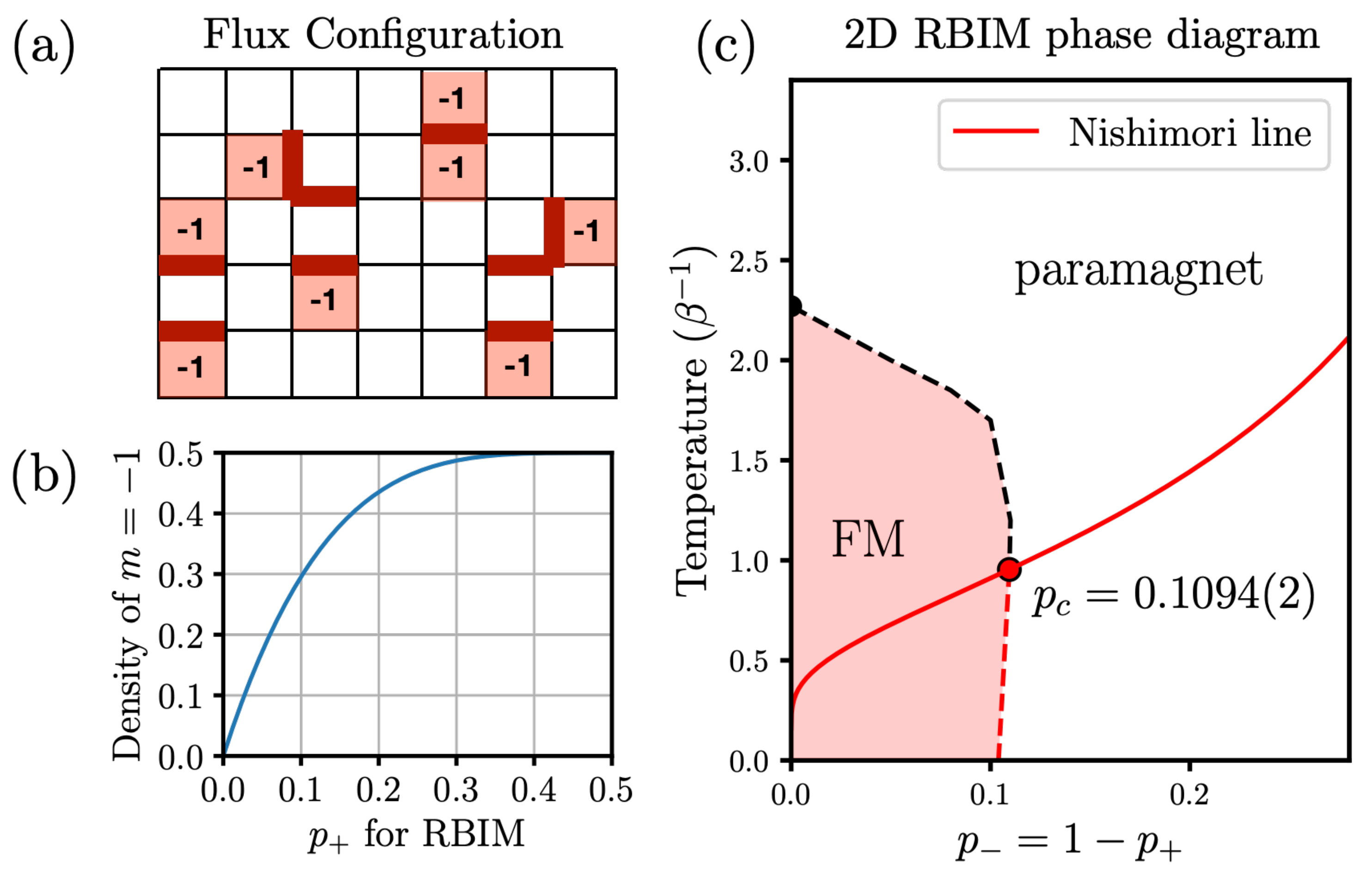}
    \caption{\label{fig:2DRBIM} (a) An example of the flux configuration where colored plaquettes have $m_p = -1$. Thick red lines correspond to antiferromagnetic bonds with $s_e = -1$. (b) The density of negative fluxes as function of $p_+$ in RBIM. (c) The phase diagrams for the 2D random bond Ising model~\cite{NishimoriPoint}. Here, the red line is called a Nishimori line~\cite{Nishimori}, where $(p_+,\beta) = ((1 + \cos \theta)/2, \tanh^{-1} (\cos \theta) )$. }
\end{figure}

The gauge-invariant structure, i.e., the equivalence class of a given measurement outcome $\bs$, can be completely specified by the flux (frustration) configuration $\bmm \equiv \{ m_p \}$. Note that there are two types of fluxes: ($i$) flux per plaquette $m_p$ and ($ii$) flux along the non-contractible loop (cycle) $m_{C}$:
\begin{equation}
    m_p \equiv \prod_{e \in p} s_e, \quad m_C \equiv \prod_{e \in C} s_e, \quad m = \pm 1 ,
\end{equation}
where $C$ is a non-contractible loop. As we assume periodic boundary condition, $\prod_p m_p = 1$ and there are $2^{N-1}$ distinct plaquette flux configurations. In addition, there are two additional fluxes coming from two distinct cycles $C_1$ and $C_2$. Therefore, there are total $2^{N+1}$ distinct equivalence classes. The counting matches since there are $2^{2N}$ possible measurement outcomes $\bs$ with $2^{N-1}$ distinct gauge transformation $\bt$ (having all $t_i = -1$ does not change $\bs$). 
For convenience, we will collectively denote two different types of fluxes as just $m_p$, and $\bmm \equiv \{ m_p \}$ denotes $N+1$ fluxes.

Given a probability distribution of $\bs$, $P_m(\bmm)$ can be directly evaluated by summing over all gauge-equivalent configurations: $P_m(\bmm) = \sum_{\bs \sim \bs_{\bmm}} P_s(\bs) = 2^{N-1} P_s(\bs_\bmm)$ where $\bs_\bmm$ is a representative element of the equivalence class defined by $\bmm$. Since the probability distribution of $\bmm$ only depends on the gauge-invariant structure, it follows that
\begin{equation}
    P_m(\bmm) = P_m^\textrm{RBIM}(\bmm)  .
\end{equation}
In fact, correlation functions of all fluxes through arbitrary loops in \eqnref{eq:correlation_2d} uniquely specify the distribution of fluxes through individual plaquettes, $P_m(\bmm)$, as proved in \appref{app:gauge}.

Mapping the measurement outcomes into flux configurations helps clarify the physics of interest. For example, at $\theta = 0$, \eqnref{eq:correlation_2d} translates into,
\begin{equation}
    \mathbb{E}[ m_p ] = 1 \quad \Leftrightarrow \quad P_m({m_p = 1}) = 1,
\end{equation}
which means that we always get a frustration-free configuration at $\theta=0$. Although there are $2^{N-1}$ possible different measurement results, $\bs$, compatible with $m_p = 1$, it is always possible to find a gauge transformation, $\bt$, which maps a given measurement outcome $\bs$ into the fully-ferromagnetic configuration $\bs^\textrm{FM}$ with $s^\textrm{FM}_{ij} = 1$.  And these gauge transformations can be found efficiently, within a time linear in the system size. 
The above reasoning implies that the ensemble of states represented by $\rho_\theta$ in \eqnref{eq:density_matrix} at $\theta = 0$, correspond to the gauge choices present in the  well-known Mattis Ising model~\cite{MATTIS1976}, where $H = -\sum_{ij} J_{ij} \sigma_i \sigma_j$ with $J_{ij} =  \epsilon_i \epsilon_j$ for random $\epsilon_i = \pm 1$. This problem is readily mapped to the pure ferromagnetic Ising model, since its random bonds have no frustration.

\subsection{Phase Diagram of 2D RBIM}

Before further pursuing the close correspondence between the ensemble of quantum states generated in our setup and the ensemble of classical thermal states generated by the distribution of the random bond Ising model, we first review the physics of the 2D random bond Ising model (RBIM).

The 2D random bond Ising model has the phase diagram illustrated in \figref{fig:2DRBIM}(c), where the two axes are the probability of negative (antiferromagnetic, AFM) bonds $p_- = 1- p_+$ and the temperature $\beta^{-1}$. There are three phases: ferromagnet (FM), paramagnet, and spin glass. For $p_-=0$ the model is simply the ferromagnetic Ising model,
and the Ising transition happens at $\beta_c^{-1} = 2.27$. At zero temperature, there is a phase transition between the ferromagnetic (FM) phase and a spin glass (SG) phase, occurring at $p_c = 0.104$~\cite{WangPreskill2003}. This transition is driven by the increase in density of  the frustrated plaquettes ($m_p = -1$), which grows  monotonically with $p_-$, as shown \figref{fig:2DRBIM}(b)~\cite{FradkinRandom}. In 2D, the SG phase is unstable at $T>0$, immediately transitioning into a paramagnetic phase. The general phase boundary between ferromagnet and paramagnet is shown by a dashed line. 

There is an interesting manifold in the parameter space where $e^{2 \beta} = p_+/p_-$, which is called the Nishimori line~\cite{Nishimori, Nishimori2} drawn as a red line in \figref{fig:2DRBIM}(c). On the Nishimori line a gauge
symmetry allows for a series of exact results~\cite{RiegerRBIM1994, ChoRBIM1997, GruzbergRBIM2001} and invariance under the RG transformation~\cite{Doussal1988}.
The line crosses the phase boundary around $p^N_c = 0.1094(2)$~\cite{NishimoriPoint}. 
Interestingly, when we make a correspondence with $\rho_\theta$, the trajectory in the 2D RBIM is given by $(p,\beta) = ((1 + \cos \theta)/2, \tanh^{-1} (\cos \theta) )$, which exactly coincides with the Nishimori line.

The FM and PM phases in the Ising model can be distinguished by the ferromagnetic susceptibility,
\begin{equation} \label{eq:RBIM_FM_order}
    \expval{\chi(\bs)}_\beta \equiv \frac{1}{N} \sum_{i,j} \qty[  \frac{1}{Z_\beta[\bs] } \sum_{\bsigma} \sigma_i \sigma_j  e^{\beta \sum_{\expval{ij}} \sigma_i s_{ij} \sigma_j } ].
\end{equation}
After averaging over disorder realizations, one obtains,
\begin{align}
    \label{eq:RBIM_average}
    \overline{ \expval{\chi(\bs)}_\beta}^\textrm{RBIM}&\equiv \sum_\bs P^\textrm{RBIM}_s(\bs) \expval{\chi(\bs)}_\beta \nonumber \\
    &\propto  \begin{cases}
        N &\,\,\,  \textrm{ (in FM phase)}\\
        \textrm{const} &\,\,\,  \textrm{ (in PM phase)}
        \end{cases} .
\end{align}

\subsection{Decoding Measurement-Prepared Quantum States}

We just showed that on average, the ensemble of thermal states from 2D RBIM exhibits ferromagnetic long-range correlations for $p_- < p_c$ along the Nishimori line. From \eqnref{eq:gauge_sym}, one can deduce that on average, the ensemble of post-measurement wavefunctions in $\rho_\theta$ should also exhibit long-range correlations for $\theta < \theta_c$ where $\theta_c = \cos^{-1}(1 - 2p_c)$. 
However, each post-measurement wavefunction is long-range ordered with arbitrary spin-directions for different sites, and we anticipate long-range correlation to be hidden in most of the post-measurement wavefunctions. In this case, would it be possible to experimentally detect this long-range correlation of the ensemble of quantum states in $\rho_\theta$?

If one is possible to obtain exponential amount of measurement outcomes, \emph{decoding} this hidden ferromagnetic correlation is a simple task. With enough data for each measurement outcome $\bs$, one can perform a direct tomography on each $\ket{\cP_\bs \psi}$, and calculate a certain observable identifying whether the resulting state is long-range ordered or not. Then, by evaluating a weighted-average of this observable over all post-measurement wavefunctions, one can talk about the behavior of the ensemble $\rho_\theta$. However, for a realistic detection (or identification) of long-range order in the ensemble of quantum states prepared by the setup in \figref{fig:setting}, we want the measurement complexity to be polynomial in the systems size.

In the following, we discuss the experimental protocol to efficiently detect long-range order in $\rho_\theta$ without any need for post selection.
To this end, let $\bsigma = \{ \sigma_i \}$ be the measurement outcomes of the qubits on the vertices in the $Z$-basis, see \figref{fig:cao}. One might naively attempt to extract the correlation function in a way similar to \eqnref{eq:RBIM_FM_order}:
\begin{align} \label{eq:QC_order}
    \expval{C_{ij} }_{\cP_\bs \psi} \equiv  \frac{ \langle \cP_\bs \psi | Z_i Z_j | \cP_\bs \psi \rangle }{   \langle \cP_\bs \psi |  \cP_\bs \psi \rangle  },
\end{align}
where an expression for this correlator is given in \eqnref{eq:Correlator}.
However, upon averaging over the measurement outcomes $\bs$, this quantity vanishes identically:
\begin{align}
    \overline{ \expval{C_{ij} }_{\cP_\bs \psi} }^{\rho_\theta} &= \sum_\bs P_s(\bs) \expval{C_{ij} }_{\cP_\bs \psi} \nonumber \\
    &= \sum_\bs  \langle \cP_\bs \psi | Z_i Z_j | \cP_\bs \psi \rangle = 0 .
\end{align}
This result can also be understood by considering the distribution of $\bsigma$ for the measurement results on the vertices, after averaging over all $\bs$, which gives a uniform distribution:
\begin{align} \label{eq:Pz}
        P_\sigma (\{\sigma \}) &= \textrm{Tr}  \qty(\rho_\theta \, | {\bm{\sigma}} \rangle \langle {\bm{\sigma}} | ) \nonumber \\
        &= \sum_\bs \abs{ \braket{\bsigma}{\cP_\bs \psi} }^2 = 1/2^N .
\end{align}
Being ignorant of the measurement outcomes on the edges, it is not possible to detect any ordering for the qubits on the vertices, regardless of the value of $\theta$.  As we shall see, it will be necessary to use the measurement outcome, $\bs$, in order to ``decode" the  ``hidden" FM order.

First consider the simplest case, with $\theta = 0$.  As we have discussed, in this case $\bs$ is always frustration-free, with no frustrated plaquettes ($m_p(\bs) = +1$). As a result, it is always possible to find a gauge transformation $\bt$ (in a time linear in $N$) such that the underlying Hamiltonian (\eqnref{eq:2d_IT_true_wavefunction}) can be mapped into the fully ferromagnetic Ising model. In other words, for any given measurement outcomes $\bs$, we can construct a gauge transformation $\bt(\bs)$, and then calculate (or measure) the following correlation function:
\begin{equation}
    C^\textrm{decode}_{ij} \equiv \sum_\bs P_s(\bs) t_i(\bs) t_j(\bs)  \expval{C_{ij} }_{\cP_\bs \psi}.
\end{equation}
This correlator should exhibit the correlations of the GHZ state $\ket{00...} + \ket{11...}$, i.e., the groundstate of the ferromagnetic Ising model.
By using $C^\textrm{decode}_{ij}$ to evaluate the ferromagnetic susceptibility $\chi^\textrm{decode} \equiv \frac{1}{N} \sum_{ij} C^\textrm{decode}_{ij}$, we can obtain $\chi^\textrm{decode} = N$.
Experimentally, for each state prepared by measuring the edge qubits, we can make a single measurement of the vertex qubits,
and average over the sequence of prepared states using the above formula.

Moreover, from measurements of (any) single vertex qubit over repeated experiments,
one would be able to 
show that the average magnetization vanishes, 
\begin{equation}
    m_i^\textrm{decode} \equiv \sum_\bs P_s(\bs) t_i(\bs) \langle Z_i \rangle_{ \cP_\bs \psi } =0. 
\end{equation}
This would imply that the prepared quantum states on the vertices are not only ferromagnetic, but also GHZ-like. The fact that $\prod_v X_v = 1$ through the experimental setup before the measurement on vertices also guarantees that it is the GHZ state with coherent superposition.

What if $\theta > 0$? In this case, $P_s(\bs)$ starts to generate frustrated configurations with non-zero probabilities, as in \figref{fig:2DRBIM}(b). 
In this case the decoding will require finding a gauge transformation which takes the (signs of the) resulting measurement outcomes, $\bs$, into those for the RBIM.

The key part of the protocol, that we detail below, is to obtain a set of gauge-equivalent configurations based on a certain RBIM distribution (which has a bias on FM bonds) characterized by $\varphi$. From various numerical simulations~\cite{Reger3D1986, WangPreskill2003,OHNO2004}, we already know that 2D RBIM exhibits a ferromagnetic ordering with diverging ferromagnetic susceptibility. Therefore, for a flux configuration $\bmm_\bs$ determined by $\bs$, if we can obtain a set of configurations likely to be generated by the RBIM and calculate a usual ferromagnetic order parameter in  each gauge and average over them, we should be able to decode the ferromagnetic ordering hidden in the randomness of the measurement outcomes $\bs$. Although we have now introduced a fancy term -  ``decoding'' -  this is exactly what we did at $\theta = 0$.

We thus propose the \emph{decoding} protocol for the 2D cluster state measured on edges as the following:

\begin{algorithm}[h!]
\caption{Decoding FM Order Parameter}\label{protocol:1}

\vspace{5pt}
\textbf{Inputs.} For each quantum experiment illustrated in \figref{fig:cao} with the cluster entangler and unitary rotation by $\theta$, the observer extracts (i.e. measures) $\bs = \{ s \}$ on edges in the $X$-basis, and $\bsigma = \{ \sigma \}$ on vertices in the $Z$-basis. 

\vspace{5pt}
\textbf{Transform} : Calculate the flux configuration $\bmm_s = \{ m_p \}$ from given measurement outcome on edges $\bs$. \\

\vspace{5pt}
\textbf{Computation} : Stochastic sampling of $\tilde{\bs}$ conditioned on the flux configuration $\bmm_s$ under the \emph{decoder}'s probability distribution: 
\begin{equation} \label{eq:sampling}
    P_{s|m}^\textrm{dec}( \tilde{\bs} |\bmm) \quad \rightarrow \quad \{ \tilde{\bs}^1, \tilde{\bs}^2, \tilde{\bs}^3, ... \} .
\end{equation}
For our decoder, we use a specific decoder distribution $P_{s}^\textrm{dec} = P_{s,\varphi}^\textrm{RBIM}$, which is defined by the independent bond distribution with $p(s) = (1 + s \cos \varphi)/2$, and given explicitly below in \eqnref{eq:decoder}. Note that $\varphi$ is not necessarily equal to the rotation angle $\theta$, although $\varphi = \theta$ would be an optimal choice to perfectly decode the hidden correlations in $\rho^\theta$ (which is accessible with post-selections).  
This step is repeated enough to generate a \emph{sequence} of different bond configurations $\bm{S} \equiv \{ \tilde{\bs}^n |\,n=1,2,... \}$.

\vspace{5pt}
\textbf{Gauge Transformation} : For each $\tilde{\bs}^n \in S$, find a gauge transformation $\bt^n$ from $\bs$ to $\tilde{\bs}^n$. This takes a time linear in the system size. Denote $T \equiv \{ \bt^n \}$.

\vspace{5pt}
\textbf{Evaluation} : Calculate the correlation function
\begin{equation} \label{eq:decode1}
    C^\textrm{decode}_{ij}(\bs, \bsigma) \equiv \frac{1}{\abs{T}} \sum_{\bt^n \in T} t^n_i(\bs) t^n_{j}(\bs) \sigma_i \sigma_j  .
\end{equation}

\vspace{5pt}
\textbf{Repeat above} : And average the correlation function

\vspace{5pt}
\end{algorithm}

An explicit expression for the decoder probability distribution function is
\begin{align}
    \label{eq:decoder}
     P_{s|m}^\textrm{dec}( \tilde{\bs} |\bmm_s) &\propto \delta(\bmm_\bs - \bmm_{\tilde{\bs}}) \cdot P_{s, \varphi}^\textrm{RBIM}(\tilde{\bs})  \nonumber \\
     &= \delta(\bmm_\bs - \bmm_{\tilde{\bs}}) \cdot \prod_{\langle i j\rangle}(1+\tilde{s}_{ij} \cos{\varphi})/2,
\end{align}
which vanishes if $\tilde{\bs} \not\sim \bs$, i.e., $\bmm_{\tilde{\bs}} \neq \bmm_{{\bs}}$.  We note that this can be re-written as,
\begin{equation} \label{eq:AFMbonds}
    P_{s, \varphi}^\textrm{RBIM}(\tilde{\bs}) = C e^{\beta_\varphi \sum_{\langle ij \rangle} \tilde{s}_{ij}} = C^\prime e^{-\beta_\varphi N_\textrm{AFM}},
\end{equation}
where $\tanh \beta_\varphi  = \cos \varphi $ and $N_\textrm{AFM}$ is the total number of anti-ferromagnetic bonds ($\tilde{s}_{ij}=-1$) in the configuration $\tilde{\bs}$.  Notice that for $\varphi \rightarrow  0$, one has $\beta_\varphi^{-1} \rightarrow 0$, and the decoder selects the configuration of $\tilde{\bs}$ which minimizes $N_\textrm{AFM}$.
For a given flux configuration, $\bmm_s$, this minimization can be achieved in polynomial time using the protocol called minimal weight perfect matching, as we discuss further, below.


We emphasize that the decoder's probability distribution reflects our initial bias (knowledge) on the underlying structure of the measurement outcomes.
By choosing   
$P^\textrm{dec}_{s|m}(\tilde{\bs}|\bmm_s) =  \delta(\bmm_\bs-\bmm_{\tilde{\bs}}) P_{s,\varphi}^\textrm{RBIM}(\tilde{\bs})$,
we are implicitly assuming that the underlying (gauge invariant) structure of
$\rho_\theta$ to be equivalent to the that of the RBIM with $p_+ = (1 + \cos \varphi)/2$. 
In fact, the optimal choice of ``bias'' is very important. If the sampling is performed by $P_{s,\varphi}^\textrm{RBIM}$ with $\varphi = \theta$, we can show that the decoded order parameter averaged over ensemble of states \emph{rigorously} equates to the ferromagnetic susceptibility of the 2D RBIM in \eqnref{eq:RBIM_average}, see \appref{app:equivalence} for the detailed proof.

To do the proposed sampling, one may perform a Monte Carlo simulation of the conditional probability distribution in \eqnref{eq:decoder},
where the Monte Carlo update is constrained to conserve the flux configuration, $\bmm_s$. With enough measurement outcomes, this procedure will reproduce the disorder average result of the RBIM, and we can decode the hidden information in the density matrix $\rho_\theta$.
While full convergence might be slow, especially at low temperatures for the Ising model, 
if $\theta$ is small, even a few experimental iterations accompanied with the decoding protocol would be able to tell whether the resulting mixed state has the long-range entanglement or not. The procedure is summarized in \figref{fig:cao}.

\begin{figure}[!t]
    \centering
    \includegraphics[width = 0.45 \textwidth]{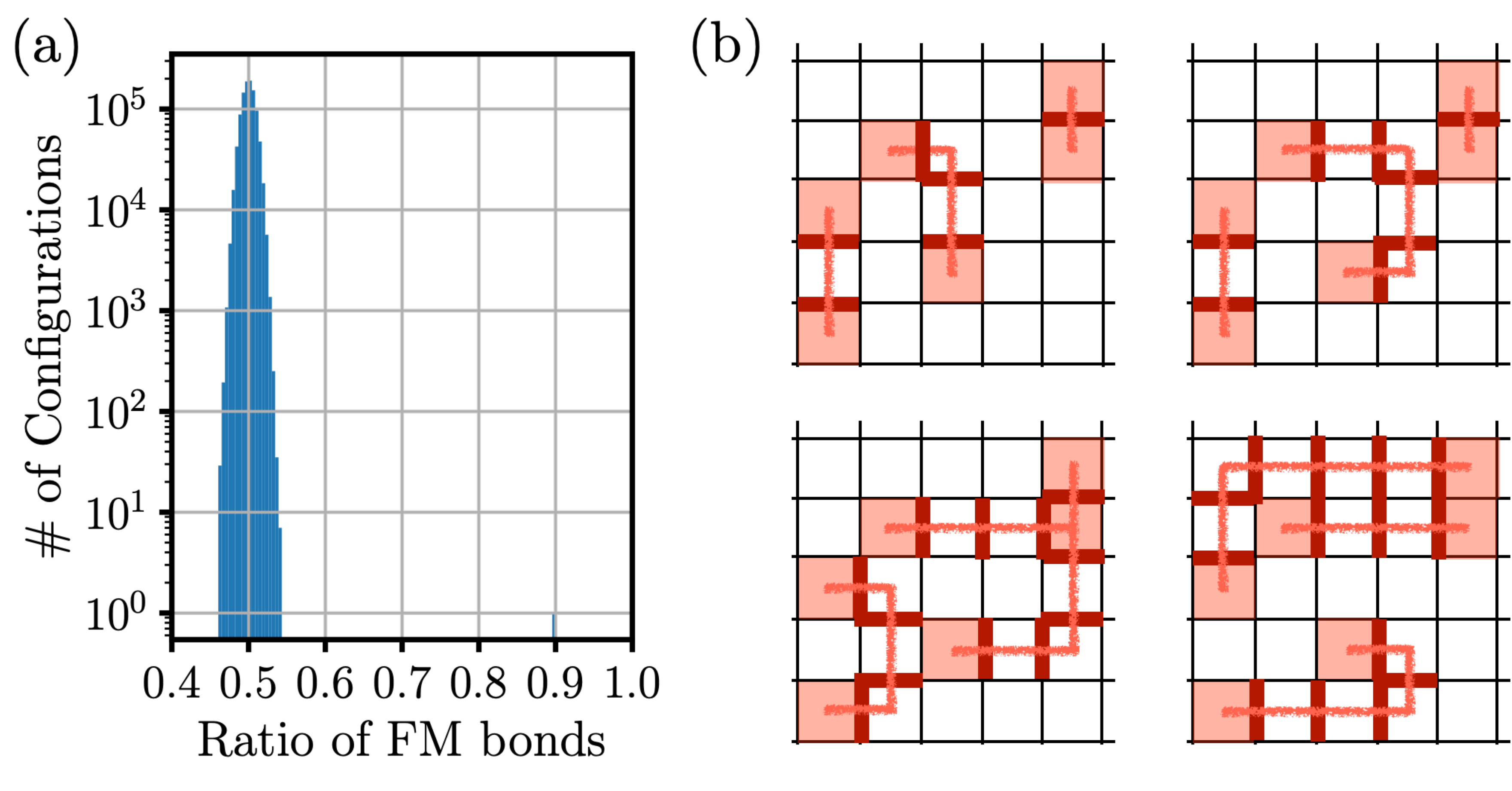}
    \caption{\label{fig:sampling_heuristics} (a) The histogram of $10^6$ gauge-equivalent configurations with respect to the number of ferromagnetic bonds, where the reference configuration is generated by $P^\textrm{RBIM}(\bs)$ with $p_+ = 0.9$. where the density of negative fluxes is $n_{-} = 0.3$ and the system size $N = 2500$. (b) Four gauge-equivalent configurations with different number of ferromagnetic bonds. Here thick red lines represent the AFM bonds, which are crossed by pink lines. 
    }
\end{figure}

\subsection{Optimal choice of Decoder}

In the previous section, we provided a  decoding protocol using \eqnref{eq:decoder} with a free parameter $\varphi$, and claimed that at $\varphi = \theta$ we can optimally decode the hidden ferromagnetic correlation in $\rho_\theta$. 
For a generic value of $\varphi$, we can prove that the decoded correlation function averaged over different measurement outcomes would approach to the following quantity (See \appref{app:equivalence}):
\begin{align} \label{eq:corr_max}
     \overline{ \langle C^{\textrm{decode,$\varphi$}}_{ij}   \rangle } ^{\rho_{\theta} } =  \overline{ \langle  C_{ij} \rangle_{\beta_\varphi } }^{\textrm{RBIM,$p_\theta$}} 
\end{align}
which is the disorder-averaged correlation function of the RBIM with the bond probability $p_\theta(s) = (1+ s \cos \theta)/2$ at the inverse temperature $\beta_\varphi = \tanh^{-1}(\cos \varphi)$.

The \eqnref{eq:corr_max} holds for any generic observable, such as magnetization. 
Once decoded observables are brought into the form in the RHS, the resulting quantity related to the decodability of the underlying ferromagnetic correlation is maximized when $\beta_\varphi = \beta$ (i.e. $\varphi = \theta$), as was pointed out by Nishimori~\cite{NishimoriDecoding}. When $\varphi \neq \theta$, the decodability can only decrease in that the decoded correlation function would falsely alarm the absence of the ferromagnetic ordering in $\rho_\theta$, although the ferromagnetic ordering is still present. This can also be visually understood: in the phase diagram of the 2D RBIM in \figref{fig:2DRBIM}(c), we observe that for a given value of $p_c$ at the Nishimori line, the critical state becomes immediately paramagnetic either for $T > T^N_c$ or $T< T^N_c$. For a given $p$ for the RBIM, the ferromagnetic nature of the thermal state is maximized at the Nishimori temperature $T^N(p)$. It implies that the decoded signal is optimal only if $\varphi = \theta$.

\begin{figure}[!t]
    \centering
    \includegraphics[width = 0.48 \textwidth]{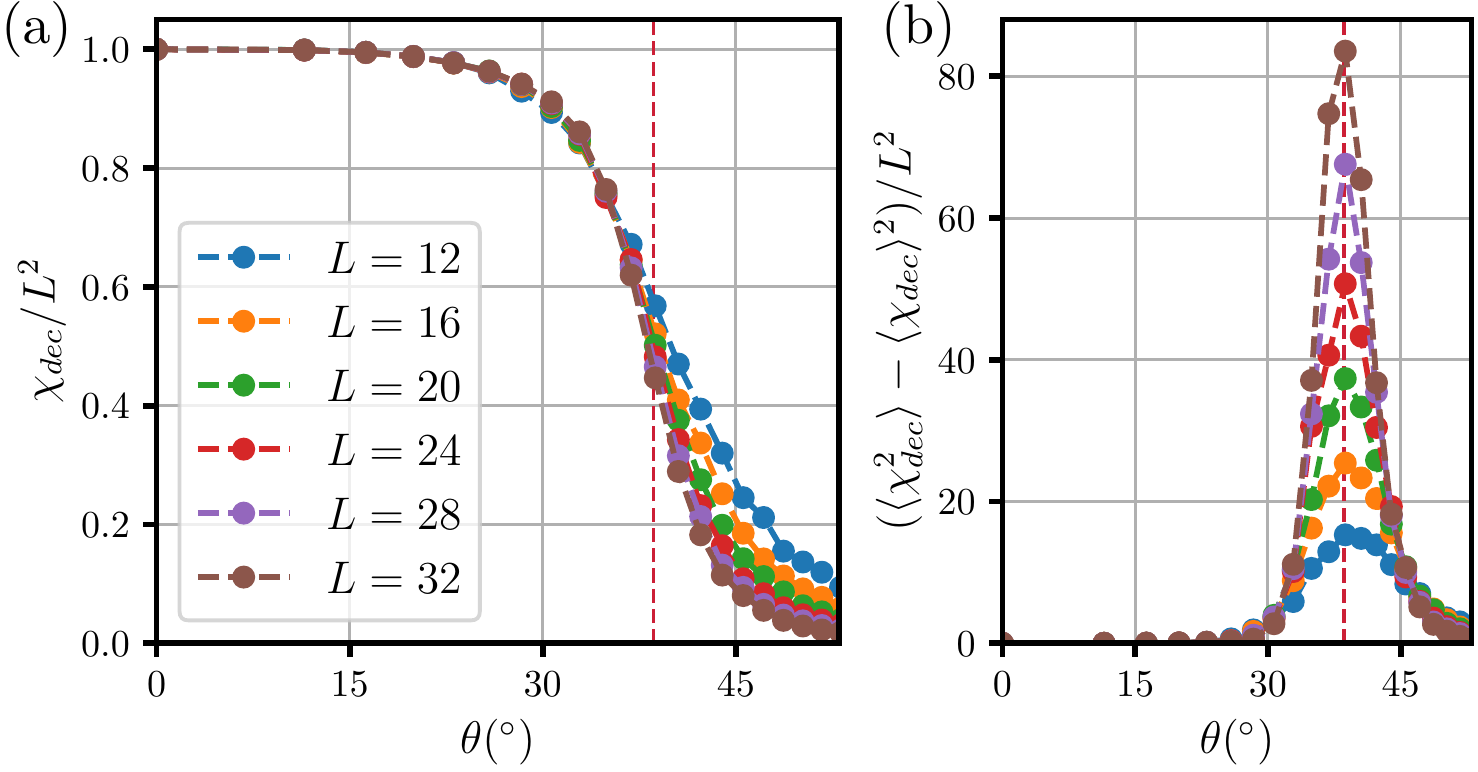}
    \caption{\label{fig:decMWPM} Simulated experiment results where we use the fast decoding protocol elaborated in \secref{sec:fast_decoding}. Here we considered $1.5 \times 10^3$ measurement outcomes ($\bs$,$\bsigma$) at $L=12,16,...,32$. The dashed red line represents the true transition point $\theta_c = 38.6^\circ$ ($p_c = 0.109$~\cite{WangPreskill2003}). (a) Normalized decoded susceptibility $\chi_\textrm{dec}/L^2$ and (b) Variance of the decoded susceptibility normalized by $L^2$. Note that $\chi_\textrm{dec}/L^2$ approaches to the squared average magnetization in the thermodynamic limit. While the crossing point of the $\chi^2/L$ is below the true transition point, the peak of the variance is roughly around the true transition point.  }
\end{figure}

\subsection{Faster Decoding} \label{sec:fast_decoding}

The above discussion established that we can decode the underlying structure perfectly as long as the optimal decoder is precisely implemented.  However, implementation requires importance sampling, using for example a Monte Carlo algorithm, and the convergence on a time scale polynomial in the system size is not rigorously guaranteed.  
As we now discuss, a non-optimal decoder with $\varphi \rightarrow 0$ can be implemented efficiently, in polynomial times scaling as $N^{3}$.  In this limit we are trying to minimize the number of anti-ferromagnetic bonds, with a constraint on
the flux configuration, as demonstrated explicitly in \eqnref{eq:AFMbonds}.  
This is equivalent to finding a solution of \emph{minimum weight matching problem} (MWMP). For example, if we have a configuration with two frustrated plaquettes $m_p = m_{p'} = -1$, a gauge-equivalent configuration that minimizes the number of AFM bonds would be the one where the bonds along the shortest path between $p$ and $p'$ are set to be AFM, and the other bonds are FM. 
In \figref{fig:sampling_heuristics}(b), the top left panel represents the configuration with minimum weight matching (there are other degenerate solutions), while the other three panels  represent configurations with the same $\bmm$ but more AFM bonds. Note that the number of AFM bonds is equal to the total length of crimson lines connecting pairs of fluxes with $m_p = -1$. Note the similarity with the error correction from syndrome detection for Pauli errors in the toric code~\cite{toMWPM2021}. The detailed formulation of the MWPM is elaborated in \appref{app:MWPM}.

In general, the MWPM problem can be solved by Edmond’s blossom
algorithm with a worst-case polynomial time complexity ${\cal O}(M^3 \log M)$~\cite{edmonds_1965}, where $M$ is the number of frustrated plaquettes ($m_p = -1$). Since the algorithm is polynomial in $M$, the MWPM problem can be efficiently solved. Of course, one may look for a faster method with a small approximation since we are already not using an optimal decoder. By using a method called Goemans-Williamson algorithm whose worst-case time complexity is given by ${\cal O}(M^2 \log M)$, one can obtain an approximate solution which is very close to the optimal solution very quickly~\cite{GoemansWilliamson}. Or, one can sparsify the original graph of frustrations (c.f. \appref{app:MWPM}) controlled by a sparsification parameter, reducing the time complexity down to ${\cal O}(M^2)$ without sacrificing accuracy too much~\cite{toMWPM2021}. 
Using a usual laptop, the run-time for the usual frustration configuration at $\theta = 25^\circ$ for $1,000$ lattice sites is about $1$ second for a rigorous MWPM algorithm, and $10^{-2}$ second for a sparsified MWPM algorithm~\cite{toMWPM2021}.\footnote{The worst-case time complexity is different from the average time complexity. Indeed, the average performance time complexity is a lot smaller than the worst-case analysis, $\sim$\,${\cal O}(M^{2.2})$ for the rigorous MWPM algorithm and $\sim$\,${\cal O}(M^{1.1})$ for the approximated MWPM algorithm~\cite{toMWPM2021}.}

We remark that in order to rigorously obtain the decoder limit $\varphi  = 0$ in \eqnref{eq:corr_max}, one has to sample over the entire list of degenerate solutions of the MWPM problem. For example, since we are on the square grid, there are already significant degeneracies in the solution due to the fact that the shortest path between two vertices $v$ and $v'$ is not unique in the square lattice. Furthermore, for a given MWPM problem, there can be multiple optimal solutions. 
In principle, in order to obtain the disorder-averaged RBIM behavior at $\varphi = 0$, which is isotropic, one should average over degenerate solutions (c.f. \eqnref{eq:decode1}) for each measurement data $(\bs,\bsigma)$. 
Instead, if we choose a single solution of the MWPM and corresponding $\tilde{\bs}$ to calculate the decoded correlation function, we would get an anisotropic correlation function. However, ferromagnetic susceptibility $\chi$ would still diverge with $N$ as long as $\theta$ is not too close to $\theta_c$. 
We thus propose the following \emph{fast decoding} protocol:

\begin{algorithm}
\caption{Fast Decoding FM Order Parameter}\label{protocol:2}

\vspace{5pt}
\textbf{Inputs.} / \textbf{Transform} : Same. \\

\vspace{5pt}
\textbf{Computation} : Find a maximally ferromagnetic bond configuration for a given flux configuration $\bmm$
\begin{enumerate}
    \item Mapping into the \emph{minimum weight matching problem} and solve using a certain classical algorithm~\cite{edmonds_1965,GoemansWilliamson, toMWPM2021}. 
    \item Assign $s_e = -1$ along the MWPM solution path, and $s_e = +1$ for all other bonds $\Rightarrow$ obtain $\tilde{\bs}$.
\end{enumerate}

\vspace{5pt}
\textbf{Gauge Transformation} / \textbf{Evaluation} : Same.

\vspace{5pt}

\end{algorithm}

\subsection{Numerical Verification of Fast Decoding}

In \figref{fig:decMWPM}, we have demonstrated  the performance of the fast decoding protocol where the gauge transformation for each measurement outcome is calculated only once using the MWPM algorithm, ignoring other degenerate solutions. In order to simulate the experimental procedure, we sampled the measurement outcomes from the distribution $P(\bs,\bsigma)$, which is given by
\begin{align}
        P_{s,\sigma}(\bsigma, \bs) &=     P_{s|\sigma}(\bs|\bsigma) P_\sigma(\bsigma).
\end{align}
The conditional probability $P_{s|\sigma}(\bs|\bsigma)$ can be obtained as
\begin{align} \label{eq:Psz}
     P_{s|\sigma}(\bs|\bsigma) &= \qty[\frac{P_{\sigma|s}(\bsigma|\bs) P_s(\bs)}{ P_\sigma(\bsigma) }] = \frac{1}{Z_0} \prod_{\langle ij \rangle} \qty[ e^{\beta s_{ij} \sigma_i \sigma_j} ]
\end{align}
where we use the Bayes rule, \eqnref{eq:Ps}, and \eqnref{eq:Pz}. The above result implies that we can sample the measurement outcomes $(\bsigma, \bs)$ using the following two steps: (1) Sample $\bsigma$ using the uniform distribution in \eqnref{eq:Pz}. (2) Conditioned on $\bsigma$, sample $\bs$ using \eqnref{eq:Psz} where the distribution decomposes into the independent distribution of each bond $s_{ij}$. Therefore, one can efficiently simulate the experimental setup and measurement outcomes without knowing exact quantum states under the quantum circuit evolution. 

For the system size $L \in [12,16,...,32]$ and rotation angle $\theta \in [0,60^\circ]$, we repeated the simulated experiment $1.5 \times 10^3$ times using the above sampling method. Then, based on the obtained set of $\{(\bsigma,\bs)_i\}$, we performed the fast decoding protocol to calculate the decoded susceptibility $\chi_\textrm{dec} \equiv \frac{1}{L^2} \sum_{ij} C^\textrm{dec}_{ij}(\bs,\bsigma)$. The entire simulations (sweeping over different $L$ and $\theta$) take about an hour using a laptop, and we have verified that the fast decoding algorithm works well, especially deep in the ferromagnetic phase. Interestingly, the uncertainty of the decoded susceptibility, i.e., $\expval{\chi^2} - \expval{\chi}^2$ diverges around the genuine critical point $p_c = 0.109$~\cite{WangPreskill2003}, signalling the true phase transition.

\subsection{Tuning away from Nishimori Line}

After optimal decoding, the post-measurement state would move along the red line in \figref{fig:2DRBIM}(a), crossing the unstable fixed point at the transition angle $\theta_c = 38.6^\circ$~\cite{NishimoriPoint}. This means that majority of the post-measurement wavefunctions in the density matrix $\rho_\theta$ exhibits long-range correlation (GHZ-ness) for $\theta < \theta_c$, and for $\theta > \theta_c$, they do not. Furthermore, by using the optimal decoding protocol, one can even study the critical behavior at $\theta = \theta_c$, which would exhibit power-law decaying correlation functions.

However, note that the fixed point is unstable; if we can slightly deviate either upward or downward along the trajectory, we can access both the pure ferromagnetic Ising and $T=0$ random bond Ising FM to spin glass universality classes. 
What ultimately affects the physics of the random bond Ising model is the density of frustrations as illustrated in \figref{fig:2DRBIM}(a). For our experimental protocols, the number of frustrated plaquettes is monotonically increasing with $\theta \in [0,\pi/2]$. 
If we can somehow manipulate the density of frustrated plaquettes in the ensemble of post-measurement wavefunctions, we can deviate from the red trajectory (Nishimori line) at a given inverse temperature.

There are several ways to achieve this.  Firstly, we could  post-select on configurations with the density of frustrated plaquettes strictly larger or smaller than the average value. As the measurement outcomes $\bs$ would generate a distribution of $\bmm_\bs$, by focusing on a certain portion of the measurement outcomes (which still scales nicely), one can move away from the Nishimori line.
Secondly, one could modify the original SPT state, starting from specific stabilizer signs to distort the probability distribution under \eqnref{eq:correlation_2d}.
This method will allow one to add frustration to the bond configurations even at $\theta = 0$ ($\beta = \infty$). For example, if we start from a product of $\ket{+}_x$ states, apply a uniform rotation along $y$-axis, and then measure in $X$-basis again, we can introduce a randomly distributed frustration in an adjustable way. In fact, adding more frustration in this method would bring our trajectory strictly below the Nishimori line in the phase diagram in \figref{fig:2DRBIM}(c).
Therefore, we have another tuning knob to access different universality classes. 
If the measured state ends up flowing into the $T=0$ universality class of the transition between random Ising model FM and spin glass, then its dynamic critical exponent would be $z\approx 3.11$~\cite{2dRBIM_crit}.

\section{Interactive Quantum Phases and Transitions} \label{sec:generalization}

In the previous section, we provided an efficient experimental protocol to study the ensemble of quantum states $\rho_\theta$ \eqref{eq:density_matrix} for the 2D cluster state where edges are rotated by angle $\theta$ and then measured in $X$-basis. 
As we revealed, the ensemble of post-measurement wavefunctions generates a completely random outputs $\bsigma$ for the vertices \eqref{eq:Pz} if we average over measurement outcomes $\bs$ on the edges, even at $\theta = 0$. 
Only after applying an appropriate transformation on $\bsigma$ based on the knowledge of $\bs$, we could efficiently identify the long-range order of the ensemble of quantum states.
Therefore, we dub the aforementioned structures in the density matrix $\rho_\theta$ as \emph{interactive} quantum phases and criticalities in that proper \emph{feedback} or  \emph{decoding} based on the measurement outcomes are required to experimentally identify interesting features in the resulting quantum state without post-selection.

\subsection{Computationally Assisted Observable}

In this subsection, we further discuss the general idea of \emph{interactive} quantum phases and their efficient identifications. Assume we prepare a quantum state by a quantum circuit consisting of a series of unitary evolutions and measurements. Let $\bs = \{ s_i \}$ be the set of measurement outcomes obtained during the evolution by the quantum circuit, where we can denote the resulting quantum circuit as ${\cal U}_\bs$. Then, the ensemble of quantum states prepared by this scheme is described by the following density matrix
\begin{equation} \label{eq:interactive}
    \rho = \sum_\bs | \,{\cal U}_\bs \psi \rangle \langle {\,\cal U}_\bs \psi | 
\end{equation}
Let $\bsigma = \{ \sigma_i \}$ be the measurement outcomes on the resulting quantum state $\ket{\,{\cal U}_\bs \psi}$, where measurement operators are not necessarily single-sited. 
Then, every time we run the quantum circuit followed by a set of measurements on the resulting states, we get a series of outcomes $\{ (\bs^1, \bsigma^1), (\bs^2, \bsigma^2), \cdots \}$.

In a conventional study of quantum phases, we say that a quantum state exhibits a certain feature if one can identify a nontrivial correlation within $\bsigma$. 
However, this usual definition hardly fits into the behavior of the ensemble of quantum states studied in \secref{sec:Random}, where we require $\bsigma$ to be either conditioned on a specific $\bs$ (post-selection) or to be transformed based on the knowledge of intermediate measurement outcomes $\bs$ (efficient decoding). 
Similarly, in the study of an ensemble of quantum states prepared with the aid of measurements, nonlinear order parameters based on both $\bsigma$ and $\bs$, such as entanglement entropy conditioned on intermediate outcomes~\cite{YaodongToapper}  or the decodability of the resulting quantum state~\cite{BarrattDecoding2022}, have been suggested as particular probes for the physics of interest.

In this regard, we propose the idea of \emph{computationally assisted observable} (CAO) for observables obtained through either quantum or classical computational decoding process using both $\bsigma$ and $\bs$, which helps identifying the \emph{interactive quantum phases and their transitions} by measuring hidden structures of the ensemble of quantum states. 
For instance, consider a density matrix $\rho(x)$ generated by a quantum circuit with measurements ${\cal U}_\bs(x)$ tuned by a parameter $x$ (c.f. \eqnref{eq:interactive}). 
Assuming we can repeat the experiment arbitrarily many times, one can fully understand the (averaged) underlying correlation structure of post-measurement wavefunctions $\{ \ket{\,{\cal U}_\bs \psi} \}$. Such a structure can undergo a transition at a certain value of the tuning parameter $x_c$.\footnote{For example, in the context of measurement-induced phase transitions, the measurement rate $p$ corresponds to the tuning parameter and whether the resulting state has an area or volume law entanglement corresponds to the structure one wants to probe.} 
If one can efficiently evaluate a quantity ${\cal O}(x)$ using measurement outcomes $\{(\bs^i,\bsigma^i)\}_{i=1}^M$, i.e. both measurements and computational time polynomial in the system size, and ${\cal O}(x)$ allows one to locate the true transition point $x_c$, then we claim that  ${\cal O}$ is the computationally assisted observable that signals the interactive quantum phases and their transitions in the density matrix $\rho(x)$.


Going back to the example discussed in \secref{sec:Random}, we emphasize the importance of the decoding protocol for practical use of the measurement-prepared quantum states. 
Naively at $\theta < \theta_c$, at every experiment we prepare a certain long-range ordered GHZ-like wavefunction although it is not conventional in that the alignment of neighboring spins is random, i.e., $\ket{\cP_\bs \psi} \sim \ket{010011..} + \ket{101100..}$. If the alignment is random every time we prepare the state through measurements, it is not useful in that we cannot detect nor make use of its long-range correlation. However, if we apply a set of single-site gates (spin-flip operation) based on the gauge transformation $\bt$ obtained through the decoding protocol, the resulting quantum state becomes \emph{identifiable} in that when we prepare a series of quantum states with different measurement outcomes, we can bring them into the similar form $\sim \ket{\bar{0}\bar{0}\bar{0}...} + \ket{\bar{1}\bar{1}\bar{1}...}$. One might argue that this \emph{identifiabiliy} is essential in any measurement-assisted setup.

Finally, we remark that post-selections are essential to study measurement-induced phases and their transitions in many literature~\cite{YaodongFisher2018,SkinnerNahum2018, Chan2019,YaodongFisher2019, Vasseur2019, XiangyuLuca2019, Gullans2020, Soonwon2020, TangZhu2020, JianLudwig2020, LopezVasseur2020, Bao2020, Rossini2020, Fan2021, Yaodong2021, BenZion2020, BarrattDecoding2022} since these features disappear once the resulting quantum states are averaged over different measurement outcomes. On the other hand, the hidden long-range correlations of our model and its phase transition behaviors can be rigorously  decoded by classical algorithms without post-selection, which provides a simple yet illustrative example of interactive quantum states along with few other cases~\cite{GullansProbe, BarrattDecoding2022, Dehghani2022, YaodongToapper}.

\subsection{Other examples}

Below, we explain \emph{interactive} nature of the measurement prepared quantum states and their transitions for the examples considered in \secref{sec:2D} and \secref{sec:3D}.

\vspace{5pt}

($1$) \textbf{2D $\mathbb{Z}_2^{(0)} \times \mathbb{Z}_2^{(1)}$ cluster state measured on vertices}: (c.f. \secref{sec:2D})
At $\theta = 0$, the measurement outcomes on the verticies are under the constraint $\mathbb{E}[ \prod_v s_v] = 1$. For the \emph{identifiable} toric code physics independent of the measurement outcomes, we want to manipulate the resulting quantum state so that it is the groundstate of the uniform toric code Hamiltonian $H_\textrm{toric} = -\sum_v A_v - \sum_p B_p$. To do so, we want the gauge where all $\tilde{s}_v = 1$. 
This decoding protocol at $\theta=0$ can be easily achieved by the following transformation: for any pair of vertices with $s_v = s_{v'} = -1$, we can find a path connecting them and perform $\tilde{Z}_e = - Z_e$, i.e., gauge transformation $t_e = -1$ along that path. Note that this transformation does not change the measurement outcomes on the vertices in between, and flip the signs of $s_v$ and $s_{v'}$ only. Formally, in this case dual to the 2D Ising model, the gauge transformation is defined by
\begin{equation} \label{eq:gauge_Ising}
    \tilde{\sigma}_e = t_e \sigma_e, \quad \tilde{s}_v = s_v \prod_{e \ni v} t_e, \quad t_e = \pm 1
\end{equation} 
By the successive application of such transformations, one can guarantee that all $\tilde{s}_v$ can be made $+1$, recovering ferromagnetic 2D gauge theory (or uniform toric code state). 
However, when $\theta > 0$, the ensemble of quantum states maps to the gauge-symmetrized version of the 2D random plaquette gauge model (RPGM), which immediately becomes trivial at any finite temperature even when there is no frustration.

\vspace{5pt}

\begin{figure}[!t]
    \centering
    \includegraphics[width = 0.48 \textwidth]{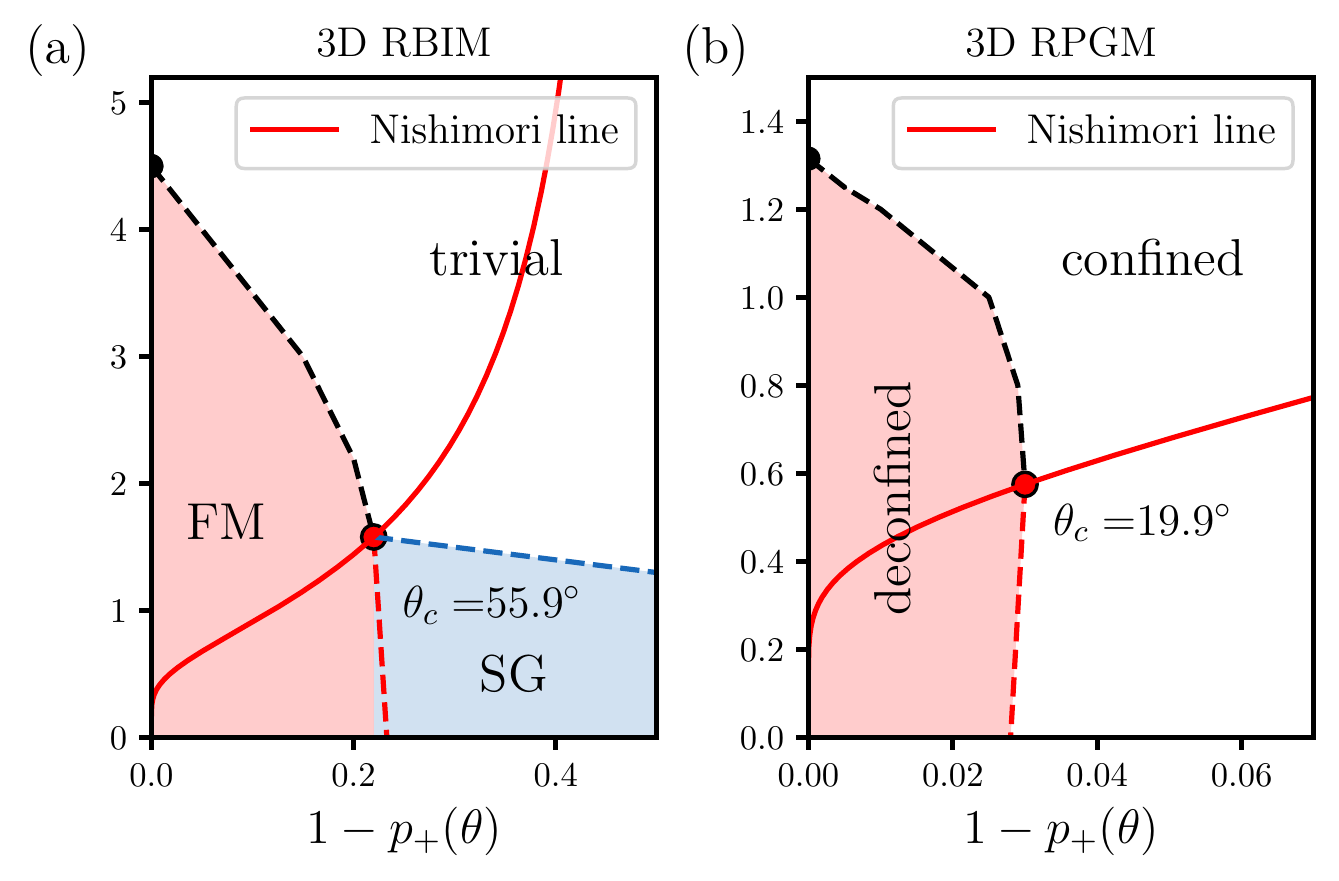}
    \caption{\label{fig:random} Phase diagrams for the (a) 3D random bond Ising model and (b) 3D random plaquette gauge model. $\theta_c$ denoted in the diagram represents the transition angle $\theta_c$ where the long-range entanglement disappears when we do not post-select outcomes. These phase boundaries are taken from~\cite{Reger3D1986, WangPreskill2003,OHNO2004}. Here, the red line is the trajectory our measurement protocol traverse without post-selection, i.e., $(p,\beta) = ((1 + \cos \theta)/2, \tanh^{-1} (\cos \theta) )$. The trajectory happens to coincide with the Nishimori line~\cite{Nishimori}. Therefore, without additional frustration from the initial stabilizer configurations, we will get a critical behavior for the unstable fixed point denoted by red circles.    }
\end{figure}

($2$) \textbf{3D $\mathbb{Z}_2^{(0)} \times \mathbb{Z}_2^{(2)}$ cluster state measured on edges}: (c.f. \secref{sec:3D}) At $\theta=0$, the resulting quantum state becomes the GHZ-like groundstate of the 3D Ising model. To convert the resulting state to be identifiable, one has to perform the gauge transformation as in the case of the 2D Ising model. The constraint is that $\prod_{e \in f} s_e = 1$ for any face $f$. As the bond configuration is frustration free, there is an efficient algorithm to convert the configuration into the fully ferromagnetic case. 

At $\theta > 0$, faces are generally frustrated, but one can show that total flux coming out from any cube or volume must be trivial. Therefore, if we draw a line on the dual lattice edges that penetrates through the frustrated faces, the line forms a closed loop. Again, we can construct an optimal decoder as in the case of 2D Ising model case. However, the classical sampling can be challenging to efficiently simulate, since it is equivalent to the simulation of elastic manifolds with boundary constraints. Still there is a rigorous classical decoding algorithm that will allow us to remove the exponential scaling of the measurement complexity. 
Our protocol would go through the Nishimori line as in \figref{fig:random}(a)~\cite{Reger3D1986, RiegerRBIM1994, Koji2000}. Again, depending on how we perturb away from the Nishimori line, we can access either 3d Ising universality or 3d random bond universality class. At the 3D random bond Ising universality, its dynamic critical exponent is given by $z \approx 2.11$~\cite{3dRBIM_crit}, which is much larger than the 3d Ising universality value.

\vspace{5pt}

($3$) \textbf{3D $\mathbb{Z}_2^{(1)} \times \mathbb{Z}_2^{(1)}$ cluster state measured on faces}: (c.f. \secref{sec:3D}) At $\theta=0$, the resulting quantum state becomes the 3D toric code state. For the state to be \emph{identifiable}, one has to perform the gauge transformation defined as
\begin{equation} \label{eq:gauge_3dIsing}
    \tilde{\sigma}_e = t_e \sigma_e, \quad \tilde{s}_f = s_f \prod_{e \in f} t_e, \quad t_e = \pm 1
\end{equation} 
which can be done very efficiently. Here, the constraint is that $\prod_{f \in S} s_f = 1$ for any closed surface $S$. It means that the line of $s_f = -1$ along the dual edges forms a closed loop $\gamma = \rd A$, where $A$ is the surface defined on the dual lattice. Then, if we apply $\prod_{e \perp A} X_e$ on the resulting quantum state, we convert the loop of $s_f=-1$ to be $\tilde{s}_f = +1$. By a successive application of this gauge transformation on all $s_f = -1$ dual loops, we recover the uniform 3D toric code state.

At $\theta > 0$, then we will get a genuine frustration where $\prod_{f \in C} s_f = -1$ for the surface of a cube $C$. Note that such frustrate cubes always occur as a pair. In this case, our optimal decoder would map the resulting ensemble of quantum states into the 3D random plaquette gauge model (RPGM)~\cite{WangPreskill2003} along the Nishimori line, where the probability of having positive plaquette Ising interaction is given as $p_+ = (1 + \cos \theta)/2$. An efficient decoder can be constructed by again solving a minimum weight perfect matching problem among frustrated cubes in three dimensions. In this case, for any paired frustrated cubes, we can connect them through the dual line, and the faces penetrated by the dual line would take a negative value for the corresponding $ZZZZ$ interaction. 
For the 3D RPGM, we remark that the phase transition behavior is closely related to the robust storage of quantum information in the surface codes~\cite{Dennis2002TQM}. The numerical study shows that the RPGM is within the perimeter law phase (deconfined) for $p<0.03$, and transitions into the area law phase (confined) for $p>0.03$~\cite{WangPreskill2003,OHNO2004}, which implies that for $\theta_c^{\textrm{3dRPGM}} = 19.9^\circ$, see \figref{fig:random}(b). As long as $\theta < \theta_c$, the preparation of the 3D topological order is robust. 
However, we note that its dynamic critical exponent has not been studied in the literature. The clear understanding of the critical theory would require a detailed future numerical work.

\section{More Cluster States} \label{sec:moreSPT}

In general, the framework we developed can be applied for any graph (cluster) state in the bipartite lattice. One interesting example is a subsystem SPT (SSPT). 
For example, consider a 2D cluster state with qubits defined on the vertices of a square lattice with $H = - \sum_{v} X_v \prod_{v' \in v} Z_{v'}$. The square lattice is bipartite, and we can decompose the lattice into two sublattices $A$ and $B$. The system has two subsystem symmetries $G_A$ and $G_B$ defined by the application of the product of $X$ operators along any diagonal direction for a corresponding sublattice.  
Upon measuring the sublattice $A$ in $X$-basis, we obtain the ground state of Xu-Moore model~\cite{XuMoore2004} on the other sublattice, described by $\hat{H}_\textrm{Xu-Moore} = - \sum_{v \in A} s_v \prod_{v' \in n(v)} Z_{v'}$.
With the subsystem symmetry $G_B$, the ground state manifold has an extensive degeneracy, and the post-measurement state  should be the superposition of exponentially-many ($\sim 2^{2L-1}$) SSB  configurations of the subsystem symmetry. 
As expected, this exponentially many superposition of SSB configuration, which is the generalization of the GHZ state, is not robust if $\theta \neq 0$. 
Indeed, the corresponding classical partition function and correlation functions are those of 2d plaquette (gonihedric) Ising model~\cite{2dgonihedric}, which is different from 2d Ising gauge theory as spins reside at vertices. This classical model exhibits an exotic correlation structure, which can be decomposed into decoupled 1d Ising models. As it maps to the stacked 1d Ising model, the system is disordered for any finite $\theta$ (or any $\beta^{-1} > 0$) and the long-range entanglement is not robust

For 3D SSPT, one can consider the cluster state whose qubits are defined on vertices and faces of the cubic lattice, described by $H= - \sum_{f} \bm{X}_{f} \prod_{v \in f} Z_v - \sum_{v} X_v \prod_{f \ni v} \bm{Z}_f$. 
The model has two symmetries. One is subsystem symmetries $G_A$ acting as $\prod_{\textrm{plane}} X$ for qubits on vertices, and there are $3L$ planes for $L \times L \times L$ sites. The other is the one-form symmetry $G_B$ acting as $\prod_{f \in S} \bm{X}_f$ for qubits on faces.
By construction, the model has $\prod_{f \in F(+_{\textrm{dual}} )} \bm{X}_{f} = 1$, where $F(+_{\textrm{dual}} )$ is the set of four faces penetrated by the cross in the dual lattice. 
Upon measuring faces in $X$-basis, one specifies $\prod_{v \in f} Z_v = s_f$ for any face, which is the ground state of the 3D-version of Xu-Moore model with a  degeneracy $\sim 2^{3L-2}$~\cite{3Dplaquette}.  Post-selecting outcomes without any frustration, the corresponding 3D Ising plaquette model exhibits a first-order phase transition at finite temperature $\beta_c \approx 0.55$~\cite{3Dplaquette}.
It implies that the long-range entanglement of the  superposition of exponentially many SSB configurations for the post-measurement state is robust upto $\theta_c \approx 60^\circ$ with post-selection $s_f = 1$. Even without any post-selection, the long-range entanglement is expected to be robust for $\beta^{-1} > 0$ with some frustrated interactions as long as the subsystem symmetry is intact, similar to \figref{fig:random}. Therefore, similar to the RBIM and RPGM, we expect the transition to occur at finite $\theta$. The exact phase diagram of 3d random plaquette Ising model (RPIM) at finite temperature is left for future study.

Upon measuring vertices in $X$-basis, we obtain $\prod_{f \ni v} \bm{Z}_f = s_v$, which gives rise to a X-cube fracton state~\cite{Xcube} in the dual lattice.  Again, the resulting state is a superposition of $\sim 2^{6L-3}$ degenerate ground states. If we thought of qubits on the faces (vertices) as edges (cubes, denoted by $\mbox{\mancube}$) of the dual lattice, we get $\prod_{\tilde{e} \in \mbox{\mancube} } \bm{Z}_{\tilde{e}} = s_{ \mbox{\mancube} }$.
Moving away from the $X$-basis with an angle $\theta$, the wavefunction is given by $\ket{\cP_{\bm{s}} \psi} \propto e^{-\beta \hat{H}_\textrm{X-cube}/2} \otimes \big|\bm{X}_f = \prod_{v \in f} s_v \big\rangle$ where $\hat{H}$ is defined as the following on the dual lattice:
\begin{equation}
    \hat{H}_\textrm{X-cube} = - \sum_{\mbox{\mancube}} s_{ \mbox{\mancube} } \Bigg( \prod_{\tilde{e} \in \mbox{\mancube} } \bm{Z}_{\tilde{e}} \Bigg) - \sum_{+} \prod_{\tilde{e} \in +} \bm{X}_{\tilde{e}}
\end{equation}
The properties of the post-measurement state at angle $\theta$ can be argued to be trivial based on the observation that the corresponding 12-body spin model has no phase transition at $T>0$. Following our previous strategy to calculate the norm of the post-measurement state, we can show that $\braket{\cP_{\bm{s}} \psi} \sim \frac{1}{2^N} ( 1 + 2^{3L} (\cos \theta)^{{\cal O}(L^2)} ) \sim c_0 +  e^{c_1 L - c_2(\theta) L^2}$ where the summation is taken for all possible intersecting planes ($2^{3L}$), i.e., elements of the subsystem symmetry $G_A$. Since the second term gets exponentially suppressed for any $\theta \neq 0$ ($c_2 > 0$) in a large system size, it implies the absence of the phase transition. Therefore, the fracton order is unstable in our scheme for any $\theta \neq 0$.

It is also worth pointing out the connection between two different Hamiltonians resulting from measuring different sublattices of a given cluster SPT. Above, measuring one sublattice gives the 3D Xu-Moore Hamiltonian, while measuring the other sublattice gives the X-cube model. As illustrated in~\cite{Xcube}, these two models are related by a generalized duality via gauging. This is precisely what is happening through the cluster entangler followed by measurements~\cite{NatMeasurement}. We note that the wavefunction of the form in \eqnref{eq:wavefunction} with the Boltzmann weight from a corresponding spin-model in~\cite{Xcube} at a certain $\beta$ would be realized through our protocol by measuring the state obtained by a fracton state coupled with aniclla qubits through cluster entangler.

\section{Summary and Outlook} \label{sec:summary}


In this work, we revealed the fate of quantum states obtained by measuring cluster states in the rotated basis $O_\theta = X \cos \theta + Z \sin \theta$, which is equivalent to applying a certain shallow depth unitary circuit to the product of $\ket{+}$ and perform measurements in the $X$-axis. We showed that any post-measurement state is expressed by a certain product state in $X$-basis under the imaginary time evolution $e^{-\beta \hat{H}}$ by the Hamiltonian depending on the measurement outcomes, where $\beta = \tanh^{-1}(\cos \theta)$. As a result, any post-measurement state has its amplitudes given by the Boltzmann weights of various corresponding classical spin models, ranging from Ising model and gauge theories~\cite{KogutRMP} to plaquette model and even beyond.

At specific angles, the post-measurement wavefunction exhibits quantum criticality, where the wavefunction exhibits spatial conformal symmetry due to the amplitude structure. Constrained by the finite amount of entanglement the shallow circuit can infuse, the resulting state has a constrained entanglement structure, giving rise to a special family of quantum critical states. A family of found quantum criticalities is called conformal quantum critical points (CQCPs), and we found that the dynamical exponent $z \geq 2$ for all the examples discussed in this work, which is consistent with the analytical bound for the dynamical exponent $z=2$ argued in several literature~\cite{z2CFT1, IsakovCQCP}.
In particular, in any dimensions, we found a family of post-measurement states whose parent Hamiltonian is generated by a so-called pivoting structure~\cite{pivot} with the phase diagram in \figref{fig:phase_diagram}. From the Kramers-Wannier duality, the CQCPs for this family of states maps into the Bose-Einstein condensation transition of hardcore bosons with $z=2$~\cite{uzunov1981, BEC, fisher1989boson}. We remark that this class of CQCPs with $z=2$ has non-local $\U(1)$ symmetry~\cite{pivot}, analogous to quantum Lifschitz transitions and famous Rohksar-Kivelson model~\cite{RKmodel, RKmodel2, z2CFT1} with $z=2$. We also found nontrivial examples with $z > 2$ where there is no extra $\U(1)$ symmetry that protects the dynamical critical exponent~\cite{IsakovCQCP}: $z\approx 2.16$ (2D) and $z\approx 2.02$ (3D) for Ising CQCPs with post-selections. 

Interestingly, we found that the post-measurement wavefunctions with $z=2$ CQCPs are all unstable in their long-range entanglement structure under $\theta \neq 0$. This is intimately tied to the observation that the cluster states are described by tensor networks with a finite bond dimension with area-law entanglement capacity, and single-sited measurements cannot change the underlying tensor-network structure. 
Indeed, for all the examples ($z=2$) presented, the groundstate at $\theta=0$ already saturates the entanglement capacity. As the post-measurement state parameterized by $\theta$ already saturates its entanglement entropy at $\theta=0$ and its entanglement entropy cannot increase with $\theta >0$, it is natural for its parent Hamiltonian to host a (multi)critical point at $\theta=0$.

Without post-selections, which is more realistic in both practical and fundamental sense, one can obtain the \emph{ensemble} of post-measurement quantum states, whose correlation structure is trivial once averaged over measurement outcomes. However, we demonstrated the presence of classical decoding protocols for given measurement outcomes, which allows one to identify a hidden long-range entanglement structure and its phase transition behavior in the experiment without exponential scaling of the measurement complexity. 
In particular, we elaborated for the 2D cluster state measured on edges in a rotated basis, where the resulting state can have a long-range correlation structure (GHZ-type) depending on the rotation angle $\theta$. 
By decoding this correlation structure into the ferromagnetic correlation of the 2D RBIM without post-selection, we provided an experimentally efficient protocol to detect the transition behavior of this long-range entanglement structure. We also generalized this idea to other ensembles of states, corresponding to random interaction models such as 3D random bond Ising model or random plaquette gauge models. 

Our results have several implications. First of all, our work establishes a clear demonstration of several important concepts: \emph{decoding} and \emph{identifiability} of a measurement-prepared ensemble of states. Any quantum state obtained by extensive number of measurements is not reproducible in a sense that one has to repeat the experiment exponential amount of times in the system size. To experimentally claim the presence of any interesting feature \emph{hidden} inside the ensemble of quantum states, one must be able to transform the resulting quantum states into an \emph{identifiable} form such that different measurement outcomes can lead to the quantum states with a shared feature that is experimentally verifiable.
In the family of ensembles of quantum states prepared by our setup in \figref{fig:setting}, we rigorously showed that decoding protocols exist to achieve this job.

With the aid of this decoding protocol, our work also answers the robustness of the measurement-based quantum state preparation in a rigorous way by mapping the problem into the concrete statistical mechanics problem. Furthermore, it provides an experimental guideline to prepare an exotic family of conformal quantum critical states. Excitingly, our framework is generalizable to various quantum phases. Although we have considered $\mathbb{Z}_2$ higher-form symmetries, in general we can consider $\mathbb{Z}_N$ symmetries for graph states with qudits, which we conjecture to give rise to general Potts model and $\mathbb{Z}_N$ gauge theories among others. Furthermore, our method would allow experimental preparation of a so-called skeleton states~\cite{gaplessMPS, PEPS_parent, skeleton} to higher dimensions, which are critical quantum states represented by tensor networks with finite bond dimensions. 

{\it Note Added:} Upon completion of the present manuscript, we became aware of an independent work studying extended long-range entangled phases and transitions from finite-depth unitaries and measurement, which will appear in the same arXiv posting~\cite{GuoYi}.

\vspace{20pt}

\acknowledgements

We thank Jeongwan Haah, Aleksander Kubica, David T. Stephen, Nat Tantivasadakarn, Ryan Thorngren, Romain Vasseur, Ruben Verresen, and Sagar Vijay for inspiring and helpful discussion with their previous works. 
We especially thank Soonwon Choi for providing insightful comments as well as coining a term \emph{computationally assisted observable} for our protocol to decode hidden long-range order.
JYL is supported by the Gordon and Betty Moore Foundation under the grant GBMF8690 and by the National Science Foundation under the grant PHY-1748958. WJ and ZB thanks for the Aspen Center for Physics, which is supported by National Science Foundation grant PHY-1607611, where part of the work was performed. WJ is supported by a grant from the Simons Foundation. This work was supported in part by the Heising-Simons Foundation (M.P.A.F.), and by the Simons Collaboration on Ultra-Quantum Matter, which is a grant from the Simons Foundation (651440, M.P.A.F.)


\begin{thebibliography}{89}%
\makeatletter
\providecommand \@ifxundefined [1]{%
 \@ifx{#1\undefined}
}%
\providecommand \@ifnum [1]{%
 \ifnum #1\expandafter \@firstoftwo
 \else \expandafter \@secondoftwo
 \fi
}%
\providecommand \@ifx [1]{%
 \ifx #1\expandafter \@firstoftwo
 \else \expandafter \@secondoftwo
 \fi
}%
\providecommand \natexlab [1]{#1}%
\providecommand \enquote  [1]{``#1''}%
\providecommand \bibnamefont  [1]{#1}%
\providecommand \bibfnamefont [1]{#1}%
\providecommand \citenamefont [1]{#1}%
\providecommand \href@noop [0]{\@secondoftwo}%
\providecommand \href [0]{\begingroup \@sanitize@url \@href}%
\providecommand \@href[1]{\@@startlink{#1}\@@href}%
\providecommand \@@href[1]{\endgroup#1\@@endlink}%
\providecommand \@sanitize@url [0]{\catcode `\\12\catcode `\$12\catcode
  `\&12\catcode `\#12\catcode `\^12\catcode `\_12\catcode `\%12\relax}%
\providecommand \@@startlink[1]{}%
\providecommand \@@endlink[0]{}%
\providecommand \url  [0]{\begingroup\@sanitize@url \@url }%
\providecommand \@url [1]{\endgroup\@href {#1}{\urlprefix }}%
\providecommand \urlprefix  [0]{URL }%
\providecommand \Eprint [0]{\href }%
\providecommand \doibase [0]{https://doi.org/}%
\providecommand \selectlanguage [0]{\@gobble}%
\providecommand \bibinfo  [0]{\@secondoftwo}%
\providecommand \bibfield  [0]{\@secondoftwo}%
\providecommand \translation [1]{[#1]}%
\providecommand \BibitemOpen [0]{}%
\providecommand \bibitemStop [0]{}%
\providecommand \bibitemNoStop [0]{.\EOS\space}%
\providecommand \EOS [0]{\spacefactor3000\relax}%
\providecommand \BibitemShut  [1]{\csname bibitem#1\endcsname}%
\let\auto@bib@innerbib\@empty
\bibitem [{\citenamefont {Li}\ \emph {et~al.}(2018)\citenamefont {Li},
  \citenamefont {Chen},\ and\ \citenamefont {Fisher}}]{YaodongFisher2018}%
  \BibitemOpen
  \bibfield  {author} {\bibinfo {author} {\bibfnamefont {Y.}~\bibnamefont
  {Li}}, \bibinfo {author} {\bibfnamefont {X.}~\bibnamefont {Chen}},\ and\
  \bibinfo {author} {\bibfnamefont {M.~P.~A.}\ \bibnamefont {Fisher}},\
  }\bibfield  {title} {\bibinfo {title} {Quantum zeno effect and the many-body
  entanglement transition},\ }\href
  {https://doi.org/10.1103/PhysRevB.98.205136} {\bibfield  {journal} {\bibinfo
  {journal} {Phys. Rev. B}\ }\textbf {\bibinfo {volume} {98}},\ \bibinfo
  {pages} {205136} (\bibinfo {year} {2018})}\BibitemShut {NoStop}%
\bibitem [{\citenamefont {Skinner}\ \emph {et~al.}(2019)\citenamefont
  {Skinner}, \citenamefont {Ruhman},\ and\ \citenamefont
  {Nahum}}]{SkinnerNahum2018}%
  \BibitemOpen
  \bibfield  {author} {\bibinfo {author} {\bibfnamefont {B.}~\bibnamefont
  {Skinner}}, \bibinfo {author} {\bibfnamefont {J.}~\bibnamefont {Ruhman}},\
  and\ \bibinfo {author} {\bibfnamefont {A.}~\bibnamefont {Nahum}},\ }\bibfield
   {title} {\bibinfo {title} {Measurement-induced phase transitions in the
  dynamics of entanglement},\ }\href
  {https://doi.org/10.1103/PhysRevX.9.031009} {\bibfield  {journal} {\bibinfo
  {journal} {Phys. Rev. X}\ }\textbf {\bibinfo {volume} {9}},\ \bibinfo {pages}
  {031009} (\bibinfo {year} {2019})}\BibitemShut {NoStop}%
\bibitem [{\citenamefont {Chan}\ \emph {et~al.}(2019)\citenamefont {Chan},
  \citenamefont {Nandkishore}, \citenamefont {Pretko},\ and\ \citenamefont
  {Smith}}]{Chan2019}%
  \BibitemOpen
  \bibfield  {author} {\bibinfo {author} {\bibfnamefont {A.}~\bibnamefont
  {Chan}}, \bibinfo {author} {\bibfnamefont {R.~M.}\ \bibnamefont
  {Nandkishore}}, \bibinfo {author} {\bibfnamefont {M.}~\bibnamefont
  {Pretko}},\ and\ \bibinfo {author} {\bibfnamefont {G.}~\bibnamefont
  {Smith}},\ }\bibfield  {title} {\bibinfo {title} {Unitary-projective
  entanglement dynamics},\ }\href {https://doi.org/10.1103/PhysRevB.99.224307}
  {\bibfield  {journal} {\bibinfo  {journal} {Phys. Rev. B}\ }\textbf {\bibinfo
  {volume} {99}},\ \bibinfo {pages} {224307} (\bibinfo {year}
  {2019})}\BibitemShut {NoStop}%
\bibitem [{\citenamefont {Li}\ \emph {et~al.}(2019)\citenamefont {Li},
  \citenamefont {Chen},\ and\ \citenamefont {Fisher}}]{YaodongFisher2019}%
  \BibitemOpen
  \bibfield  {author} {\bibinfo {author} {\bibfnamefont {Y.}~\bibnamefont
  {Li}}, \bibinfo {author} {\bibfnamefont {X.}~\bibnamefont {Chen}},\ and\
  \bibinfo {author} {\bibfnamefont {M.~P.~A.}\ \bibnamefont {Fisher}},\
  }\bibfield  {title} {\bibinfo {title} {Measurement-driven entanglement
  transition in hybrid quantum circuits},\ }\href
  {https://doi.org/10.1103/PhysRevB.100.134306} {\bibfield  {journal} {\bibinfo
   {journal} {Phys. Rev. B}\ }\textbf {\bibinfo {volume} {100}},\ \bibinfo
  {pages} {134306} (\bibinfo {year} {2019})}\BibitemShut {NoStop}%
\bibitem [{\citenamefont {Vasseur}\ \emph {et~al.}(2019)\citenamefont
  {Vasseur}, \citenamefont {Potter}, \citenamefont {You},\ and\ \citenamefont
  {Ludwig}}]{Vasseur2019}%
  \BibitemOpen
  \bibfield  {author} {\bibinfo {author} {\bibfnamefont {R.}~\bibnamefont
  {Vasseur}}, \bibinfo {author} {\bibfnamefont {A.~C.}\ \bibnamefont {Potter}},
  \bibinfo {author} {\bibfnamefont {Y.-Z.}\ \bibnamefont {You}},\ and\ \bibinfo
  {author} {\bibfnamefont {A.~W.~W.}\ \bibnamefont {Ludwig}},\ }\bibfield
  {title} {\bibinfo {title} {Entanglement transitions from holographic random
  tensor networks},\ }\href {https://doi.org/10.1103/PhysRevB.100.134203}
  {\bibfield  {journal} {\bibinfo  {journal} {Phys. Rev. B}\ }\textbf {\bibinfo
  {volume} {100}},\ \bibinfo {pages} {134203} (\bibinfo {year}
  {2019})}\BibitemShut {NoStop}%
\bibitem [{\citenamefont {Cao}\ \emph {et~al.}(2019)\citenamefont {Cao},
  \citenamefont {Tilloy},\ and\ \citenamefont {Luca}}]{XiangyuLuca2019}%
  \BibitemOpen
  \bibfield  {author} {\bibinfo {author} {\bibfnamefont {X.}~\bibnamefont
  {Cao}}, \bibinfo {author} {\bibfnamefont {A.}~\bibnamefont {Tilloy}},\ and\
  \bibinfo {author} {\bibfnamefont {A.~D.}\ \bibnamefont {Luca}},\ }\bibfield
  {title} {\bibinfo {title} {{Entanglement in a fermion chain under continuous
  monitoring}},\ }\href {https://doi.org/10.21468/SciPostPhys.7.2.024}
  {\bibfield  {journal} {\bibinfo  {journal} {SciPost Phys.}\ }\textbf
  {\bibinfo {volume} {7}},\ \bibinfo {pages} {24} (\bibinfo {year}
  {2019})}\BibitemShut {NoStop}%
\bibitem [{\citenamefont {Gullans}\ and\ \citenamefont
  {Huse}(2020{\natexlab{a}})}]{Gullans2020}%
  \BibitemOpen
  \bibfield  {author} {\bibinfo {author} {\bibfnamefont {M.~J.}\ \bibnamefont
  {Gullans}}\ and\ \bibinfo {author} {\bibfnamefont {D.~A.}\ \bibnamefont
  {Huse}},\ }\bibfield  {title} {\bibinfo {title} {Dynamical purification phase
  transition induced by quantum measurements},\ }\href
  {https://doi.org/10.1103/PhysRevX.10.041020} {\bibfield  {journal} {\bibinfo
  {journal} {Phys. Rev. X}\ }\textbf {\bibinfo {volume} {10}},\ \bibinfo
  {pages} {041020} (\bibinfo {year} {2020}{\natexlab{a}})}\BibitemShut
  {NoStop}%
\bibitem [{\citenamefont {Choi}\ \emph {et~al.}(2020)\citenamefont {Choi},
  \citenamefont {Bao}, \citenamefont {Qi},\ and\ \citenamefont
  {Altman}}]{Soonwon2020}%
  \BibitemOpen
  \bibfield  {author} {\bibinfo {author} {\bibfnamefont {S.}~\bibnamefont
  {Choi}}, \bibinfo {author} {\bibfnamefont {Y.}~\bibnamefont {Bao}}, \bibinfo
  {author} {\bibfnamefont {X.-L.}\ \bibnamefont {Qi}},\ and\ \bibinfo {author}
  {\bibfnamefont {E.}~\bibnamefont {Altman}},\ }\bibfield  {title} {\bibinfo
  {title} {Quantum error correction in scrambling dynamics and
  measurement-induced phase transition},\ }\href
  {https://doi.org/10.1103/PhysRevLett.125.030505} {\bibfield  {journal}
  {\bibinfo  {journal} {Phys. Rev. Lett.}\ }\textbf {\bibinfo {volume} {125}},\
  \bibinfo {pages} {030505} (\bibinfo {year} {2020})}\BibitemShut {NoStop}%
\bibitem [{\citenamefont {Tang}\ and\ \citenamefont {Zhu}(2020)}]{TangZhu2020}%
  \BibitemOpen
  \bibfield  {author} {\bibinfo {author} {\bibfnamefont {Q.}~\bibnamefont
  {Tang}}\ and\ \bibinfo {author} {\bibfnamefont {W.}~\bibnamefont {Zhu}},\
  }\bibfield  {title} {\bibinfo {title} {Measurement-induced phase transition:
  A case study in the nonintegrable model by density-matrix renormalization
  group calculations},\ }\href
  {https://doi.org/10.1103/PhysRevResearch.2.013022} {\bibfield  {journal}
  {\bibinfo  {journal} {Phys. Rev. Research}\ }\textbf {\bibinfo {volume}
  {2}},\ \bibinfo {pages} {013022} (\bibinfo {year} {2020})}\BibitemShut
  {NoStop}%
\bibitem [{\citenamefont {Jian}\ \emph {et~al.}(2020)\citenamefont {Jian},
  \citenamefont {You}, \citenamefont {Vasseur},\ and\ \citenamefont
  {Ludwig}}]{JianLudwig2020}%
  \BibitemOpen
  \bibfield  {author} {\bibinfo {author} {\bibfnamefont {C.-M.}\ \bibnamefont
  {Jian}}, \bibinfo {author} {\bibfnamefont {Y.-Z.}\ \bibnamefont {You}},
  \bibinfo {author} {\bibfnamefont {R.}~\bibnamefont {Vasseur}},\ and\ \bibinfo
  {author} {\bibfnamefont {A.~W.~W.}\ \bibnamefont {Ludwig}},\ }\bibfield
  {title} {\bibinfo {title} {Measurement-induced criticality in random quantum
  circuits},\ }\href {https://doi.org/10.1103/PhysRevB.101.104302} {\bibfield
  {journal} {\bibinfo  {journal} {Phys. Rev. B}\ }\textbf {\bibinfo {volume}
  {101}},\ \bibinfo {pages} {104302} (\bibinfo {year} {2020})}\BibitemShut
  {NoStop}%
\bibitem [{\citenamefont {Lopez-Piqueres}\ \emph {et~al.}(2020)\citenamefont
  {Lopez-Piqueres}, \citenamefont {Ware},\ and\ \citenamefont
  {Vasseur}}]{LopezVasseur2020}%
  \BibitemOpen
  \bibfield  {author} {\bibinfo {author} {\bibfnamefont {J.}~\bibnamefont
  {Lopez-Piqueres}}, \bibinfo {author} {\bibfnamefont {B.}~\bibnamefont
  {Ware}},\ and\ \bibinfo {author} {\bibfnamefont {R.}~\bibnamefont
  {Vasseur}},\ }\bibfield  {title} {\bibinfo {title} {Mean-field entanglement
  transitions in random tree tensor networks},\ }\href
  {https://doi.org/10.1103/PhysRevB.102.064202} {\bibfield  {journal} {\bibinfo
   {journal} {Phys. Rev. B}\ }\textbf {\bibinfo {volume} {102}},\ \bibinfo
  {pages} {064202} (\bibinfo {year} {2020})}\BibitemShut {NoStop}%
\bibitem [{\citenamefont {Bao}\ \emph {et~al.}(2020)\citenamefont {Bao},
  \citenamefont {Choi},\ and\ \citenamefont {Altman}}]{Bao2020}%
  \BibitemOpen
  \bibfield  {author} {\bibinfo {author} {\bibfnamefont {Y.}~\bibnamefont
  {Bao}}, \bibinfo {author} {\bibfnamefont {S.}~\bibnamefont {Choi}},\ and\
  \bibinfo {author} {\bibfnamefont {E.}~\bibnamefont {Altman}},\ }\bibfield
  {title} {\bibinfo {title} {Theory of the phase transition in random unitary
  circuits with measurements},\ }\href
  {https://doi.org/10.1103/PhysRevB.101.104301} {\bibfield  {journal} {\bibinfo
   {journal} {Phys. Rev. B}\ }\textbf {\bibinfo {volume} {101}},\ \bibinfo
  {pages} {104301} (\bibinfo {year} {2020})}\BibitemShut {NoStop}%
\bibitem [{\citenamefont {Rossini}\ and\ \citenamefont
  {Vicari}(2020)}]{Rossini2020}%
  \BibitemOpen
  \bibfield  {author} {\bibinfo {author} {\bibfnamefont {D.}~\bibnamefont
  {Rossini}}\ and\ \bibinfo {author} {\bibfnamefont {E.}~\bibnamefont
  {Vicari}},\ }\bibfield  {title} {\bibinfo {title} {Measurement-induced
  dynamics of many-body systems at quantum criticality},\ }\href
  {https://doi.org/10.1103/PhysRevB.102.035119} {\bibfield  {journal} {\bibinfo
   {journal} {Phys. Rev. B}\ }\textbf {\bibinfo {volume} {102}},\ \bibinfo
  {pages} {035119} (\bibinfo {year} {2020})}\BibitemShut {NoStop}%
\bibitem [{\citenamefont {Fan}\ \emph {et~al.}(2021)\citenamefont {Fan},
  \citenamefont {Vijay}, \citenamefont {Vishwanath},\ and\ \citenamefont
  {You}}]{Fan2021}%
  \BibitemOpen
  \bibfield  {author} {\bibinfo {author} {\bibfnamefont {R.}~\bibnamefont
  {Fan}}, \bibinfo {author} {\bibfnamefont {S.}~\bibnamefont {Vijay}}, \bibinfo
  {author} {\bibfnamefont {A.}~\bibnamefont {Vishwanath}},\ and\ \bibinfo
  {author} {\bibfnamefont {Y.-Z.}\ \bibnamefont {You}},\ }\bibfield  {title}
  {\bibinfo {title} {Self-organized error correction in random unitary circuits
  with measurement},\ }\href {https://doi.org/10.1103/PhysRevB.103.174309}
  {\bibfield  {journal} {\bibinfo  {journal} {Phys. Rev. B}\ }\textbf {\bibinfo
  {volume} {103}},\ \bibinfo {pages} {174309} (\bibinfo {year}
  {2021})}\BibitemShut {NoStop}%
\bibitem [{\citenamefont {Li}\ \emph {et~al.}(2021)\citenamefont {Li},
  \citenamefont {Chen}, \citenamefont {Ludwig},\ and\ \citenamefont
  {Fisher}}]{Yaodong2021}%
  \BibitemOpen
  \bibfield  {author} {\bibinfo {author} {\bibfnamefont {Y.}~\bibnamefont
  {Li}}, \bibinfo {author} {\bibfnamefont {X.}~\bibnamefont {Chen}}, \bibinfo
  {author} {\bibfnamefont {A.~W.~W.}\ \bibnamefont {Ludwig}},\ and\ \bibinfo
  {author} {\bibfnamefont {M.~P.~A.}\ \bibnamefont {Fisher}},\ }\bibfield
  {title} {\bibinfo {title} {Conformal invariance and quantum nonlocality in
  critical hybrid circuits},\ }\href
  {https://doi.org/10.1103/PhysRevB.104.104305} {\bibfield  {journal} {\bibinfo
   {journal} {Phys. Rev. B}\ }\textbf {\bibinfo {volume} {104}},\ \bibinfo
  {pages} {104305} (\bibinfo {year} {2021})}\BibitemShut {NoStop}%
\bibitem [{\citenamefont {Ben-Zion}\ \emph {et~al.}(2020)\citenamefont
  {Ben-Zion}, \citenamefont {McGreevy},\ and\ \citenamefont
  {Grover}}]{BenZion2020}%
  \BibitemOpen
  \bibfield  {author} {\bibinfo {author} {\bibfnamefont {D.}~\bibnamefont
  {Ben-Zion}}, \bibinfo {author} {\bibfnamefont {J.}~\bibnamefont {McGreevy}},\
  and\ \bibinfo {author} {\bibfnamefont {T.}~\bibnamefont {Grover}},\
  }\bibfield  {title} {\bibinfo {title} {Disentangling quantum matter with
  measurements},\ }\href {https://doi.org/10.1103/PhysRevB.101.115131}
  {\bibfield  {journal} {\bibinfo  {journal} {Phys. Rev. B}\ }\textbf {\bibinfo
  {volume} {101}},\ \bibinfo {pages} {115131} (\bibinfo {year}
  {2020})}\BibitemShut {NoStop}%
\bibitem [{\citenamefont {{Barratt}}\ \emph {et~al.}(2022)\citenamefont
  {{Barratt}}, \citenamefont {{Agrawal}}, \citenamefont {{Potter}},
  \citenamefont {{Gopalakrishnan}},\ and\ \citenamefont
  {{Vasseur}}}]{BarrattDecoding2022}%
  \BibitemOpen
  \bibfield  {author} {\bibinfo {author} {\bibfnamefont {F.}~\bibnamefont
  {{Barratt}}}, \bibinfo {author} {\bibfnamefont {U.}~\bibnamefont
  {{Agrawal}}}, \bibinfo {author} {\bibfnamefont {A.~C.}\ \bibnamefont
  {{Potter}}}, \bibinfo {author} {\bibfnamefont {S.}~\bibnamefont
  {{Gopalakrishnan}}},\ and\ \bibinfo {author} {\bibfnamefont {R.}~\bibnamefont
  {{Vasseur}}},\ }\bibfield  {title} {\bibinfo {title} {{Transitions in the
  learnability of global charges from local measurements}},\ }\href@noop {}
  {\bibfield  {journal} {\bibinfo  {journal} {arXiv e-prints}\ ,\ \bibinfo
  {eid} {arXiv:2206.12429}} (\bibinfo {year} {2022})}\BibitemShut {NoStop}%
\bibitem [{\citenamefont {Briegel}\ and\ \citenamefont
  {Raussendorf}(2001)}]{1Dcluster_GHZ}%
  \BibitemOpen
  \bibfield  {author} {\bibinfo {author} {\bibfnamefont {H.~J.}\ \bibnamefont
  {Briegel}}\ and\ \bibinfo {author} {\bibfnamefont {R.}~\bibnamefont
  {Raussendorf}},\ }\bibfield  {title} {\bibinfo {title} {Persistent
  entanglement in arrays of interacting particles},\ }\href
  {https://doi.org/10.1103/PhysRevLett.86.910} {\bibfield  {journal} {\bibinfo
  {journal} {Phys. Rev. Lett.}\ }\textbf {\bibinfo {volume} {86}},\ \bibinfo
  {pages} {910} (\bibinfo {year} {2001})}\BibitemShut {NoStop}%
\bibitem [{\citenamefont {Raussendorf}\ \emph {et~al.}(2005)\citenamefont
  {Raussendorf}, \citenamefont {Bravyi},\ and\ \citenamefont
  {Harrington}}]{2Dcluster}%
  \BibitemOpen
  \bibfield  {author} {\bibinfo {author} {\bibfnamefont {R.}~\bibnamefont
  {Raussendorf}}, \bibinfo {author} {\bibfnamefont {S.}~\bibnamefont
  {Bravyi}},\ and\ \bibinfo {author} {\bibfnamefont {J.}~\bibnamefont
  {Harrington}},\ }\bibfield  {title} {\bibinfo {title} {Long-range quantum
  entanglement in noisy cluster states},\ }\href
  {https://doi.org/10.1103/PhysRevA.71.062313} {\bibfield  {journal} {\bibinfo
  {journal} {Phys. Rev. A}\ }\textbf {\bibinfo {volume} {71}},\ \bibinfo
  {pages} {062313} (\bibinfo {year} {2005})}\BibitemShut {NoStop}%
\bibitem [{\citenamefont {Aguado}\ \emph {et~al.}(2008)\citenamefont {Aguado},
  \citenamefont {Brennen}, \citenamefont {Verstraete},\ and\ \citenamefont
  {Cirac}}]{2Dcluster_toric}%
  \BibitemOpen
  \bibfield  {author} {\bibinfo {author} {\bibfnamefont {M.}~\bibnamefont
  {Aguado}}, \bibinfo {author} {\bibfnamefont {G.~K.}\ \bibnamefont {Brennen}},
  \bibinfo {author} {\bibfnamefont {F.}~\bibnamefont {Verstraete}},\ and\
  \bibinfo {author} {\bibfnamefont {J.~I.}\ \bibnamefont {Cirac}},\ }\bibfield
  {title} {\bibinfo {title} {Creation, manipulation, and detection of abelian
  and non-abelian anyons in optical lattices},\ }\href
  {https://doi.org/10.1103/PhysRevLett.101.260501} {\bibfield  {journal}
  {\bibinfo  {journal} {Phys. Rev. Lett.}\ }\textbf {\bibinfo {volume} {101}},\
  \bibinfo {pages} {260501} (\bibinfo {year} {2008})}\BibitemShut {NoStop}%
\bibitem [{\citenamefont {Piroli}\ \emph {et~al.}(2021)\citenamefont {Piroli},
  \citenamefont {Styliaris},\ and\ \citenamefont {Cirac}}]{Piroli2021}%
  \BibitemOpen
  \bibfield  {author} {\bibinfo {author} {\bibfnamefont {L.}~\bibnamefont
  {Piroli}}, \bibinfo {author} {\bibfnamefont {G.}~\bibnamefont {Styliaris}},\
  and\ \bibinfo {author} {\bibfnamefont {J.~I.}\ \bibnamefont {Cirac}},\
  }\bibfield  {title} {\bibinfo {title} {Quantum circuits assisted by local
  operations and classical communication: Transformations and phases of
  matter},\ }\href {https://doi.org/10.1103/PhysRevLett.127.220503} {\bibfield
  {journal} {\bibinfo  {journal} {Phys. Rev. Lett.}\ }\textbf {\bibinfo
  {volume} {127}},\ \bibinfo {pages} {220503} (\bibinfo {year}
  {2021})}\BibitemShut {NoStop}%
\bibitem [{\citenamefont {Bolt}\ \emph
  {et~al.}(2016{\natexlab{a}})\citenamefont {Bolt}, \citenamefont
  {Duclos-Cianci}, \citenamefont {Poulin},\ and\ \citenamefont
  {Stace}}]{3dCluster_fracton1}%
  \BibitemOpen
  \bibfield  {author} {\bibinfo {author} {\bibfnamefont {A.}~\bibnamefont
  {Bolt}}, \bibinfo {author} {\bibfnamefont {G.}~\bibnamefont {Duclos-Cianci}},
  \bibinfo {author} {\bibfnamefont {D.}~\bibnamefont {Poulin}},\ and\ \bibinfo
  {author} {\bibfnamefont {T.~M.}\ \bibnamefont {Stace}},\ }\bibfield  {title}
  {\bibinfo {title} {Foliated quantum error-correcting codes},\ }\href
  {https://doi.org/10.1103/PhysRevLett.117.070501} {\bibfield  {journal}
  {\bibinfo  {journal} {Phys. Rev. Lett.}\ }\textbf {\bibinfo {volume} {117}},\
  \bibinfo {pages} {070501} (\bibinfo {year} {2016}{\natexlab{a}})}\BibitemShut
  {NoStop}%
\bibitem [{\citenamefont {Williamson}\ and\ \citenamefont
  {Devakul}(2021)}]{3dCluster_fracton2}%
  \BibitemOpen
  \bibfield  {author} {\bibinfo {author} {\bibfnamefont {D.~J.}\ \bibnamefont
  {Williamson}}\ and\ \bibinfo {author} {\bibfnamefont {T.}~\bibnamefont
  {Devakul}},\ }\bibfield  {title} {\bibinfo {title} {Type-ii fractons from
  coupled spin chains and layers},\ }\href
  {https://doi.org/10.1103/PhysRevB.103.155140} {\bibfield  {journal} {\bibinfo
   {journal} {Phys. Rev. B}\ }\textbf {\bibinfo {volume} {103}},\ \bibinfo
  {pages} {155140} (\bibinfo {year} {2021})}\BibitemShut {NoStop}%
\bibitem [{\citenamefont {{Verresen}}\ \emph {et~al.}(2021)\citenamefont
  {{Verresen}}, \citenamefont {{Tantivasadakarn}},\ and\ \citenamefont
  {{Vishwanath}}}]{NatRydberg}%
  \BibitemOpen
  \bibfield  {author} {\bibinfo {author} {\bibfnamefont {R.}~\bibnamefont
  {{Verresen}}}, \bibinfo {author} {\bibfnamefont {N.}~\bibnamefont
  {{Tantivasadakarn}}},\ and\ \bibinfo {author} {\bibfnamefont
  {A.}~\bibnamefont {{Vishwanath}}},\ }\bibfield  {title} {\bibinfo {title}
  {{Efficiently preparing Schr{\"o}dinger's cat, fractons and non-Abelian
  topological order in quantum devices}},\ }\href@noop {} {\bibfield  {journal}
  {\bibinfo  {journal} {arXiv e-prints}\ ,\ \bibinfo {eid} {arXiv:2112.03061}}
  (\bibinfo {year} {2021})}\BibitemShut {NoStop}%
\bibitem [{\citenamefont {Calderbank}\ and\ \citenamefont
  {Shor}(1996)}]{CSScode}%
  \BibitemOpen
  \bibfield  {author} {\bibinfo {author} {\bibfnamefont {A.~R.}\ \bibnamefont
  {Calderbank}}\ and\ \bibinfo {author} {\bibfnamefont {P.~W.}\ \bibnamefont
  {Shor}},\ }\bibfield  {title} {\bibinfo {title} {Good quantum
  error-correcting codes exist},\ }\href
  {https://doi.org/10.1103/PhysRevA.54.1098} {\bibfield  {journal} {\bibinfo
  {journal} {Phys. Rev. A}\ }\textbf {\bibinfo {volume} {54}},\ \bibinfo
  {pages} {1098} (\bibinfo {year} {1996})}\BibitemShut {NoStop}%
\bibitem [{\citenamefont {Steane}(1996)}]{CSScode2}%
  \BibitemOpen
  \bibfield  {author} {\bibinfo {author} {\bibfnamefont {A.}~\bibnamefont
  {Steane}},\ }\bibfield  {title} {\bibinfo {title} {Multiple-particle
  interference and quantum error correction},\ }\href
  {https://doi.org/10.1098/rspa.1996.0136} {\bibfield  {journal} {\bibinfo
  {journal} {Proceedings of the Royal Society of London. Series A:
  Mathematical, Physical and Engineering Sciences}\ }\textbf {\bibinfo {volume}
  {452}},\ \bibinfo {pages} {2551} (\bibinfo {year} {1996})}\BibitemShut
  {NoStop}%
\bibitem [{\citenamefont {Bolt}\ \emph
  {et~al.}(2016{\natexlab{b}})\citenamefont {Bolt}, \citenamefont
  {Duclos-Cianci}, \citenamefont {Poulin},\ and\ \citenamefont
  {Stace}}]{ClusterCSS}%
  \BibitemOpen
  \bibfield  {author} {\bibinfo {author} {\bibfnamefont {A.}~\bibnamefont
  {Bolt}}, \bibinfo {author} {\bibfnamefont {G.}~\bibnamefont {Duclos-Cianci}},
  \bibinfo {author} {\bibfnamefont {D.}~\bibnamefont {Poulin}},\ and\ \bibinfo
  {author} {\bibfnamefont {T.~M.}\ \bibnamefont {Stace}},\ }\bibfield  {title}
  {\bibinfo {title} {Foliated quantum error-correcting codes},\ }\href
  {https://doi.org/10.1103/PhysRevLett.117.070501} {\bibfield  {journal}
  {\bibinfo  {journal} {Phys. Rev. Lett.}\ }\textbf {\bibinfo {volume} {117}},\
  \bibinfo {pages} {070501} (\bibinfo {year} {2016}{\natexlab{b}})}\BibitemShut
  {NoStop}%
\bibitem [{\citenamefont {Lu}\ \emph {et~al.}(2022)\citenamefont {Lu},
  \citenamefont {Lessa}, \citenamefont {Kim},\ and\ \citenamefont
  {Hsieh}}]{Lu2022}%
  \BibitemOpen
  \bibfield  {author} {\bibinfo {author} {\bibfnamefont {T.-C.}\ \bibnamefont
  {Lu}}, \bibinfo {author} {\bibfnamefont {L.~A.}\ \bibnamefont {Lessa}},
  \bibinfo {author} {\bibfnamefont {I.~H.}\ \bibnamefont {Kim}},\ and\ \bibinfo
  {author} {\bibfnamefont {T.~H.}\ \bibnamefont {Hsieh}},\ }\href
  {https://doi.org/10.48550/ARXIV.2206.13527} {\bibinfo {title} {Measurement as
  a shortcut to long-range entangled quantum matter}} (\bibinfo {year}
  {2022})\BibitemShut {NoStop}%
\bibitem [{\citenamefont {Ardonne}\ \emph {et~al.}(2004)\citenamefont
  {Ardonne}, \citenamefont {Fendley},\ and\ \citenamefont {Fradkin}}]{z2CFT1}%
  \BibitemOpen
  \bibfield  {author} {\bibinfo {author} {\bibfnamefont {E.}~\bibnamefont
  {Ardonne}}, \bibinfo {author} {\bibfnamefont {P.}~\bibnamefont {Fendley}},\
  and\ \bibinfo {author} {\bibfnamefont {E.}~\bibnamefont {Fradkin}},\
  }\bibfield  {title} {\bibinfo {title} {Topological order and conformal
  quantum critical points},\ }\href
  {https://doi.org/https://doi.org/10.1016/j.aop.2004.01.004} {\bibfield
  {journal} {\bibinfo  {journal} {Annals of Physics}\ }\textbf {\bibinfo
  {volume} {310}},\ \bibinfo {pages} {493} (\bibinfo {year}
  {2004})}\BibitemShut {NoStop}%
\bibitem [{\citenamefont {Fradkin}\ and\ \citenamefont {Moore}(2006)}]{z2CFT2}%
  \BibitemOpen
  \bibfield  {author} {\bibinfo {author} {\bibfnamefont {E.}~\bibnamefont
  {Fradkin}}\ and\ \bibinfo {author} {\bibfnamefont {J.~E.}\ \bibnamefont
  {Moore}},\ }\bibfield  {title} {\bibinfo {title} {Entanglement entropy of 2d
  conformal quantum critical points: Hearing the shape of a quantum drum},\
  }\href {https://doi.org/10.1103/PhysRevLett.97.050404} {\bibfield  {journal}
  {\bibinfo  {journal} {Phys. Rev. Lett.}\ }\textbf {\bibinfo {volume} {97}},\
  \bibinfo {pages} {050404} (\bibinfo {year} {2006})}\BibitemShut {NoStop}%
\bibitem [{\citenamefont {{Cotler}}\ \emph {et~al.}(2021)\citenamefont
  {{Cotler}}, \citenamefont {{Mark}}, \citenamefont {{Huang}}, \citenamefont
  {{Hernandez}}, \citenamefont {{Choi}}, \citenamefont {{Shaw}}, \citenamefont
  {{Endres}},\ and\ \citenamefont {{Choi}}}]{CotlerChoi}%
  \BibitemOpen
  \bibfield  {author} {\bibinfo {author} {\bibfnamefont {J.~S.}\ \bibnamefont
  {{Cotler}}}, \bibinfo {author} {\bibfnamefont {D.~K.}\ \bibnamefont
  {{Mark}}}, \bibinfo {author} {\bibfnamefont {H.-Y.}\ \bibnamefont {{Huang}}},
  \bibinfo {author} {\bibfnamefont {F.}~\bibnamefont {{Hernandez}}}, \bibinfo
  {author} {\bibfnamefont {J.}~\bibnamefont {{Choi}}}, \bibinfo {author}
  {\bibfnamefont {A.~L.}\ \bibnamefont {{Shaw}}}, \bibinfo {author}
  {\bibfnamefont {M.}~\bibnamefont {{Endres}}},\ and\ \bibinfo {author}
  {\bibfnamefont {S.}~\bibnamefont {{Choi}}},\ }\bibfield  {title} {\bibinfo
  {title} {{Emergent quantum state designs from individual many-body
  wavefunctions}},\ }\href@noop {} {\bibfield  {journal} {\bibinfo  {journal}
  {arXiv e-prints}\ ,\ \bibinfo {eid} {arXiv:2103.03536}} (\bibinfo {year}
  {2021})},\ \Eprint {https://arxiv.org/abs/2103.03536} {arXiv:2103.03536
  [quant-ph]} \BibitemShut {NoStop}%
\bibitem [{\citenamefont {{Choi}}\ \emph {et~al.}(2021)\citenamefont {{Choi}},
  \citenamefont {{Shaw}}, \citenamefont {{Madjarov}}, \citenamefont {{Xie}},
  \citenamefont {{Finkelstein}}, \citenamefont {{Covey}}, \citenamefont
  {{Cotler}}, \citenamefont {{Mark}}, \citenamefont {{Huang}}, \citenamefont
  {{Kale}}, \citenamefont {{Pichler}}, \citenamefont {{Brand{\~a}o}},
  \citenamefont {{Choi}},\ and\ \citenamefont {{Endres}}}]{ChoiShaw2021}%
  \BibitemOpen
  \bibfield  {author} {\bibinfo {author} {\bibfnamefont {J.}~\bibnamefont
  {{Choi}}}, \bibinfo {author} {\bibfnamefont {A.~L.}\ \bibnamefont {{Shaw}}},
  \bibinfo {author} {\bibfnamefont {I.~S.}\ \bibnamefont {{Madjarov}}},
  \bibinfo {author} {\bibfnamefont {X.}~\bibnamefont {{Xie}}}, \bibinfo
  {author} {\bibfnamefont {R.}~\bibnamefont {{Finkelstein}}}, \bibinfo {author}
  {\bibfnamefont {J.~P.}\ \bibnamefont {{Covey}}}, \bibinfo {author}
  {\bibfnamefont {J.~S.}\ \bibnamefont {{Cotler}}}, \bibinfo {author}
  {\bibfnamefont {D.~K.}\ \bibnamefont {{Mark}}}, \bibinfo {author}
  {\bibfnamefont {H.-Y.}\ \bibnamefont {{Huang}}}, \bibinfo {author}
  {\bibfnamefont {A.}~\bibnamefont {{Kale}}}, \bibinfo {author} {\bibfnamefont
  {H.}~\bibnamefont {{Pichler}}}, \bibinfo {author} {\bibfnamefont
  {F.~G.~S.~L.}\ \bibnamefont {{Brand{\~a}o}}}, \bibinfo {author}
  {\bibfnamefont {S.}~\bibnamefont {{Choi}}},\ and\ \bibinfo {author}
  {\bibfnamefont {M.}~\bibnamefont {{Endres}}},\ }\bibfield  {title} {\bibinfo
  {title} {{Emergent Quantum Randomness and Benchmarking from Hamiltonian
  Many-body Dynamics}},\ }\href@noop {} {\bibfield  {journal} {\bibinfo
  {journal} {arXiv e-prints}\ ,\ \bibinfo {eid} {arXiv:2103.03535}} (\bibinfo
  {year} {2021})},\ \Eprint {https://arxiv.org/abs/2103.03535}
  {arXiv:2103.03535 [quant-ph]} \BibitemShut {NoStop}%
\bibitem [{\citenamefont {Nishimori}(1981)}]{Nishimori}%
  \BibitemOpen
  \bibfield  {author} {\bibinfo {author} {\bibfnamefont {H.}~\bibnamefont
  {Nishimori}},\ }\bibfield  {title} {\bibinfo {title} {{Internal Energy,
  Specific Heat and Correlation Function of the Bond-Random Ising Model}},\
  }\href {https://doi.org/10.1143/PTP.66.1169} {\bibfield  {journal} {\bibinfo
  {journal} {Progress of Theoretical Physics}\ }\textbf {\bibinfo {volume}
  {66}},\ \bibinfo {pages} {1169} (\bibinfo {year} {1981})}\BibitemShut
  {NoStop}%
\bibitem [{\citenamefont {Nishimori}(1986)}]{Nishimori2}%
  \BibitemOpen
  \bibfield  {author} {\bibinfo {author} {\bibfnamefont {H.}~\bibnamefont
  {Nishimori}},\ }\bibfield  {title} {\bibinfo {title} {Geometry-induced phase
  transition in the $\pm$j ising model},\ }\href
  {https://doi.org/10.1143/JPSJ.55.3305} {\bibfield  {journal} {\bibinfo
  {journal} {Journal of the Physical Society of Japan}\ }\textbf {\bibinfo
  {volume} {55}},\ \bibinfo {pages} {3305} (\bibinfo {year}
  {1986})}\BibitemShut {NoStop}%
\bibitem [{\citenamefont {Honecker}\ \emph {et~al.}(2001)\citenamefont
  {Honecker}, \citenamefont {Picco},\ and\ \citenamefont
  {Pujol}}]{NishimoriPoint}%
  \BibitemOpen
  \bibfield  {author} {\bibinfo {author} {\bibfnamefont {A.}~\bibnamefont
  {Honecker}}, \bibinfo {author} {\bibfnamefont {M.}~\bibnamefont {Picco}},\
  and\ \bibinfo {author} {\bibfnamefont {P.}~\bibnamefont {Pujol}},\ }\bibfield
   {title} {\bibinfo {title} {Universality class of the nishimori point in the
  2d $\ifmmode\pm\else\textpm\fi{}\mathit{J}$ random-bond ising model},\ }\href
  {https://doi.org/10.1103/PhysRevLett.87.047201} {\bibfield  {journal}
  {\bibinfo  {journal} {Phys. Rev. Lett.}\ }\textbf {\bibinfo {volume} {87}},\
  \bibinfo {pages} {047201} (\bibinfo {year} {2001})}\BibitemShut {NoStop}%
\bibitem [{\citenamefont {Chen}\ \emph {et~al.}(2014)\citenamefont {Chen},
  \citenamefont {Lu},\ and\ \citenamefont {Vishwanath}}]{Chen2014}%
  \BibitemOpen
  \bibfield  {author} {\bibinfo {author} {\bibfnamefont {X.}~\bibnamefont
  {Chen}}, \bibinfo {author} {\bibfnamefont {Y.-M.}\ \bibnamefont {Lu}},\ and\
  \bibinfo {author} {\bibfnamefont {A.}~\bibnamefont {Vishwanath}},\ }\bibfield
   {title} {\bibinfo {title} {Symmetry-protected topological phases from
  decorated domain walls},\ }\href {https://doi.org/10.1038/ncomms4507}
  {\bibfield  {journal} {\bibinfo  {journal} {Nature Communications}\ }\textbf
  {\bibinfo {volume} {5}},\ \bibinfo {pages} {3507} (\bibinfo {year}
  {2014})}\BibitemShut {NoStop}%
\bibitem [{\citenamefont {Witten}(1982)}]{WITTEN1982}%
  \BibitemOpen
  \bibfield  {author} {\bibinfo {author} {\bibfnamefont {E.}~\bibnamefont
  {Witten}},\ }\bibfield  {title} {\bibinfo {title} {Constraints on
  supersymmetry breaking},\ }\href
  {https://doi.org/https://doi.org/10.1016/0550-3213(82)90071-2} {\bibfield
  {journal} {\bibinfo  {journal} {Nuclear Physics B}\ }\textbf {\bibinfo
  {volume} {202}},\ \bibinfo {pages} {253} (\bibinfo {year}
  {1982})}\BibitemShut {NoStop}%
\bibitem [{\citenamefont {Wouters}\ \emph {et~al.}(2021)\citenamefont
  {Wouters}, \citenamefont {Katsura},\ and\ \citenamefont
  {Schuricht}}]{Wouters2021}%
  \BibitemOpen
  \bibfield  {author} {\bibinfo {author} {\bibfnamefont {J.}~\bibnamefont
  {Wouters}}, \bibinfo {author} {\bibfnamefont {H.}~\bibnamefont {Katsura}},\
  and\ \bibinfo {author} {\bibfnamefont {D.}~\bibnamefont {Schuricht}},\
  }\bibfield  {title} {\bibinfo {title} {{Interrelations among frustration-free
  models via Witten's conjugation}},\ }\href
  {https://doi.org/10.21468/SciPostPhysCore.4.4.027} {\bibfield  {journal}
  {\bibinfo  {journal} {SciPost Phys. Core}\ }\textbf {\bibinfo {volume} {4}},\
  \bibinfo {pages} {027} (\bibinfo {year} {2021})}\BibitemShut {NoStop}%
\bibitem [{\citenamefont {{Tantivasadakarn}}\ \emph
  {et~al.}(2021{\natexlab{a}})\citenamefont {{Tantivasadakarn}}, \citenamefont
  {{Thorngren}}, \citenamefont {{Vishwanath}},\ and\ \citenamefont
  {{Verresen}}}]{pivot}%
  \BibitemOpen
  \bibfield  {author} {\bibinfo {author} {\bibfnamefont {N.}~\bibnamefont
  {{Tantivasadakarn}}}, \bibinfo {author} {\bibfnamefont {R.}~\bibnamefont
  {{Thorngren}}}, \bibinfo {author} {\bibfnamefont {A.}~\bibnamefont
  {{Vishwanath}}},\ and\ \bibinfo {author} {\bibfnamefont {R.}~\bibnamefont
  {{Verresen}}},\ }\bibfield  {title} {\bibinfo {title} {{Pivot Hamiltonians as
  generators of symmetry and entanglement}},\ }\href@noop {} {\bibfield
  {journal} {\bibinfo  {journal} {arXiv e-prints}\ ,\ \bibinfo {eid}
  {arXiv:2110.07599}} (\bibinfo {year} {2021}{\natexlab{a}})}\BibitemShut
  {NoStop}%
\bibitem [{\citenamefont {Fern{\'a}ndez-Gonz{\'a}lez}\ \emph
  {et~al.}(2015)\citenamefont {Fern{\'a}ndez-Gonz{\'a}lez}, \citenamefont
  {Schuch}, \citenamefont {Wolf}, \citenamefont {Cirac},\ and\ \citenamefont
  {P{\'e}rez-Garc{\'\i}a}}]{parentHam}%
  \BibitemOpen
  \bibfield  {author} {\bibinfo {author} {\bibfnamefont {C.}~\bibnamefont
  {Fern{\'a}ndez-Gonz{\'a}lez}}, \bibinfo {author} {\bibfnamefont
  {N.}~\bibnamefont {Schuch}}, \bibinfo {author} {\bibfnamefont {M.~M.}\
  \bibnamefont {Wolf}}, \bibinfo {author} {\bibfnamefont {J.~I.}\ \bibnamefont
  {Cirac}},\ and\ \bibinfo {author} {\bibfnamefont {D.}~\bibnamefont
  {P{\'e}rez-Garc{\'\i}a}},\ }\bibfield  {title} {\bibinfo {title} {Frustration
  free gapless hamiltonians for matrix product states},\ }\href
  {https://doi.org/10.1007/s00220-014-2173-z} {\bibfield  {journal} {\bibinfo
  {journal} {Communications in Mathematical Physics}\ }\textbf {\bibinfo
  {volume} {333}},\ \bibinfo {pages} {299} (\bibinfo {year}
  {2015})}\BibitemShut {NoStop}%
\bibitem [{\citenamefont {Chen}\ \emph {et~al.}(2013)\citenamefont {Chen},
  \citenamefont {Gu}, \citenamefont {Liu},\ and\ \citenamefont
  {Wen}}]{Chen2013}%
  \BibitemOpen
  \bibfield  {author} {\bibinfo {author} {\bibfnamefont {X.}~\bibnamefont
  {Chen}}, \bibinfo {author} {\bibfnamefont {Z.-C.}\ \bibnamefont {Gu}},
  \bibinfo {author} {\bibfnamefont {Z.-X.}\ \bibnamefont {Liu}},\ and\ \bibinfo
  {author} {\bibfnamefont {X.-G.}\ \bibnamefont {Wen}},\ }\bibfield  {title}
  {\bibinfo {title} {Symmetry protected topological orders and the group
  cohomology of their symmetry group},\ }\href
  {https://doi.org/10.1103/PhysRevB.87.155114} {\bibfield  {journal} {\bibinfo
  {journal} {Phys. Rev. B}\ }\textbf {\bibinfo {volume} {87}},\ \bibinfo
  {pages} {155114} (\bibinfo {year} {2013})}\BibitemShut {NoStop}%
\bibitem [{\citenamefont {Verstraete}\ \emph {et~al.}(2006)\citenamefont
  {Verstraete}, \citenamefont {Wolf}, \citenamefont {Perez-Garcia},\ and\
  \citenamefont {Cirac}}]{PEPS_cirac}%
  \BibitemOpen
  \bibfield  {author} {\bibinfo {author} {\bibfnamefont {F.}~\bibnamefont
  {Verstraete}}, \bibinfo {author} {\bibfnamefont {M.~M.}\ \bibnamefont
  {Wolf}}, \bibinfo {author} {\bibfnamefont {D.}~\bibnamefont {Perez-Garcia}},\
  and\ \bibinfo {author} {\bibfnamefont {J.~I.}\ \bibnamefont {Cirac}},\
  }\bibfield  {title} {\bibinfo {title} {Criticality, the area law, and the
  computational power of projected entangled pair states},\ }\href
  {https://doi.org/10.1103/PhysRevLett.96.220601} {\bibfield  {journal}
  {\bibinfo  {journal} {Phys. Rev. Lett.}\ }\textbf {\bibinfo {volume} {96}},\
  \bibinfo {pages} {220601} (\bibinfo {year} {2006})}\BibitemShut {NoStop}%
\bibitem [{\citenamefont {Wolf}\ \emph {et~al.}(2006)\citenamefont {Wolf},
  \citenamefont {Ortiz}, \citenamefont {Verstraete},\ and\ \citenamefont
  {Cirac}}]{gaplessMPS}%
  \BibitemOpen
  \bibfield  {author} {\bibinfo {author} {\bibfnamefont {M.~M.}\ \bibnamefont
  {Wolf}}, \bibinfo {author} {\bibfnamefont {G.}~\bibnamefont {Ortiz}},
  \bibinfo {author} {\bibfnamefont {F.}~\bibnamefont {Verstraete}},\ and\
  \bibinfo {author} {\bibfnamefont {J.~I.}\ \bibnamefont {Cirac}},\ }\bibfield
  {title} {\bibinfo {title} {Quantum phase transitions in matrix product
  systems},\ }\href {https://doi.org/10.1103/PhysRevLett.97.110403} {\bibfield
  {journal} {\bibinfo  {journal} {Phys. Rev. Lett.}\ }\textbf {\bibinfo
  {volume} {97}},\ \bibinfo {pages} {110403} (\bibinfo {year}
  {2006})}\BibitemShut {NoStop}%
\bibitem [{\citenamefont {Gaiotto}\ \emph {et~al.}(2015)\citenamefont
  {Gaiotto}, \citenamefont {Kapustin}, \citenamefont {Seiberg},\ and\
  \citenamefont {Willett}}]{Kapustin2014}%
  \BibitemOpen
  \bibfield  {author} {\bibinfo {author} {\bibfnamefont {D.}~\bibnamefont
  {Gaiotto}}, \bibinfo {author} {\bibfnamefont {A.}~\bibnamefont {Kapustin}},
  \bibinfo {author} {\bibfnamefont {N.}~\bibnamefont {Seiberg}},\ and\ \bibinfo
  {author} {\bibfnamefont {B.}~\bibnamefont {Willett}},\ }\bibfield  {title}
  {\bibinfo {title} {Generalized global symmetries},\ }\href
  {https://doi.org/10.1007/JHEP02(2015)172} {\bibfield  {journal} {\bibinfo
  {journal} {Journal of High Energy Physics}\ }\textbf {\bibinfo {volume}
  {2015}},\ \bibinfo {pages} {172} (\bibinfo {year} {2015})}\BibitemShut
  {NoStop}%
\bibitem [{\citenamefont {{Kitaev}}(2006)}]{kitaev2006}%
  \BibitemOpen
  \bibfield  {author} {\bibinfo {author} {\bibfnamefont {A.}~\bibnamefont
  {{Kitaev}}},\ }\bibfield  {title} {\bibinfo {title} {{Anyons in an exactly
  solved model and beyond}},\ }\href
  {https://doi.org/10.1016/j.aop.2005.10.005} {\bibfield  {journal} {\bibinfo
  {journal} {Annals of Physics}\ }\textbf {\bibinfo {volume} {321}},\ \bibinfo
  {pages} {2} (\bibinfo {year} {2006})}\BibitemShut {NoStop}%
\bibitem [{\citenamefont {Uzunov}(1981)}]{uzunov1981}%
  \BibitemOpen
  \bibfield  {author} {\bibinfo {author} {\bibfnamefont {D.}~\bibnamefont
  {Uzunov}},\ }\bibfield  {title} {\bibinfo {title} {On the zero temperature
  critical behaviour of the nonideal bose gas},\ }\href
  {https://doi.org/https://doi.org/10.1016/0375-9601(81)90602-2} {\bibfield
  {journal} {\bibinfo  {journal} {Physics Letters A}\ }\textbf {\bibinfo
  {volume} {87}},\ \bibinfo {pages} {11} (\bibinfo {year} {1981})}\BibitemShut
  {NoStop}%
\bibitem [{\citenamefont {Fisher}\ and\ \citenamefont {Hohenberg}(1988)}]{BEC}%
  \BibitemOpen
  \bibfield  {author} {\bibinfo {author} {\bibfnamefont {D.~S.}\ \bibnamefont
  {Fisher}}\ and\ \bibinfo {author} {\bibfnamefont {P.~C.}\ \bibnamefont
  {Hohenberg}},\ }\bibfield  {title} {\bibinfo {title} {Dilute bose gas in two
  dimensions},\ }\href {https://doi.org/10.1103/PhysRevB.37.4936} {\bibfield
  {journal} {\bibinfo  {journal} {Phys. Rev. B}\ }\textbf {\bibinfo {volume}
  {37}},\ \bibinfo {pages} {4936} (\bibinfo {year} {1988})}\BibitemShut
  {NoStop}%
\bibitem [{\citenamefont {Fisher}\ \emph {et~al.}(1989)\citenamefont {Fisher},
  \citenamefont {Weichman}, \citenamefont {Grinstein},\ and\ \citenamefont
  {Fisher}}]{fisher1989boson}%
  \BibitemOpen
  \bibfield  {author} {\bibinfo {author} {\bibfnamefont {M.~P.}\ \bibnamefont
  {Fisher}}, \bibinfo {author} {\bibfnamefont {P.~B.}\ \bibnamefont
  {Weichman}}, \bibinfo {author} {\bibfnamefont {G.}~\bibnamefont
  {Grinstein}},\ and\ \bibinfo {author} {\bibfnamefont {D.~S.}\ \bibnamefont
  {Fisher}},\ }\bibfield  {title} {\bibinfo {title} {Boson localization and the
  superfluid-insulator transition},\ }\href@noop {} {\bibfield  {journal}
  {\bibinfo  {journal} {Physical Review B}\ }\textbf {\bibinfo {volume} {40}},\
  \bibinfo {pages} {546} (\bibinfo {year} {1989})}\BibitemShut {NoStop}%
\bibitem [{\citenamefont {Henley}(2004)}]{Henley2004}%
  \BibitemOpen
  \bibfield  {author} {\bibinfo {author} {\bibfnamefont {C.~L.}\ \bibnamefont
  {Henley}},\ }\bibfield  {title} {\bibinfo {title} {From classical to quantum
  dynamics at rokhsar{\textendash}kivelson points},\ }\href
  {https://doi.org/10.1088/0953-8984/16/11/045} {\bibfield  {journal} {\bibinfo
   {journal} {Journal of Physics: Condensed Matter}\ }\textbf {\bibinfo
  {volume} {16}},\ \bibinfo {pages} {S891} (\bibinfo {year}
  {2004})}\BibitemShut {NoStop}%
\bibitem [{\citenamefont {Nightingale}\ and\ \citenamefont
  {Bl\"ote}(2000)}]{IsingCQCP}%
  \BibitemOpen
  \bibfield  {author} {\bibinfo {author} {\bibfnamefont {M.~P.}\ \bibnamefont
  {Nightingale}}\ and\ \bibinfo {author} {\bibfnamefont {H.~W.~J.}\
  \bibnamefont {Bl\"ote}},\ }\bibfield  {title} {\bibinfo {title} {Monte carlo
  computation of correlation times of independent relaxation modes at
  criticality},\ }\href {https://doi.org/10.1103/PhysRevB.62.1089} {\bibfield
  {journal} {\bibinfo  {journal} {Phys. Rev. B}\ }\textbf {\bibinfo {volume}
  {62}},\ \bibinfo {pages} {1089} (\bibinfo {year} {2000})}\BibitemShut
  {NoStop}%
\bibitem [{\citenamefont {Isakov}\ \emph {et~al.}(2011)\citenamefont {Isakov},
  \citenamefont {Fendley}, \citenamefont {Ludwig}, \citenamefont {Trebst},\
  and\ \citenamefont {Troyer}}]{IsakovCQCP}%
  \BibitemOpen
  \bibfield  {author} {\bibinfo {author} {\bibfnamefont {S.~V.}\ \bibnamefont
  {Isakov}}, \bibinfo {author} {\bibfnamefont {P.}~\bibnamefont {Fendley}},
  \bibinfo {author} {\bibfnamefont {A.~W.~W.}\ \bibnamefont {Ludwig}}, \bibinfo
  {author} {\bibfnamefont {S.}~\bibnamefont {Trebst}},\ and\ \bibinfo {author}
  {\bibfnamefont {M.}~\bibnamefont {Troyer}},\ }\bibfield  {title} {\bibinfo
  {title} {Dynamics at and near conformal quantum critical points},\ }\href
  {https://doi.org/10.1103/PhysRevB.83.125114} {\bibfield  {journal} {\bibinfo
  {journal} {Phys. Rev. B}\ }\textbf {\bibinfo {volume} {83}},\ \bibinfo
  {pages} {125114} (\bibinfo {year} {2011})}\BibitemShut {NoStop}%
\bibitem [{\citenamefont {Bl\"ote}\ and\ \citenamefont {Deng}(2002)}]{2dIsing}%
  \BibitemOpen
  \bibfield  {author} {\bibinfo {author} {\bibfnamefont {H.~W.~J.}\
  \bibnamefont {Bl\"ote}}\ and\ \bibinfo {author} {\bibfnamefont
  {Y.}~\bibnamefont {Deng}},\ }\bibfield  {title} {\bibinfo {title} {Cluster
  monte carlo simulation of the transverse ising model},\ }\href
  {https://doi.org/10.1103/PhysRevE.66.066110} {\bibfield  {journal} {\bibinfo
  {journal} {Phys. Rev. E}\ }\textbf {\bibinfo {volume} {66}},\ \bibinfo
  {pages} {066110} (\bibinfo {year} {2002})}\BibitemShut {NoStop}%
\bibitem [{\citenamefont {Castelnovo}\ \emph {et~al.}(2005)\citenamefont
  {Castelnovo}, \citenamefont {Chamon}, \citenamefont {Mudry},\ and\
  \citenamefont {Pujol}}]{CASTELNOVO2005}%
  \BibitemOpen
  \bibfield  {author} {\bibinfo {author} {\bibfnamefont {C.}~\bibnamefont
  {Castelnovo}}, \bibinfo {author} {\bibfnamefont {C.}~\bibnamefont {Chamon}},
  \bibinfo {author} {\bibfnamefont {C.}~\bibnamefont {Mudry}},\ and\ \bibinfo
  {author} {\bibfnamefont {P.}~\bibnamefont {Pujol}},\ }\bibfield  {title}
  {\bibinfo {title} {From quantum mechanics to classical statistical physics:
  Generalized rokhsar--kivelson hamiltonians and the ``stochastic matrix form''
  decomposition},\ }\href
  {https://doi.org/https://doi.org/10.1016/j.aop.2005.01.006} {\bibfield
  {journal} {\bibinfo  {journal} {Annals of Physics}\ }\textbf {\bibinfo
  {volume} {318}},\ \bibinfo {pages} {316} (\bibinfo {year}
  {2005})}\BibitemShut {NoStop}%
\bibitem [{\citenamefont {Wegner}(1971)}]{wegner1971duality}%
  \BibitemOpen
  \bibfield  {author} {\bibinfo {author} {\bibfnamefont {F.~J.}\ \bibnamefont
  {Wegner}},\ }\bibfield  {title} {\bibinfo {title} {Duality in generalized
  ising models and phase transitions without local order parameters},\
  }\href@noop {} {\bibfield  {journal} {\bibinfo  {journal} {Journal of
  Mathematical Physics}\ }\textbf {\bibinfo {volume} {12}},\ \bibinfo {pages}
  {2259} (\bibinfo {year} {1971})}\BibitemShut {NoStop}%
\bibitem [{\citenamefont {Balian}\ \emph {et~al.}(1975)\citenamefont {Balian},
  \citenamefont {Drouffe},\ and\ \citenamefont {Itzykson}}]{balian1975gauge}%
  \BibitemOpen
  \bibfield  {author} {\bibinfo {author} {\bibfnamefont {R.}~\bibnamefont
  {Balian}}, \bibinfo {author} {\bibfnamefont {J.}~\bibnamefont {Drouffe}},\
  and\ \bibinfo {author} {\bibfnamefont {C.}~\bibnamefont {Itzykson}},\
  }\bibfield  {title} {\bibinfo {title} {Gauge fields on a lattice. iii.
  strong-coupling expansions and transition points},\ }\href@noop {} {\bibfield
   {journal} {\bibinfo  {journal} {Physical Review D}\ }\textbf {\bibinfo
  {volume} {11}},\ \bibinfo {pages} {2104} (\bibinfo {year}
  {1975})}\BibitemShut {NoStop}%
\bibitem [{\citenamefont {Creutz}\ \emph {et~al.}(1979)\citenamefont {Creutz},
  \citenamefont {Jacobs},\ and\ \citenamefont {Rebbi}}]{creutz1979experiments}%
  \BibitemOpen
  \bibfield  {author} {\bibinfo {author} {\bibfnamefont {M.}~\bibnamefont
  {Creutz}}, \bibinfo {author} {\bibfnamefont {L.}~\bibnamefont {Jacobs}},\
  and\ \bibinfo {author} {\bibfnamefont {C.}~\bibnamefont {Rebbi}},\ }\bibfield
   {title} {\bibinfo {title} {Experiments with a gauge-invariant ising
  system},\ }\href@noop {} {\bibfield  {journal} {\bibinfo  {journal} {Physical
  Review Letters}\ }\textbf {\bibinfo {volume} {42}},\ \bibinfo {pages} {1390}
  (\bibinfo {year} {1979})}\BibitemShut {NoStop}%
\bibitem [{\citenamefont {Binder}\ and\ \citenamefont
  {Young}(1986)}]{BinderYoung1986}%
  \BibitemOpen
  \bibfield  {author} {\bibinfo {author} {\bibfnamefont {K.}~\bibnamefont
  {Binder}}\ and\ \bibinfo {author} {\bibfnamefont {A.~P.}\ \bibnamefont
  {Young}},\ }\bibfield  {title} {\bibinfo {title} {Spin glasses: Experimental
  facts, theoretical concepts, and open questions},\ }\href
  {https://doi.org/10.1103/RevModPhys.58.801} {\bibfield  {journal} {\bibinfo
  {journal} {Rev. Mod. Phys.}\ }\textbf {\bibinfo {volume} {58}},\ \bibinfo
  {pages} {801} (\bibinfo {year} {1986})}\BibitemShut {NoStop}%
\bibitem [{\citenamefont {Mattis}(1976)}]{MATTIS1976}%
  \BibitemOpen
  \bibfield  {author} {\bibinfo {author} {\bibfnamefont {D.}~\bibnamefont
  {Mattis}},\ }\bibfield  {title} {\bibinfo {title} {Solvable spin systems with
  random interactions},\ }\href
  {https://doi.org/https://doi.org/10.1016/0375-9601(76)90396-0} {\bibfield
  {journal} {\bibinfo  {journal} {Physics Letters A}\ }\textbf {\bibinfo
  {volume} {56}},\ \bibinfo {pages} {421} (\bibinfo {year} {1976})}\BibitemShut
  {NoStop}%
\bibitem [{\citenamefont {Wang}\ \emph {et~al.}(2003)\citenamefont {Wang},
  \citenamefont {Harrington},\ and\ \citenamefont
  {Preskill}}]{WangPreskill2003}%
  \BibitemOpen
  \bibfield  {author} {\bibinfo {author} {\bibfnamefont {C.}~\bibnamefont
  {Wang}}, \bibinfo {author} {\bibfnamefont {J.}~\bibnamefont {Harrington}},\
  and\ \bibinfo {author} {\bibfnamefont {J.}~\bibnamefont {Preskill}},\
  }\bibfield  {title} {\bibinfo {title} {Confinement-higgs transition in a
  disordered gauge theory and the accuracy threshold for quantum memory},\
  }\href {https://doi.org/https://doi.org/10.1016/S0003-4916(02)00019-2}
  {\bibfield  {journal} {\bibinfo  {journal} {Annals of Physics}\ }\textbf
  {\bibinfo {volume} {303}},\ \bibinfo {pages} {31} (\bibinfo {year}
  {2003})}\BibitemShut {NoStop}%
\bibitem [{\citenamefont {Fradkin}\ \emph {et~al.}(1978)\citenamefont
  {Fradkin}, \citenamefont {Huberman},\ and\ \citenamefont
  {Shenker}}]{FradkinRandom}%
  \BibitemOpen
  \bibfield  {author} {\bibinfo {author} {\bibfnamefont {E.}~\bibnamefont
  {Fradkin}}, \bibinfo {author} {\bibfnamefont {B.~A.}\ \bibnamefont
  {Huberman}},\ and\ \bibinfo {author} {\bibfnamefont {S.~H.}\ \bibnamefont
  {Shenker}},\ }\bibfield  {title} {\bibinfo {title} {Gauge symmetries in
  random magnetic systems},\ }\href {https://doi.org/10.1103/PhysRevB.18.4789}
  {\bibfield  {journal} {\bibinfo  {journal} {Phys. Rev. B}\ }\textbf {\bibinfo
  {volume} {18}},\ \bibinfo {pages} {4789} (\bibinfo {year}
  {1978})}\BibitemShut {NoStop}%
\bibitem [{\citenamefont {Rieger}\ and\ \citenamefont
  {Young}(1994)}]{RiegerRBIM1994}%
  \BibitemOpen
  \bibfield  {author} {\bibinfo {author} {\bibfnamefont {H.}~\bibnamefont
  {Rieger}}\ and\ \bibinfo {author} {\bibfnamefont {A.~P.}\ \bibnamefont
  {Young}},\ }\bibfield  {title} {\bibinfo {title} {Zero-temperature quantum
  phase transition of a two-dimensional ising spin glass},\ }\href
  {https://doi.org/10.1103/PhysRevLett.72.4141} {\bibfield  {journal} {\bibinfo
   {journal} {Phys. Rev. Lett.}\ }\textbf {\bibinfo {volume} {72}},\ \bibinfo
  {pages} {4141} (\bibinfo {year} {1994})}\BibitemShut {NoStop}%
\bibitem [{\citenamefont {Cho}\ and\ \citenamefont
  {Fisher}(1997)}]{ChoRBIM1997}%
  \BibitemOpen
  \bibfield  {author} {\bibinfo {author} {\bibfnamefont {S.}~\bibnamefont
  {Cho}}\ and\ \bibinfo {author} {\bibfnamefont {M.~P.~A.}\ \bibnamefont
  {Fisher}},\ }\bibfield  {title} {\bibinfo {title} {Criticality in the
  two-dimensional random-bond ising model},\ }\href
  {https://doi.org/10.1103/PhysRevB.55.1025} {\bibfield  {journal} {\bibinfo
  {journal} {Phys. Rev. B}\ }\textbf {\bibinfo {volume} {55}},\ \bibinfo
  {pages} {1025} (\bibinfo {year} {1997})}\BibitemShut {NoStop}%
\bibitem [{\citenamefont {Gruzberg}\ \emph {et~al.}(2001)\citenamefont
  {Gruzberg}, \citenamefont {Read},\ and\ \citenamefont
  {Ludwig}}]{GruzbergRBIM2001}%
  \BibitemOpen
  \bibfield  {author} {\bibinfo {author} {\bibfnamefont {I.~A.}\ \bibnamefont
  {Gruzberg}}, \bibinfo {author} {\bibfnamefont {N.}~\bibnamefont {Read}},\
  and\ \bibinfo {author} {\bibfnamefont {A.~W.~W.}\ \bibnamefont {Ludwig}},\
  }\bibfield  {title} {\bibinfo {title} {Random-bond ising model in two
  dimensions: The nishimori line and supersymmetry},\ }\href
  {https://doi.org/10.1103/PhysRevB.63.104422} {\bibfield  {journal} {\bibinfo
  {journal} {Phys. Rev. B}\ }\textbf {\bibinfo {volume} {63}},\ \bibinfo
  {pages} {104422} (\bibinfo {year} {2001})}\BibitemShut {NoStop}%
\bibitem [{\citenamefont {Le~Doussal}\ and\ \citenamefont
  {Harris}(1988)}]{Doussal1988}%
  \BibitemOpen
  \bibfield  {author} {\bibinfo {author} {\bibfnamefont {P.}~\bibnamefont
  {Le~Doussal}}\ and\ \bibinfo {author} {\bibfnamefont {A.~B.}\ \bibnamefont
  {Harris}},\ }\bibfield  {title} {\bibinfo {title} {Location of the ising
  spin-glass multicritical point on nishimori's line},\ }\href
  {https://doi.org/10.1103/PhysRevLett.61.625} {\bibfield  {journal} {\bibinfo
  {journal} {Phys. Rev. Lett.}\ }\textbf {\bibinfo {volume} {61}},\ \bibinfo
  {pages} {625} (\bibinfo {year} {1988})}\BibitemShut {NoStop}%
\bibitem [{\citenamefont {Reger}\ and\ \citenamefont
  {Zippelius}(1986)}]{Reger3D1986}%
  \BibitemOpen
  \bibfield  {author} {\bibinfo {author} {\bibfnamefont {J.~D.}\ \bibnamefont
  {Reger}}\ and\ \bibinfo {author} {\bibfnamefont {A.}~\bibnamefont
  {Zippelius}},\ }\bibfield  {title} {\bibinfo {title} {Three-dimensional
  random-bond ising model: Phase diagram and critical properties},\ }\href
  {https://doi.org/10.1103/PhysRevLett.57.3225} {\bibfield  {journal} {\bibinfo
   {journal} {Phys. Rev. Lett.}\ }\textbf {\bibinfo {volume} {57}},\ \bibinfo
  {pages} {3225} (\bibinfo {year} {1986})}\BibitemShut {NoStop}%
\bibitem [{\citenamefont {Ohno}\ \emph {et~al.}(2004)\citenamefont {Ohno},
  \citenamefont {Arakawa}, \citenamefont {Ichinose},\ and\ \citenamefont
  {Matsui}}]{OHNO2004}%
  \BibitemOpen
  \bibfield  {author} {\bibinfo {author} {\bibfnamefont {T.}~\bibnamefont
  {Ohno}}, \bibinfo {author} {\bibfnamefont {G.}~\bibnamefont {Arakawa}},
  \bibinfo {author} {\bibfnamefont {I.}~\bibnamefont {Ichinose}},\ and\
  \bibinfo {author} {\bibfnamefont {T.}~\bibnamefont {Matsui}},\ }\bibfield
  {title} {\bibinfo {title} {Phase structure of the random-plaquette z2 gauge
  model: accuracy threshold for a toric quantum memory},\ }\href
  {https://doi.org/https://doi.org/10.1016/j.nuclphysb.2004.07.003} {\bibfield
  {journal} {\bibinfo  {journal} {Nuclear Physics B}\ }\textbf {\bibinfo
  {volume} {697}},\ \bibinfo {pages} {462} (\bibinfo {year}
  {2004})}\BibitemShut {NoStop}%
\bibitem [{\citenamefont {Nishimori}(1993)}]{NishimoriDecoding}%
  \BibitemOpen
  \bibfield  {author} {\bibinfo {author} {\bibfnamefont {H.}~\bibnamefont
  {Nishimori}},\ }\bibfield  {title} {\bibinfo {title} {Optimum decoding
  temperature for error-correcting codes},\ }\href
  {https://doi.org/10.1143/JPSJ.62.2973} {\bibfield  {journal} {\bibinfo
  {journal} {Journal of the Physical Society of Japan}\ }\textbf {\bibinfo
  {volume} {62}},\ \bibinfo {pages} {2973} (\bibinfo {year}
  {1993})}\BibitemShut {NoStop}%
\bibitem [{\citenamefont {Higgott}(2021)}]{toMWPM2021}%
  \BibitemOpen
  \bibfield  {author} {\bibinfo {author} {\bibfnamefont {O.}~\bibnamefont
  {Higgott}},\ }\href {https://doi.org/10.48550/ARXIV.2105.13082} {\bibinfo
  {title} {Pymatching: A python package for decoding quantum codes with
  minimum-weight perfect matching}} (\bibinfo {year} {2021})\BibitemShut
  {NoStop}%
\bibitem [{\citenamefont {Edmonds}(1965)}]{edmonds_1965}%
  \BibitemOpen
  \bibfield  {author} {\bibinfo {author} {\bibfnamefont {J.}~\bibnamefont
  {Edmonds}},\ }\bibfield  {title} {\bibinfo {title} {Paths, trees, and
  flowers},\ }\href {https://doi.org/10.4153/CJM-1965-045-4} {\bibfield
  {journal} {\bibinfo  {journal} {Canadian Journal of Mathematics}\ }\textbf
  {\bibinfo {volume} {17}},\ \bibinfo {pages} {449–467} (\bibinfo {year}
  {1965})}\BibitemShut {NoStop}%
\bibitem [{\citenamefont {Goemans}\ and\ \citenamefont
  {Williamson}(1995)}]{GoemansWilliamson}%
  \BibitemOpen
  \bibfield  {author} {\bibinfo {author} {\bibfnamefont {M.~X.}\ \bibnamefont
  {Goemans}}\ and\ \bibinfo {author} {\bibfnamefont {D.~P.}\ \bibnamefont
  {Williamson}},\ }\bibfield  {title} {\bibinfo {title} {Improved approximation
  algorithms for maximum cut and satisfiability problems using semidefinite
  programming},\ }\href {https://doi.org/10.1145/227683.227684} {\bibfield
  {journal} {\bibinfo  {journal} {J. ACM}\ }\textbf {\bibinfo {volume} {42}},\
  \bibinfo {pages} {1115–1145} (\bibinfo {year} {1995})}\BibitemShut
  {NoStop}%
\bibitem [{\citenamefont {Luo}\ \emph {et~al.}(2001)\citenamefont {Luo},
  \citenamefont {Sch\"ulke},\ and\ \citenamefont {Zheng}}]{2dRBIM_crit}%
  \BibitemOpen
  \bibfield  {author} {\bibinfo {author} {\bibfnamefont {H.~J.}\ \bibnamefont
  {Luo}}, \bibinfo {author} {\bibfnamefont {L.}~\bibnamefont {Sch\"ulke}},\
  and\ \bibinfo {author} {\bibfnamefont {B.}~\bibnamefont {Zheng}},\ }\bibfield
   {title} {\bibinfo {title} {Short-time critical dynamics of the
  two-dimensional random-bond ising model},\ }\href
  {https://doi.org/10.1103/PhysRevE.64.036123} {\bibfield  {journal} {\bibinfo
  {journal} {Phys. Rev. E}\ }\textbf {\bibinfo {volume} {64}},\ \bibinfo
  {pages} {036123} (\bibinfo {year} {2001})}\BibitemShut {NoStop}%
\bibitem [{Yao()}]{YaodongToapper}%
  \BibitemOpen
  \href@noop {} {}\bibinfo {note} {Y. Li, Y. Zou, P. Glorioso, E. Altman, and
  M. P. A. Fisher, To Appear}\BibitemShut {NoStop}%
\bibitem [{\citenamefont {Gullans}\ and\ \citenamefont
  {Huse}(2020{\natexlab{b}})}]{GullansProbe}%
  \BibitemOpen
  \bibfield  {author} {\bibinfo {author} {\bibfnamefont {M.~J.}\ \bibnamefont
  {Gullans}}\ and\ \bibinfo {author} {\bibfnamefont {D.~A.}\ \bibnamefont
  {Huse}},\ }\bibfield  {title} {\bibinfo {title} {Scalable probes of
  measurement-induced criticality},\ }\href
  {https://doi.org/10.1103/PhysRevLett.125.070606} {\bibfield  {journal}
  {\bibinfo  {journal} {Phys. Rev. Lett.}\ }\textbf {\bibinfo {volume} {125}},\
  \bibinfo {pages} {070606} (\bibinfo {year} {2020}{\natexlab{b}})}\BibitemShut
  {NoStop}%
\bibitem [{\citenamefont {{Dehghani}}\ \emph {et~al.}(2022)\citenamefont
  {{Dehghani}}, \citenamefont {{Lavasani}}, \citenamefont {{Hafezi}},\ and\
  \citenamefont {{Gullans}}}]{Dehghani2022}%
  \BibitemOpen
  \bibfield  {author} {\bibinfo {author} {\bibfnamefont {H.}~\bibnamefont
  {{Dehghani}}}, \bibinfo {author} {\bibfnamefont {A.}~\bibnamefont
  {{Lavasani}}}, \bibinfo {author} {\bibfnamefont {M.}~\bibnamefont
  {{Hafezi}}},\ and\ \bibinfo {author} {\bibfnamefont {M.~J.}\ \bibnamefont
  {{Gullans}}},\ }\bibfield  {title} {\bibinfo {title} {{Neural-Network
  Decoders for Measurement Induced Phase Transitions}},\ }\href@noop {}
  {\bibfield  {journal} {\bibinfo  {journal} {arXiv e-prints}\ ,\ \bibinfo
  {eid} {arXiv:2204.10904}} (\bibinfo {year} {2022})},\ \Eprint
  {https://arxiv.org/abs/2204.10904} {arXiv:2204.10904 [quant-ph]} \BibitemShut
  {NoStop}%
\bibitem [{\citenamefont {Hukushima}(2000)}]{Koji2000}%
  \BibitemOpen
  \bibfield  {author} {\bibinfo {author} {\bibfnamefont {K.}~\bibnamefont
  {Hukushima}},\ }\bibfield  {title} {\bibinfo {title} {Random fixed point of
  three-dimensional random-bond ising models},\ }\href
  {https://doi.org/10.1143/JPSJ.69.631} {\bibfield  {journal} {\bibinfo
  {journal} {Journal of the Physical Society of Japan}\ }\textbf {\bibinfo
  {volume} {69}},\ \bibinfo {pages} {631} (\bibinfo {year} {2000})}\BibitemShut
  {NoStop}%
\bibitem [{\citenamefont {Xiong}\ \emph {et~al.}(2010)\citenamefont {Xiong},
  \citenamefont {Zhong}, \citenamefont {Yuan},\ and\ \citenamefont
  {Fan}}]{3dRBIM_crit}%
  \BibitemOpen
  \bibfield  {author} {\bibinfo {author} {\bibfnamefont {W.}~\bibnamefont
  {Xiong}}, \bibinfo {author} {\bibfnamefont {F.}~\bibnamefont {Zhong}},
  \bibinfo {author} {\bibfnamefont {W.}~\bibnamefont {Yuan}},\ and\ \bibinfo
  {author} {\bibfnamefont {S.}~\bibnamefont {Fan}},\ }\bibfield  {title}
  {\bibinfo {title} {Critical behavior of a three-dimensional random-bond ising
  model using finite-time scaling with extensive monte carlo
  renormalization-group method},\ }\href
  {https://doi.org/10.1103/PhysRevE.81.051132} {\bibfield  {journal} {\bibinfo
  {journal} {Phys. Rev. E}\ }\textbf {\bibinfo {volume} {81}},\ \bibinfo
  {pages} {051132} (\bibinfo {year} {2010})}\BibitemShut {NoStop}%
\bibitem [{\citenamefont {Dennis}\ \emph {et~al.}(2002)\citenamefont {Dennis},
  \citenamefont {Kitaev}, \citenamefont {Landahl},\ and\ \citenamefont
  {Preskill}}]{Dennis2002TQM}%
  \BibitemOpen
  \bibfield  {author} {\bibinfo {author} {\bibfnamefont {E.}~\bibnamefont
  {Dennis}}, \bibinfo {author} {\bibfnamefont {A.}~\bibnamefont {Kitaev}},
  \bibinfo {author} {\bibfnamefont {A.}~\bibnamefont {Landahl}},\ and\ \bibinfo
  {author} {\bibfnamefont {J.}~\bibnamefont {Preskill}},\ }\bibfield  {title}
  {\bibinfo {title} {Topological quantum memory},\ }\href
  {https://doi.org/10.1063/1.1499754} {\bibfield  {journal} {\bibinfo
  {journal} {Journal of Mathematical Physics}\ }\textbf {\bibinfo {volume}
  {43}},\ \bibinfo {pages} {4452} (\bibinfo {year} {2002})}\BibitemShut
  {NoStop}%
\bibitem [{\citenamefont {Xu}\ and\ \citenamefont {Moore}(2004)}]{XuMoore2004}%
  \BibitemOpen
  \bibfield  {author} {\bibinfo {author} {\bibfnamefont {C.}~\bibnamefont
  {Xu}}\ and\ \bibinfo {author} {\bibfnamefont {J.~E.}\ \bibnamefont {Moore}},\
  }\bibfield  {title} {\bibinfo {title} {Strong-weak coupling self-duality in
  the two-dimensional quantum phase transition of $p+ip$ superconducting
  arrays},\ }\href {https://doi.org/10.1103/PhysRevLett.93.047003} {\bibfield
  {journal} {\bibinfo  {journal} {Phys. Rev. Lett.}\ }\textbf {\bibinfo
  {volume} {93}},\ \bibinfo {pages} {047003} (\bibinfo {year}
  {2004})}\BibitemShut {NoStop}%
\bibitem [{\citenamefont {Mueller}\ \emph {et~al.}(2017)\citenamefont
  {Mueller}, \citenamefont {Johnston},\ and\ \citenamefont
  {Janke}}]{2dgonihedric}%
  \BibitemOpen
  \bibfield  {author} {\bibinfo {author} {\bibfnamefont {M.}~\bibnamefont
  {Mueller}}, \bibinfo {author} {\bibfnamefont {D.~A.}\ \bibnamefont
  {Johnston}},\ and\ \bibinfo {author} {\bibfnamefont {W.}~\bibnamefont
  {Janke}},\ }\bibfield  {title} {\bibinfo {title} {Exact solutions to
  plaquette ising models with free and periodic boundaries},\ }\href
  {https://doi.org/https://doi.org/10.1016/j.nuclphysb.2016.11.005} {\bibfield
  {journal} {\bibinfo  {journal} {Nuclear Physics B}\ }\textbf {\bibinfo
  {volume} {914}},\ \bibinfo {pages} {388} (\bibinfo {year}
  {2017})}\BibitemShut {NoStop}%
\bibitem [{\citenamefont {Mueller}\ \emph {et~al.}(2014)\citenamefont
  {Mueller}, \citenamefont {Janke},\ and\ \citenamefont
  {Johnston}}]{3Dplaquette}%
  \BibitemOpen
  \bibfield  {author} {\bibinfo {author} {\bibfnamefont {M.}~\bibnamefont
  {Mueller}}, \bibinfo {author} {\bibfnamefont {W.}~\bibnamefont {Janke}},\
  and\ \bibinfo {author} {\bibfnamefont {D.~A.}\ \bibnamefont {Johnston}},\
  }\bibfield  {title} {\bibinfo {title} {Nonstandard finite-size scaling at
  first-order phase transitions},\ }\href
  {https://doi.org/10.1103/PhysRevLett.112.200601} {\bibfield  {journal}
  {\bibinfo  {journal} {Phys. Rev. Lett.}\ }\textbf {\bibinfo {volume} {112}},\
  \bibinfo {pages} {200601} (\bibinfo {year} {2014})}\BibitemShut {NoStop}%
\bibitem [{\citenamefont {Vijay}\ \emph {et~al.}(2016)\citenamefont {Vijay},
  \citenamefont {Haah},\ and\ \citenamefont {Fu}}]{Xcube}%
  \BibitemOpen
  \bibfield  {author} {\bibinfo {author} {\bibfnamefont {S.}~\bibnamefont
  {Vijay}}, \bibinfo {author} {\bibfnamefont {J.}~\bibnamefont {Haah}},\ and\
  \bibinfo {author} {\bibfnamefont {L.}~\bibnamefont {Fu}},\ }\bibfield
  {title} {\bibinfo {title} {Fracton topological order, generalized lattice
  gauge theory, and duality},\ }\href
  {https://doi.org/10.1103/PhysRevB.94.235157} {\bibfield  {journal} {\bibinfo
  {journal} {Phys. Rev. B}\ }\textbf {\bibinfo {volume} {94}},\ \bibinfo
  {pages} {235157} (\bibinfo {year} {2016})}\BibitemShut {NoStop}%
\bibitem [{\citenamefont {{Tantivasadakarn}}\ \emph
  {et~al.}(2021{\natexlab{b}})\citenamefont {{Tantivasadakarn}}, \citenamefont
  {{Thorngren}}, \citenamefont {{Vishwanath}},\ and\ \citenamefont
  {{Verresen}}}]{NatMeasurement}%
  \BibitemOpen
  \bibfield  {author} {\bibinfo {author} {\bibfnamefont {N.}~\bibnamefont
  {{Tantivasadakarn}}}, \bibinfo {author} {\bibfnamefont {R.}~\bibnamefont
  {{Thorngren}}}, \bibinfo {author} {\bibfnamefont {A.}~\bibnamefont
  {{Vishwanath}}},\ and\ \bibinfo {author} {\bibfnamefont {R.}~\bibnamefont
  {{Verresen}}},\ }\bibfield  {title} {\bibinfo {title} {{Long-range
  entanglement from measuring symmetry-protected topological phases}},\
  }\href@noop {} {\bibfield  {journal} {\bibinfo  {journal} {arXiv e-prints}\
  ,\ \bibinfo {eid} {arXiv:2112.01519}} (\bibinfo {year}
  {2021}{\natexlab{b}})}\BibitemShut {NoStop}%
\bibitem [{\citenamefont {Kogut}(1979)}]{KogutRMP}%
  \BibitemOpen
  \bibfield  {author} {\bibinfo {author} {\bibfnamefont {J.~B.}\ \bibnamefont
  {Kogut}},\ }\bibfield  {title} {\bibinfo {title} {An introduction to lattice
  gauge theory and spin systems},\ }\href
  {https://doi.org/10.1103/RevModPhys.51.659} {\bibfield  {journal} {\bibinfo
  {journal} {Rev. Mod. Phys.}\ }\textbf {\bibinfo {volume} {51}},\ \bibinfo
  {pages} {659} (\bibinfo {year} {1979})}\BibitemShut {NoStop}%
\bibitem [{\citenamefont {Rokhsar}\ and\ \citenamefont
  {Kivelson}(1988)}]{RKmodel}%
  \BibitemOpen
  \bibfield  {author} {\bibinfo {author} {\bibfnamefont {D.~S.}\ \bibnamefont
  {Rokhsar}}\ and\ \bibinfo {author} {\bibfnamefont {S.~A.}\ \bibnamefont
  {Kivelson}},\ }\bibfield  {title} {\bibinfo {title} {Superconductivity and
  the quantum hard-core dimer gas},\ }\href
  {https://doi.org/10.1103/PhysRevLett.61.2376} {\bibfield  {journal} {\bibinfo
   {journal} {Phys. Rev. Lett.}\ }\textbf {\bibinfo {volume} {61}},\ \bibinfo
  {pages} {2376} (\bibinfo {year} {1988})}\BibitemShut {NoStop}%
\bibitem [{\citenamefont {Moessner}\ \emph {et~al.}(2001)\citenamefont
  {Moessner}, \citenamefont {Sondhi},\ and\ \citenamefont
  {Fradkin}}]{RKmodel2}%
  \BibitemOpen
  \bibfield  {author} {\bibinfo {author} {\bibfnamefont {R.}~\bibnamefont
  {Moessner}}, \bibinfo {author} {\bibfnamefont {S.~L.}\ \bibnamefont
  {Sondhi}},\ and\ \bibinfo {author} {\bibfnamefont {E.}~\bibnamefont
  {Fradkin}},\ }\bibfield  {title} {\bibinfo {title} {Short-ranged resonating
  valence bond physics, quantum dimer models, and ising gauge theories},\
  }\href {https://doi.org/10.1103/PhysRevB.65.024504} {\bibfield  {journal}
  {\bibinfo  {journal} {Phys. Rev. B}\ }\textbf {\bibinfo {volume} {65}},\
  \bibinfo {pages} {024504} (\bibinfo {year} {2001})}\BibitemShut {NoStop}%
\bibitem [{\citenamefont {{Perez-Garcia}}\ \emph {et~al.}(2007)\citenamefont
  {{Perez-Garcia}}, \citenamefont {{Verstraete}}, \citenamefont {{Cirac}},\
  and\ \citenamefont {{Wolf}}}]{PEPS_parent}%
  \BibitemOpen
  \bibfield  {author} {\bibinfo {author} {\bibfnamefont {D.}~\bibnamefont
  {{Perez-Garcia}}}, \bibinfo {author} {\bibfnamefont {F.}~\bibnamefont
  {{Verstraete}}}, \bibinfo {author} {\bibfnamefont {J.~I.}\ \bibnamefont
  {{Cirac}}},\ and\ \bibinfo {author} {\bibfnamefont {M.~M.}\ \bibnamefont
  {{Wolf}}},\ }\bibfield  {title} {\bibinfo {title} {{PEPS as unique ground
  states of local Hamiltonians}},\ }\href@noop {} {\bibfield  {journal}
  {\bibinfo  {journal} {arXiv e-prints}\ ,\ \bibinfo {pages} {arXiv:0707.2260}}
  (\bibinfo {year} {2007})}\BibitemShut {NoStop}%
\bibitem [{\citenamefont {Jones}\ \emph {et~al.}(2021)\citenamefont {Jones},
  \citenamefont {Bibo}, \citenamefont {Jobst}, \citenamefont {Pollmann},
  \citenamefont {Smith},\ and\ \citenamefont {Verresen}}]{skeleton}%
  \BibitemOpen
  \bibfield  {author} {\bibinfo {author} {\bibfnamefont {N.~G.}\ \bibnamefont
  {Jones}}, \bibinfo {author} {\bibfnamefont {J.}~\bibnamefont {Bibo}},
  \bibinfo {author} {\bibfnamefont {B.}~\bibnamefont {Jobst}}, \bibinfo
  {author} {\bibfnamefont {F.}~\bibnamefont {Pollmann}}, \bibinfo {author}
  {\bibfnamefont {A.}~\bibnamefont {Smith}},\ and\ \bibinfo {author}
  {\bibfnamefont {R.}~\bibnamefont {Verresen}},\ }\bibfield  {title} {\bibinfo
  {title} {Skeleton of matrix-product-state-solvable models connecting
  topological phases of matter},\ }\href
  {https://doi.org/10.1103/PhysRevResearch.3.033265} {\bibfield  {journal}
  {\bibinfo  {journal} {Phys. Rev. Research}\ }\textbf {\bibinfo {volume}
  {3}},\ \bibinfo {pages} {033265} (\bibinfo {year} {2021})}\BibitemShut
  {NoStop}%
\bibitem [{Guo()}]{GuoYi}%
  \BibitemOpen
  \href@noop {} {}\bibinfo {note} {G.-Y. Zhu, N. Tantivasadakarn, A.
  Vishwanath, S. Trebst, R. Verresen, To Appear}\BibitemShut {NoStop}%
\bibitem [{\citenamefont {Bieche}\ \emph {et~al.}(1980)\citenamefont {Bieche},
  \citenamefont {Uhry}, \citenamefont {Maynard},\ and\ \citenamefont
  {Rammal}}]{Bieche1980}%
  \BibitemOpen
  \bibfield  {author} {\bibinfo {author} {\bibfnamefont {L.}~\bibnamefont
  {Bieche}}, \bibinfo {author} {\bibfnamefont {J.~P.}\ \bibnamefont {Uhry}},
  \bibinfo {author} {\bibfnamefont {R.}~\bibnamefont {Maynard}},\ and\ \bibinfo
  {author} {\bibfnamefont {R.}~\bibnamefont {Rammal}},\ }\bibfield  {title}
  {\bibinfo {title} {On the ground states of the frustration model of a spin
  glass by a matching method of graph theory},\ }\href
  {https://doi.org/10.1088/0305-4470/13/8/005} {\bibfield  {journal} {\bibinfo
  {journal} {Journal of Physics A: Mathematical and General}\ }\textbf
  {\bibinfo {volume} {13}},\ \bibinfo {pages} {2553} (\bibinfo {year}
  {1980})}\BibitemShut {NoStop}%
\end{thebibliography}

%

\newpage 

\appendix

\begin{widetext}

\section{Correlation functions} \label{app:lemma}

In this appendix, we demonstrate the following lemma in the main text and its consequence on calculating the correlation function and norm of the post-measurement wavefunction.
\medskip 

\noindent {\bf Lemma} Consider a stabilizer $d$-dim SPT state with $G^{(n)}_1 \times G^{(d-n-1)}_2$ mixed anomaly, where the superscript represents they are $n$ and $(d-n-1)$ form symmetries. Here, $G_1$ and $G_2$ act on different sublattices. Then, the expectation value of an operator defined on a certain sublattice does not vanish only if the operator is a symmetry action on the corresponding sublattice.

\subsection{Understanding SPT and Lemma}

First, let us demonstrate the simplest example in 1D cluster state defined by the following Hamiltonian
\begin{equation}\label{eq:app_1dcluster}
    H = - \sum_n Z_{n-1} X_n Z_{n+1},
\end{equation}
whose ground state is  an 1D SPT protected by $G = G_1 \times G_2 = \mathbb{Z}_2 \times \mathbb{Z}_2$, defined by product of $X$s in even and odd sites respectively. More precisely, we have two generators for $G$: $g_1 = \prod_{n=1}^N X_{2n-1}$ and $g_2 = \prod_{n=1}^N X_{2n}$.
Here, $X$ measures $\mathbb{Z}_2$ charge, while $Z$ creates $\mathbb{Z}_2$ charge. (Think about $\U(1)$ symmetry, whose transformation is $e^{i \theta Q}$. In this sense, $X$ indeed measures charge) Note that $ZZ$ measures whether the $G_1$ domain wall exists, and whenever there is a $G_1$ domain wall, the term enforces $X$ in between to take nontrivial value, i.e., nontrivial $G_2 = \mathbb{Z}_2$ charge. Note that 0d $G_2$ SPT is distinguished by the charge. Therefore, we call such an SPT phase to have a \emph{mixed anomaly} between $G_1 = \mathbb{Z}^{(0)}_2$ and $G_2 = \mathbb{Z}_2^{(0)}$ (superscript implies they are 0-form symmetries).
Note that it can be interpreted in the other way -- instead of prescribing an energy penalty for certain configurations, each term can be interpreted as \emph{creating} and annihilation certain configurations. To elaborate further, $Z$ term creates non-trivial SPT by flipping $X$-basis, while $X$ term creates two domain walls next to it (and annihilate). Therefore, $ZXZ$ term can be interpreted as creating two domain walls next to $X$, and then each $Z$ creates a non-trivial SPT at each domain wall. The ground state would be the superposition of all configurations that is invariant under the action of such operations, which fits into the decorated domain wall construction picture~\cite{Chen2014}. The later perspective will be used for the generalization.

For this 1D cluster state, we want to demonstrate the Lemma. Let $L_1$ ($L_2$) be the odd (even) sublattice where $G_1$ ($G_2$) is acting on.
Without loss of generality, consider an operator $O$ defined on $L_1$. Then, we can show that its expectation value under $\ket{\psi}$ disappears if it involves $Z_{2n+1}$ or $Y_{2n+1}$:
\begin{align}
    \bra{\psi} O_i \ket{\psi} &= \bra{\psi} (Z_{2n} X_{2n+1} Z_{2n+2}) O_i (Z_{2n} X_{2n+1} Z_{2n+2}) \ket{\psi} = - \bra{\psi} O_i \ket{\psi} = 0
\end{align}
Here, we can pull $Z_{2n}X_{2n+1}Z_{2n+2}$ from $\ket{\psi}$ since the ground state is stabilized by it. Furthermore, as two $Z$s on $L_2$ in the stabilizer simply commutes with any operator defined on $L_1$, while $X_{2n+1}$ anti-commutes with $Z_{2n+1}$ or $Y_{2n+1}$ inside $O$. Therefore, for the expectation value of $O$ to not vanish, it has to be made of either $I$ or $X$ operators. Now, we claim that it is one of two cases: either $O = I$ or $O = \prod_{n=1}^N X_{2n+1}$. Assume that $O \neq I, \prod X$. Then, there must be a neighboring two odd sites $2m-1$ and $2m+1$ where $O\big|_{2m-1} = I$ while $O\big|_{2m+1} = X$. Then, 
\begin{align}
    \bra{\psi} O_i \ket{\psi} &= \bra{\psi} (Z_{2m-1} X_{2m} Z_{2m+1}) O_i (Z_{2m-1} X_{2m} Z_{2m+1}) \ket{\psi} = - \bra{\psi} O_i \ket{\psi} = 0
\end{align}
This concludes the proof, i.e., $\expval{O} = 0$ for $O$ defined on $L_1$ if $O$ is not the element of $G_1 = \{ I, \prod_{n=1}^N X_{2n+1} \}$. In fact, this is not a coincidence. For a given commuting stabilizer Hamiltonian, the ground state has a vanishing expectation value against any operator that anti-commutes with any stabilizer. However, if a given operator commutes with all stabilizers, if simply means that the operator is nothing but a symmetry action of a given Hamiltonian.

Now we want to generalize the claim for higher dimensional cluster state SPTs with $G^{(n)}_1 \times G^{(d-n-1)}_2$ mixed anomaly, where $G_1$ is an $n$-form symmetry acting on the sublattice $L_1$ and $G_2$ is a $(d-n-1)$-form symmetry acting on the sublattice $L_2$. Such a cluster state SPT Hamiltonian consists of two local terms:
\begin{equation}
    H = - \sum_i h_{1,i} - \sum_j h_{2,j}, \qquad h_{1,i} = {\cal O}^\textrm{charge}_{1,i} {\cal O}^\textrm{d.w.}_{2,i} \qquad h_{2,j} = {\cal O}^\textrm{d.w.}_{1,j} {\cal O}^\textrm{charge}_{2,j}
\end{equation}
where ${\cal O}_1^\textrm{charge}$ (${\cal O}_1^\textrm{d.w.}$) creates the charge (domain wall) of the symmetry $G_1^{(n)}$ in a symmetric fashion.

For example, if $G_1$ is a $\mathbb{Z}_2$ 1-form symmetry, the generators of $G_1$ is given by $h_\gamma =  \prod_{e \in \gamma_1} X_e$ for any closed loop $\gamma_1$   defined on $L_1$. This is the symmetry that randomly creates or annihilates $Z=-1$ closed loops. 
For this symmetry, ${\cal O}^\textrm{charge}_1 = \prod_{e \in \gamma^\perp} Z_e$ for the loop $\gamma^\perp$ defined in the dual lattice of $L_1$ creates the 1-form charged object in a symmetric fashion; more precisely, an open string (or non-contractible loop) is a charged object, but when we create a charged object as a term in the Hamiltonian, we always create a \emph{pair} of them in the $\mathbb{Z}_2$ symmetric case so we combine two open strings to make it a loop. This is analogous to the situation in the 1D cluster Hamiltonian, where we have a $ZXZ$ term with two $Z$s. On the other hand, ${\cal O}^\textrm{d.w.}_1 = X_e$ creates a pair of 1-form symmetry domain wall, i.e., the discontinuity in $Z=-1$ strings perpendicular to the edges.

In general, if a Pauli operator $O$ defined on the sublattice $L_1$ does not to vanish for this SPT, it has to satisfy
\begin{equation}
    \forall i,\,\,\,  h_{1,i} O h_{1,i} =  O \quad  \textrm{ and } \quad \forall j,\,\,\,  h_{2,j} O h_{2,j} = O 
\end{equation}
These conditions imply that  $O$ is invariant under the conjugation by ${\cal O}_{1,i}^\textrm{d.w.}$ and ${\cal O}_{1,i}^\textrm{charge}$. In fact, this already implies that $O$ is in fact a symmetry of the system. Since $O$ is restricted to $L_1$, it means that $O \in G_1$.

\subsection{Application of Lemma}

Below, we evaluate the correlation function for various post-measurement cluster states, which can be thought of as $\cP_{\bm{s}} \ket{\psi}$ where the measurement-projection operator $\cP_{\bm{s}}$ on the sublattice $L_i$ is defined as 
\begin{equation}  
    \cP_{\bm{s}} = \prod_{n \in L_i} P_n, \quad  P_i = \frac{1}{2} \qty[I + s_i(X_i n_x  + Y_i n_y + Z_i n_z))],\quad P_i^2 = P_i
\end{equation}
where $s_i$ is the measurement outcome. 

\subsubsection{1D Cluster State} \label{app:1dIsing}
In this case, the even-sited operator's expectation value with respect to $\ket{\psi}$ is non-vanishing only if the operator is identity or product of $X$s in all even sites. Then, we can calculate the correlation 
\begin{align}
      \bra{\cP_{\bm{s}} \psi}  Z_{1} Z_{2n+1} \ket{\cP_{\bm{s}} \psi}  & = \bra{\psi}  Z_{1} Z_{2n+1} \ket{\cP_{\bm{s}} \psi}  \nonumber \\
      &= \bra{\psi}  X_{2} X_{4} \dots X_{2n} \ket{\cP_{\bm{s}} \psi}  \nonumber \\
      &= \frac{1}{2^N} \qty( \Big[\prod_{m=1}^n s_{2m}\Big] (\Xfactor)^n + \Big[\prod_{m=n+1}^N s_{2m}\Big] (\Xfactor)^{N-n} )   \nonumber \\
      &= \frac{\qty( \Big[\prod_{m=1}^n s_{2m}\Big] (\Xfactor)^n + \Big[\prod_{m=n+1}^N s_{2m}\Big] (\Xfactor)^{N-n} )}{\qty[ 1 + \Big[\prod_{m=1}^N s_{2m}\Big] (\Xfactor)^{N} ]} \braket{ \cP_{\bm{s}} \psi} 
\end{align}
since $[\cP_{\bm{s}}, Z_{2n+1}] = 0$ and $Z_1 Z_{2n+1} \ket{\psi} = X_{2} \dots X_{2n} \ket{\psi}$. 
Here we used that 
\begin{align}
    X_2 X_4 \dots X_{2n} \cP_{\bm{s}} & = (\textrm{Vanishing terms}) \nonumber \\
    &+ \frac{1}{2^N} \qty( \Big[\prod_{m=1}^n s_{2m}\Big] \cos^{n} \theta \cdot I + \Big[\prod_{m=n+1}^N s_{2m}\Big] \cos^{N-n} \theta \cdot  \prod_{n=1}^N X_{2n} )
\end{align}
Note that the norm of $\ket{\cP_{\bm{s}} \psi}$ can be similarly calculated as
\begin{align}
    \braket{ \cP_{\bm{s}} \psi} 
    &=  \bra{\psi} \cP_{\bm{s}} \ket{\psi} = \frac{1}{2^N} \qty(1 + \Big[\prod_{m=1}^N s_{2m}\Big] (\cos \theta)^N ) =   \frac{1}{(2 \cosh \beta)^N}  \sum_{ \{\sigma \} } e^{s_{2n} \beta \sum_n \sigma_{2n-1} \sigma_{2n+1}}, \quad \tanh \beta = \cos \theta
\end{align}
for a system with $2N$ sites. Here I used that $e^{a \sigma_i \sigma_j} = \cosh a (1 + \sigma_i \sigma_j \tanh a )$ for $\sigma_{i,j} = \pm 1$. We can also calculate other correlation functions involving $X$ or $Y$ by rewriting them in terms of operators defined on even sites. For example, note that the expectation value of $X_{2n+1}$ can be calculated as 
\begin{align}
      \bra{\cP_{\bm{s}} \psi} X_{2n+1} \ket{\cP_{\bm{s}} \psi}  & = \bra{\psi}  X_{2n+1} \ket{\cP_{\bm{s}} \psi}  \nonumber \\
      &= \bra{\psi}  Z_{2n} Z_{2n+2} \ket{\cP_{\bm{s}} \psi}  \nonumber \\
      &= \frac{1}{2^N} \qty[ \eta_1^2 + \eta_2^2 \eta_1^{N-2} ]  
\end{align}
where $\eta_1 \equiv \Zfactor \sin \phi$ and $\eta_2 \equiv \Zfactor \cos \phi$. Expectation values of more complicated operators can be calculated in a similar way.


\subsubsection{2D Cluster State, measurement on vertices} \label{app:2dIsingGauge}

A 2D cluster state becomes a 2D toric code state when measured on vertices in $X$-basis. The Lemma implies that the expectation value of operators defined on vertices disappear unless it is an element of the 0-form symmetry $\mathbb{Z}_2^{(0)}$, i.e., a product of $X_i$ over the all vertices. Then, the post-measurement wavefunction norm is given by 
\begin{align}
    \braket{ \cP_{\bm{s}} \psi} 
    &=  \bra{\psi} \cP_{\bm{s}} \ket{\psi} = \frac{1}{2^N} \qty(1 + \Big[\prod_{v} s_v \Big] (\cos \theta)^N )  
\end{align}
To detect the spontaneous symmetry breaking of the 1-form symmetry, one can measure the product of $\bm{Z}$s along the boundary of a certain region $S$ as the following:
\begin{align}
      \bra{\cP_{\bm{s}} \psi}  \prod_{e \in {\rd S}} \bm{Z}_e  \ket{\cP_{\bm{s}} \psi}  & = \bra{\psi}  \prod_{e \in  {\rd S}} \bm{Z}_e  \ket{\cP_{\bm{s}} \psi} \nonumber \\
      &= \bra{\psi}  \prod_{v \in  S} X_v \ket{\cP_{\bm{s}} \psi}  \nonumber \\
      &= \frac{1}{2^N} \qty( \Big[\prod_{v \in S} s_v \Big] (\cos \theta)^n + \Big[\prod_{v \notin S} s_v \Big] (\cos \theta)^{N-n} )   
\end{align}
In the limit $N \rightarrow \infty$, the correlation function measuring the 1-form symmetry breaking becomes:
\begin{equation} \label{eq:2d_oneform}
   \expval{\prod_{e \in  {\rd S}} \bm{Z}_e }_{\cP_{\bm{s}} \psi} \propto \qty(\Xfactor)^{\abs{S}}
\end{equation} 
This 1-form symmetry breaking looks robust only up to a finite region. Such 1-form symmetry becomes restored in a long-enough length scale, and the resulting state must become 1-form symmetric.

\subsubsection{2D Cluster State, measurement on edges} \label{app:2dIsingModel}
 
The Lemma implies that the expectation value of operators defined on edges disappear unless it is an element of the 1-form symmetry $\mathbb{Z}_2^{(1)}$, i.e., a product of $\bm{X}_i$ on the closed loop. Let $C$ be the set of closed loops one can draw on a given lattice. Then, the post-measurement wavefunction norm is given by 
\begin{equation}
    \bra{  \psi}\bm{\cP_{\bm{s}}}  \ket{ \psi} = \frac{1}{2^N} \sum_{l \in C} \Big[\prod_{e \in l} s_{e} \Big] (\Xfactor)^{\abs{l}}
\end{equation}
as when we expand $\bm{\cP_{\bm{s}}}$, terms vanish unless it forms a product of $\bm{X}$ along a closed loop. For given two vertices $v$ and $v'$, let $p$ be a path between them. Then, the correlation between two vertices is given as 
\begin{align}
    \frac{\bra{\bm{P} \psi}Z_{v} Z_{v'} \ket{\bm{P} \psi}}{\bra{  \psi}\bm{P}  \ket{ \psi}} & = \frac{\bra{ \psi} \prod_{l \in p} \bm{X}_l \ket{\bm{P} \psi}}{\bra{  \psi}\bm{P}  \ket{ \psi}} \nonumber \\
    &= \frac{1}{\bra{  \psi}\bm{P}  \ket{ \psi}} \qty[ \frac{1}{2^N} \qty( \sum_{\bar{p}\,\textrm{s.t.}\,p+\bar{p} \in C} \Big[\prod_{e \in \bar{p}} s_{e} \Big] (\Xfactor)^{\abs{ \bar{p} }} ) ] 
\end{align}
which is nothing but an expression for the 2D Ising model with the sign of the Ising interaction at the edge $e$ is given by $s_e$. From this structure, we can immediately infer that the amplitudes of the wavefunction in the $Z$ basis should be proportional to the Boltzmann weights.

\subsubsection{3D Cluster State with \texorpdfstring{$\mathbb{Z}_2^{(0)} \times \mathbb{Z}_2^{(2)}$}{Z2(0) x Z2(2)}: measurements on vertices} \label{app:3dIsing2form}

Here we prepare a 3D cluster state defined on the cubic lattice where qubits reside at both vertices ($V$) and edges ($E$). This cluster state has the mixed anomaly of 0-form and 2-form symmetries, and by measuring vertices in $X$-basis, one can get the 3D toric code state. The stabilizer Hamiltonian is given by 
\begin{equation}
    H = - \sum_{v  }  X_v \prod_{e \in n(v)}  \bm{Z}_e   - \sum_{e } \qty( \bm{X}_e \prod_{v \in \rd e} Z_v )
\end{equation}
where $n(v)$ is the set of edges neighboring the vertex $v$. Here, all operators commute and we have two symmetries: 
\begin{align}
    \textrm{0-form: } & g = \prod_{v } X_v \nonumber \\
    \textrm{2-form: } & h_\gamma =  \prod_{e \in  \gamma} \bm{X}_e, \quad \textrm{$\gamma$ is the loop along the bonds} 
\end{align}

When measured on vertices, the above 3D cluster state becomes a 3D toric code state, spontaneously breaking the 2-form symmetry. The Lemma implies that the expectation value of operators defined on edges disappear unless it is an element of the 0-form symmetry $\mathbb{Z}_2^{(0)}$, i.e., a product of $X_i$ over the all vertices. Then, the post-measurement wavefunction norm is given by 
\begin{align}
    \braket{ \cP_{\bm{s}} \psi} 
    &=  \bra{\psi} \cP_{\bm{s}} \ket{\psi} = \frac{1}{2^N} \qty(1 + \Big[\prod_{v} s_v \Big] (\cos \theta)^N )  
\end{align}
To detect the spontaneous symmetry breaking of the 2-form symmetry, one can measure the product of $\bm{Z}$s along the boundary of a certain region $V$ as the following:
\begin{align}
      \bra{\cP_{\bm{s}} \psi}  \prod_{e \in {\rd V}} \bm{Z}_e  \ket{\cP_{\bm{s}} \psi}  & = \bra{\psi}  \prod_{e \in  {\rd V}} \bm{Z}_e  \ket{\cP_{\bm{s}} \psi} \nonumber \\
      &= \bra{\psi}  \prod_{v \in  V} X_v \ket{\cP_{\bm{s}} \psi}  \nonumber \\
      &= \frac{1}{2^N} \qty( \Big[\prod_{v \in V} s_v \Big] (\cos \theta)^n + \Big[\prod_{v \notin V} s_v \Big] (\cos \theta)^{N-n} )   
\end{align}
In the limit $N \rightarrow \infty$, the correlation function measuring the 2-form symmetry breaking becomes:
\begin{equation}  
   \expval{\prod_{e \in  {\rd V}} \bm{Z}_e }_{\cP_{\bm{s}} \psi} \propto \qty(\Xfactor)^{\abs{V}}
\end{equation} 
This 2-form symmetry breaking looks robust only up to a finite region for $\theta \neq 0$, implying that the resulting 3D topological order is unstable under the deviation from  the $X$-basis measurements. The corresponding classical model is a Ising membrane theory (or Ising 2-form symmetric theory), which is defined by 
\begin{equation}
    H(\{\sigma \}) = - \sum_v \prod_{\tilde{f} \in n(v)} \sigma_{\tilde{f}}
\end{equation}
where $\tilde{f}$ is the face of the dual cubic lattice, and $n(v)$ is the set of dual faces neighboring to the vertex $v$ in the original cubic lattice. Note that each term is that the product of Ising spins defined on six faces.  In fact, this is a natural extension of Ising model (Ising $0$-form symmetric theory) and Ising gauge theory (Ising $1$-form symmetric theory). Note that this classical partition function is also exactly solvable.

\subsubsection{3D Cluster State with \texorpdfstring{$\mathbb{Z}_2^{(0)} \times \mathbb{Z}_2^{(2)}$}{Z2(0) x Z2(2)}: measurements on edges} \label{app:3d_IT}

Here, we studied the same model as above but measuring edges. The Lemma implies that the expectation value of operators defined on edges disappear unless it is an element of the 2-form symmetry $\mathbb{Z}_2^{(2)}$, i.e., a product of $\bm{X}_i$ on the closed loop. Let $C$ be the set of closed loops one can draw on the cubic lattce. Then, the post-measurement wavefunction norm is given by 
\begin{equation}
    \bra{  \psi}\bm{\cP_{\bm{s}}}  \ket{ \psi} = \frac{1}{2^N} \sum_{l \in C} \Big[\prod_{e \in l} s_{e} \Big] (\Xfactor)^{\abs{l}}
\end{equation}
as when we expand $\bm{\cP_{\bm{s}}}$, terms vanish unless it forms a product of $\bm{X}$ along a closed loop. For given two vertices $v$ and $v'$, let $p$ be a path between them. Then, the correlation between two vertices is given as 
\begin{align}
    \frac{\bra{\bm{P} \psi}Z_{v} Z_{v'} \ket{\bm{P} \psi}}{\bra{  \psi}\bm{P}  \ket{ \psi}} & = \frac{\bra{ \psi} \prod_{l \in p} \bm{X}_l \ket{\bm{P} \psi}}{\bra{  \psi}\bm{P}  \ket{ \psi}} \nonumber \\
    &= \frac{1}{\bra{  \psi}\bm{P}  \ket{ \psi}} \qty[ \frac{1}{2^N} \qty( \sum_{\bar{p}\,\textrm{s.t.}\,p+\bar{p} \in C} \Big[\prod_{e \in \bar{p}} s_{e} \Big] (\Xfactor)^{\abs{ \bar{p} }} ) ] 
\end{align}
which is nothing but an expression for the 3D Ising model with the sign of the Ising interaction at the edge $e$ is given by $s_e$. From this structure, we can immediately infer that the amplitudes of the wavefunction in the $Z$ basis should be proportional to the Boltzmann weights.

\subsubsection{3D Cluster State with \texorpdfstring{$\mathbb{Z}_2^{(1)} \times \mathbb{Z}_2^{(1)}$}{Z2(1) x Z2(1)}: measurements on edges} \label{app:3D_IGT}

Another 3D cluster Hamiltonian is written as the following:
\begin{equation}
    H_{\text{3D SPT}} = - \sum_{e}  X_e \prod_{f \ni e}  \bm{Z}_{f}   - \sum_{f} \bm{X}_{f} \prod_{e \in f}  {Z}_{e}  
\end{equation}
where $f$ runs for all faces of the cubic lattice. Bolded symbols act on faces, and unbolded symbols act on edges. Note that by multiplying stabilizers, we obtain that $\prod_{f \in c} \bm{X}_f = 1$ for any cube $c$ and $\prod_{e \ni v} X_e = 1$ for any vertex $v$.
Here, generators of two 1-form symmetries are defined on two-dimensional surfaces as the following: 
\begin{align} 
    \textrm{$\mathbb{Z}_2^{(1)}$ 1-form: } & h_{\rd V} \equiv \prod_{f \in \rd V} \bm{X}_{f} \nonumber \\
    \textrm{$\mathbb{Z}_2^{(1)}$ 1-form: } & g_{\rd V} \equiv \prod_{e \perp \rd \tilde{V}} {X}_{e}
\end{align}
where $V$ is a certain three-dimensional volume enclosed by cubic faces, and $\tilde{V}$ is a infinitesimally inflated version of $V$ which intersects with edges emanating from $V$. Therefore, $\rd V$ is a set of  faces, while $\rd \tilde{V}$ is a set of edges. Without loss of generality, if we measure all faces in $X$-basis, then we obtain that the resulting state has $\prod_{e \in f} Z_e = 1$ and $\prod_{e \ni v} X_e = 1$ for all $f$ and $v$, which gives the 3D toric code ground state.

When the measurement direction deviates from the $\hat{x}$-direction, the plaquette terms $B_p=\prod_{l\in p}Z_l$ are no longer stabilizers. Nevertheless, the star terms $A_s=\prod_{l\ni s}X_l$ is still a stabilizer, as we illustrated in Fig. \ref{fig:star}.
\begin{figure}
    \centering
    \includegraphics{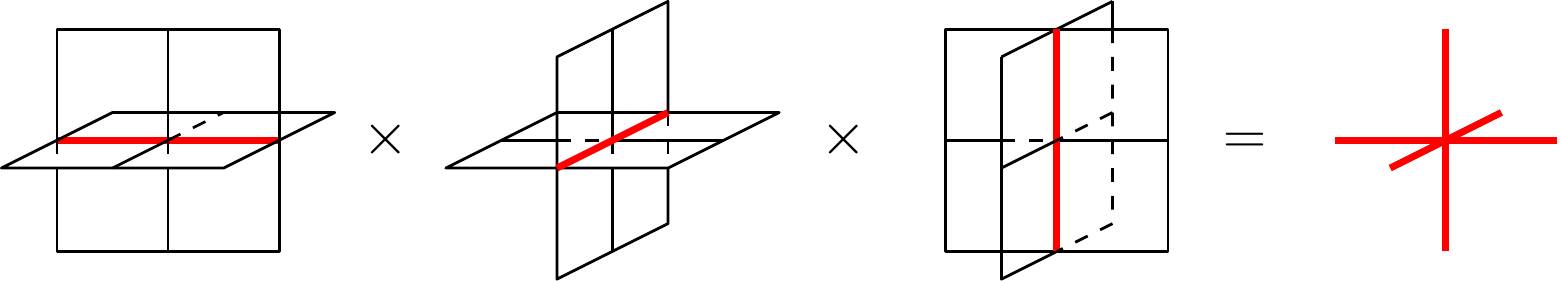}
    \caption{The stabilizer $A_s=\prod_{l\ni s}X_l$ at each site $s$. It commutes with any single-site measurements on sites. It is a product of six stabilizers, $X_l\prod_{p\ni l} Z_l$, with $l\in s$. In the figure, there is a Pauli-$X$ on each red link, and a Pauli-$Z$ on each plaquette.}
    \label{fig:star}
\end{figure}

Assume we measured faces with angles. The norm of the wavefunction is given by 
\begin{align}
      \braket{\cP_{\bm{s}} \psi}  & = \bra{\psi}  \prod_{e \in  {\rd S}} {Z}_e  \ket{\cP_{\bm{s}} \psi} = \bra{\psi}  \prod_{f \in  S} \bm{X}_f \ket{\cP_{\bm{s}} \psi}  \nonumber \\
      &=\frac{1}{2^N} \sum_{V} \Big[\prod_{f \in \rd V} s_{f} \Big] (\Xfactor)^{\abs{\rd V}}
\end{align}
where the summation of $V$ is  over any three-dimensional volume. In fact, one can notice that this is the partition function of 3D Ising gauge theory with plaquette signs given by  $\{ s_f \}$. To detect the spontaneous symmetry breaking of the 1-form symmetry (topological orders) one can measure the product of ${Z}$s along the boundary of a certain surface $S$ as the following:
\begin{align}
      \bra{\cP_{\bm{s}} \psi}  \prod_{e \in {\rd S}} {Z}_e  \ket{\cP_{\bm{s}} \psi}  & = \bra{\psi}  \prod_{e \in  {\rd S}} {Z}_e  \ket{\cP_{\bm{s}} \psi} = \bra{\psi}  \prod_{f \in  S} \bm{X}_f \ket{\cP_{\bm{s}} \psi}  \nonumber \\
      &=\frac{1}{2^N} \sum_{V \textrm{ s.t. }\rd V = S} \Big[\prod_{f \in \rd V} s_{f} \Big] (\Xfactor)^{\abs{\rd V}}
\end{align}
which agrees with the expressions for the loop correlation functions in 3D Ising gauge theory. Depend on $\theta$, the loop correlation decay exponentially either by $\abs{\rd S}$ or $\abs{S}$.

\section{Post-measurement wavefunctions} \label{app:wavefunction}

One can directly write down the wavefunction after measurements in \eqnref{eq:wavefunction} using the decorated domain-wall construction. The cluster state wavefunction is written as the equal superposition of all domain wall configurations with charges attached accordingly:
\begin{equation}
    \ket{\psi} = \frac{1}{\sqrt{2^{N-1}} } \sum_{\{ d_{2n} \}}  \ket{\{ d_{2n} \} }_\textrm{ddw}
\end{equation}
where the subscript ddw stands for that the state is a decorated domain wall basis, where domains (charges) are defined on odd (even) sites. For example, the basis without a domain wall, $\ket{\{d_{2n} = 1\}}$ would be the GHZ state on odd sites (and accordingly, all $\ket{+}$ states on even sites):
\begin{equation}
    \frac{1}{\sqrt{2}} \Big[ \ket{\uparrow \uparrow \cdots} + \ket{\downarrow \downarrow \cdots} \Big]_\textrm{odd} \otimes \Big[ \ket{+}^{\otimes N} \Big]_\textrm{even}
\end{equation}
where the odd sites define spin configurations and even sites are charged based on whether the domain wall exists in neighboring odd sites. The summation is over $2^{N-1}$ configurations since domain walls are under the constraint $\prod d_{2n} = 1$.
Also, the domain-wall basis is the cat state of two different spin configurations giving the same domain-wall configuration.
Then, the wavefunction norm can be calculated as
\begin{align}
    \braket{ \cP_{\bm{s}} \psi} & = \frac{1}{2^N} \bra{\psi} \prod_n (1 + s_{2n} \Xfactor X_{2n}) \ket{\psi} \nonumber \\
    & = \frac{1}{2^{2N-1}} \sum_{\{ d_{2n} \}}    \prod_n (1 + s_{2n} \Xfactor d_{2n} )   \nonumber \\
    & = \frac{1}{2^N} (1 + \prod_{m=1}^N s_{2m} (\cos \theta)^N ) = {\cal Z} 
\end{align} 
where we used the fact that in the expansion of ${\cal P}$, any terms involving $Y$ or $Z$ disappears under the Lemma. 
The measured wavefunction can be written as  
\begin{equation}
    \ket{\cP_{\bm{s}} \psi} =  \sum_{\{ d_{2n} \}} C(\{ d_{2n} \}) \ket{\{ \sigma(\{d_{2n}\}) \} } \otimes \ket{M}   
\end{equation}
where now $\ket{M} = \otimes_{n=1}^N \ket{M_{s_{2n} }}$ stands for the measured component on even sites. The structure of the $Z$ already implies that 
\begin{align}
    \abs{ C(\{ d_{2n} \}) } &=  \frac{1}{\sqrt{2^{2N-1} }} \qty({\prod_{n=1}^N (1 + s_{2n} \Xfactor d_{2n} ) } )^{1/2} \nonumber \\
    &= \frac{1}{\sqrt{2^{2N-1}}} \qty[\cosh (\beta/2)]^{-N}  e^{\frac{\beta}{2} \sum_n s_{2n}  d_{2n} }  
\end{align}
where we used $e^{a \sigma_i \sigma_j} = \cosh(a) (1 + \tanh(a) \sigma_i \sigma_j )$. Although the calculated magnitude agrees with the square root of the Boltzmann weight, in order to calculate the phase factor, one has to proceed with details. First, decompose $\ket{\pm}$ state into the measurement basis: 
\begin{equation}
    \ket{\pm} = a_\pm \ket{M_+} +  b_\pm \ket{M_-}
\end{equation}
Then, we can obtain that $a_\pm = \braket{M_+}{\pm}$ and $b_\pm = \braket{M_-}{\pm}$, where $\ket{\pm} = [1,\pm 1]^T/\sqrt{2}$. Here, for a different parametrization $\hat{n} = (\cos \theta \cos \phi, \cos \theta \sin \phi, \sin \theta)$, we get 
\begin{equation}
    \ket{M_{\pm}} = \frac{1}{\sqrt{2(1 \pm  \sin \theta)}}\mqty( \sin (\theta) \pm 1 \\ e^{i\phi} \cos(\theta) )
\end{equation}

Then, note that for a measurement outcome $\{ s_{2n} \}$, the projection is defined as 
\begin{equation}
    \cP_{\bm{s}} \mapsto \otimes_{n=1}^N \ket{M_{s_{2n}}} \bra{M_{s_{2n}}}
\end{equation}
Then, we see that 
\begin{align}
        C(\{ d_{2n} \})   = \frac{\bra{M} \otimes_{n=1}^N \ket{d_{2n}}}{\sqrt{2^{N-1}}}  = \frac{\prod_{n=1}^N \braket{M_{s_{2n}}}{d_{2n}} }{\sqrt{2^{N-1}}}
\end{align}
Note that 
\begin{align}
     \braket{M_{s_{2n}}}{d_{2n}} & = \frac{s_{2n}}{2\sqrt{1+s_{2n} \sin \theta}} \qty[ s_{2n} \sin \theta + 1 + s_{2n} d_{2n} e^{i\phi} \cos \theta ] \nonumber \\
     & = \frac{s_{2n}}{2(\cos \theta/2 + s_{2n} \sin \theta/2)} \qty[  (\cos \theta/2 + s_{2n} \sin \theta/2)^2 + s_{2n} d_{2n} e^{i\phi} (\cos^2 \theta/2 - \sin^2 \theta/2)  ] \nonumber \\
     & = \frac{s_{2n}}{2} \qty(\cos \theta/2 + s_{2n} \sin \theta/2) + \frac{e^{i\phi}d_{2n}}{2} \qty(\cos \theta/2 - s_{2n} \sin \theta/2) \nonumber \\
     & = \frac{s_{2n}}{2} \qty[ (1 + s_{2n} d_{2n} e^{i\phi}) \cos (\theta/2) +  (s_{2n} - d_{2n} e^{i\phi}) \sin (\theta/2) ]
\end{align}
For $\phi=0$, one can show that 
\begin{align}
    \braket{M_{s_{2n}}}{d_{2n}} & = \begin{cases}
        s_{2n} \cos (\theta/2) = s_{2n} \sqrt{(1 + \cos \theta)/2},& \qquad  \text{if } s_{2n} d_{2n} > 0, \phi=0\\
        \sin (\theta/2) =  \sqrt{(1 - \cos \theta)/2},& \qquad \text{if } s_{2n} d_{2n} < 0, \phi=0 
        \end{cases} \nonumber \\
     & = \varphi_{2n} \sqrt{(1 + s_{2n} d_{2n} \cos \theta)/2}, \qquad \varphi_{2n} \equiv (-1)^{(1-s_{2n})(1-d_{2n})/4}, \qquad \textrm{if } \phi = 0 \nonumber \\
     \prod_n \braket{M_{s_{2n}}}{d_{2n}} &= \qty(\prod_n \varphi_{2n}) \sqrt{ \frac{\prod_n (1 + s_{2n} d_{2n} \cos \theta ) }{2^N} },
\end{align}
which agrees with the expression obtained from the norm $\braket{\cP_{\bm{s}} \psi}$. Now, for $\phi \neq 0$, we obtains that 
\begin{equation}
     \braket{M_{s_{2n}}}{d_{2n}}  = e^{i \eta_{2n}} \sqrt{\frac{1 +  s_{2n} d_{2n} \cos \theta \cos \phi}{2}}, \qquad  \eta_{2n} = \textrm{Arg}( \sin \theta + s_{2n} + s_{2n} d_{2n} \cos \theta e^{i\phi} ) 
\end{equation}
which agrees with the norm calculation (here the prefactor for $X$ is $\cos \theta \cos \phi$).

Although these phase factors can affect the expectation values when we measure correlations of $Y$ or $Z$ operators, they should not change any essential physics. The result implies that if we measure in $xz$-plane without $y$-component, the wavefunction is real. Then, it simply implies that the wavefunction weight would be given by a Gibbs weight. However, even if there is a phase factor, it would simply correspond to some basis rotation; as long as we do not measure correlations in $Y$ or $Z$, such basis rotation along the $X$-axis should not matter. After such rotation, we should obtain a real wavefunction weight again.

\section{Parent Hamiltonians}
We use the strategy described in \secref{subsec:1dparent}~\cite{WITTEN1982, Wouters2021, pivot} to generate the parent Hamiltonians for our post-measurement states in higher dimensions. In the cases described below, minor modifications are needed compared to the derivation for 1D.

\subsection{2d measurement on edges}\label{app:2dlinks_parent}

After the measurement, the wavefunction is
\begin{align}
    |\Psi\rangle = M_\beta|\Psi_0\rangle,~~
    |\Psi_0\rangle = \left|\left\{X_i=\prod_{l\ni i} s_l\right\}\right\rangle,~~M_\beta = \prod_{\langle ij\rangle}e^{\frac{\beta}{2} s_{ij}Z_iZ_j} .
\end{align}
One choice of its parent Hamiltonians is $H=\sum_i H_i$, and 
\begin{align}
    H_i =& -\cos\theta (1+\cos^2\theta)\sum_{\langle ij\rangle}s_{ij}Z_iZ_j+\frac{2}{3}\cos^2\theta B_i-X_i\prod_{l\ni i}s_l\left(1+\cos^4\theta \prod_{l\ni i}s_l\prod_{j\in n(i)}Z_j-\frac{\cos^2 \theta}{3}B_i\right),\nonumber \\
    B_i=&\sum_{j,k\in n(i), j\neq k}s_{ij}s_{ik}Z_jZ_k.
    \label{eq:2dparent_Is}
\end{align}
where $n(i)$ is the set of sites neighboring the site $i$.
We derive the Hamiltonian \eqref{eq:2dparent_Is} by first performing a Kramers-Wannier duality of the wavefunction, which becomes
\begin{align}
    |\Psi'\rangle = M_\beta'|\Psi_0'\rangle,~~|\Psi_0'\rangle =\left|\left\{ \prod_{l\ni i}s_l X_l'=1\right\}\right\rangle,~~M_\beta'=\prod_le^{\frac{\beta}{2}s_lZ_l'}.
\end{align}
Adapting the strategy in Subsection \ref{subsec:1dparent}, the parent Hamiltonian for this state is found to be
\begin{align}
    H' = \frac{1}{3}\sum_l \sum_{j=2}^4 (M_\beta' \Gamma_{1j}'(l)M_\beta')^\dagger (M_\beta' \Gamma_{1j}'(l)M_\beta'),
\end{align}
where 
\begin{align}
        \Gamma_{12}'(l)=\fh\left(s_1s_2X_1'^lX_2'^l-s_3s_4X_3'^lX_4'^l\right),~~\Gamma_{12}'(l)^\dagger\Gamma_{12}'(l)=\fh\left(1-\prod_{i=1}^4s_iX_i'^l\right), 
\end{align}
and similarly for $\Gamma_{13}$,$\Gamma_{14}$. The result is $H'=\sum_i H_i'$, where
\begin{align}
    H_i'=&-\cos\theta(1+\cos^2\theta)\sum_{l\ni i}s_lZ_l'+\frac{2}{3}\cos^2\theta B_i'-A_i'\left(1+\cos^4\theta \prod_{l\ni i}s_lZ_l'+\frac{\cos^2 \theta}{3}B_i'\right),\nonumber \\
    A_i'=&\prod_{l\ni i}s_lX_l',~~B_i'=\sum_{l\ni i,m\ni i,l\neq m}s_ls_mZ_l'Z_m'.
\end{align}
After reversing the KW duality, we obtain the aforementioned parent Hamiltonian.

\subsection{3d measurement on plaquettes of the  \texorpdfstring{$\mathbb{Z}_2^{(1)}\times \mathbb{Z}_2^{(1)}$}{Z2(1) x Z2(1)} SPT}\label{app:3dparent11}

One choice of the parent Hamiltonian $H=\sum_lH_l$ is
\begin{align} \label{eq:3dparent}
    2H_l=&-\cos\theta (1+\cos^2\theta)\sum_{p\ni l}s_p B_p +\frac{2}{3}\cos^2\theta \mathbf B_l-X_l\prod_{p\ni l}s_p\left[1+\cos^4\theta \prod_{p\ni l}s_p\mathbf{Z}_p-\frac{\cos^2\theta}{3}\mathbf B_l\right]\nonumber \\
    B_p=&\prod_{l\in p}Z_p,~~\mathbf B_l = \sum_{p\ni l,q\ni l, p\neq q}s_ps_q B_pB_q,
\end{align}
together with the gauge constraint $\prod_{l\in s}X_l=1$ for each vertex $s$.

We derive it in this way. The state post-measurement is
\begin{align}
    |\Psi\rangle = M_\beta |\Psi_0\rangle,~~|\Psi_0\rangle = |\{X_l=\prod_{p\ni l}s_p\}\rangle, ~~M_\beta = \prod_{p}e^{\beta s_p Z_1^pZ_2^pZ_3^pZ_p^4},
\end{align}
where $Z_i^p$, $i=1,2,3,4$ act on the links on the boundary of the plaquette $p$. 
Under a Kramers-Wannier duality from links to plaquettes, the state is mapped to
\begin{align}
    |\Psi'\rangle =  M_\beta' |\Psi_0'\rangle,~~|\Psi_0'\rangle = |\{X'^l_1X'^l_2 X'^l_3X'^l_4 = s_1^ls_2^ls_3^ls_4^l\}\rangle,~~M_\beta'=\prod_p e^{\beta s_p Z'_p},
\end{align}
where $X'^l_i, i=1,2,3,4$ represent the Pauli-$X$ operators that act on the four plaquettes neighboring the link $l$, and $s^l_i,i=1,2,3,4$ are the measurement outcomes on those plaquettes. 
$|\Psi_0'\rangle$ is the ground state of the Hamiltonian
\begin{align}
    H_0'=\sum_l\frac{1-s_1^ls_2^ls_3^ls_4^lX'^l_1X'^l_2 X'^l_3X'^l_4}{2},
\end{align}
which can be rewritten as
\begin{align}
    H_0'=&\frac{1}{3}\sum_l\left[\Gamma_{12}'^\dagger \Gamma_{12}'(l)+\Gamma_{13}'^\dagger \Gamma_{13}'(l)+\Gamma_{14}'^\dagger \Gamma_{14}'(l)\right].
\end{align}
where 
\begin{align}
    \Gamma'_{12}(l)=\frac{1}{2}(s_1^ls_2^lX_1'^lX_2'^l-s_3^ls_4^lX_3'^lX_4'^l),
\end{align}
and similarly for $\Gamma_{13}'^l$ and $\Gamma_{14}'^l$. In particular, $\Gamma_{ij}'(l)|\Psi_0'\rangle=0$, $ij=12,13,14$.
Then it follows that the state $|\Psi'\rangle$ is the ground state of the following Hamiltonian,
\begin{align}
    H_0'=&\frac{1}{3}\sum_l\sum_{j=2,3,4}\left(M_\beta'\Gamma_{1j}'(l)M_\beta'^{-1}\right)^\dagger \left(M_\beta'\Gamma_{1j}'(l)M_\beta'^{-1} \right).
\end{align}

More explicitly,
\begin{align}
\Gamma_{12}'(l)=\fh\left(s_1s_2e^{\beta(s_1Z_1'+s_2Z_2')}X_1'X_2'-s_3s_4e^{\beta(s_3Z_3'+s_4Z_4')}X_3'X_4'\right),
\end{align}
where the Pauli operators act on the four plaquettes around the link $l$.
\begin{align}
2\Gamma_{12}'(l)^\dagger \Gamma_{12}'(l) =&-\cos\theta \left(1+\cos^2\theta\right)\sum_{p=1}^4 s_p Z_p'+2\cos^2\theta (s_1s_2Z_1'Z_2'+s_3s_4Z_3'Z_4')\nonumber \\
&-\left(\prod_{p=1}^4s_pX_p'\right)\Bigg[1+\cos^4\theta \prod_{p=1}^4s_pZ_p' +\cos^2\theta (s_1s_2Z_1'Z_2'+s_3s_4Z_3'Z_4')\nonumber \\
&-\cos^2\theta(s_1s_2Z_1'Z_3'+s_1s_4Z_1'Z_4'+s_2s_3Z_2'Z_3'+s_2s_4Z_2'Z_4')\bigg].
\label{eq:3DdualH}
\end{align}
Similar results hold for $\Gamma_{1j}'^\dagger(l)\Gamma_{1j}'(l)$ for $j=3,4$. Here, we have scaled \eqnref{eq:3DdualH} by an overall constant $\cosh^2\beta$, such that at $\theta=\frac{\pi}{2}$, the parent Hamiltonian is $-\frac{1}{2}\sum_l X_l\prod_{p\ni l}s_p$. 

Then it follows that 
\begin{align}
    2H_l'= &-\cos\theta(1+\cos^2\theta)\sum_{p=1}^4s_pZ_p'^l+\frac{2}{3}\cos^2\theta \mathbf B_l'-A_l'\left[1+\cos^4\theta\prod_{p=1}^4s_pZ_p'^l-\frac{\cos^2\theta }{3}\mathbf B_l'\right] \nonumber \\
    A_l'=&\prod_{p\ni l}s_pX_p',~~\mathbf B_l'=\sum_{p\ni l, q\ni l,p\neq q}s_ps_qZ_p'Z_q'.
\end{align}

These local terms, after reversing KW duality, becomes \eqnref{eq:3dparent}.

\subsection{3d measurement on sites of the \texorpdfstring{$\mathbb{Z}_2^{(0)}\times \mathbb{Z}_2^{(2)}$}{Z2(2) x Z2(0)} SPT}\label{app:3d02SPT}

A computation similar as in 2d case leads us to the following parent Hamiltonian, up to an unimportant constant and total prefactor, $H=\sum_l H_l$,
\begin{align}
    H_l = -X_l\prod_{v\in l}s_v +\cos^2\theta X_l\prod_{v\in l}\prod_{m\ni v}Z_m -\cos\theta \sum_{v\in l}s_v\prod_{l\ni v}Z_l,
\end{align}
together with gauge contraints, $\prod_{l\in p}X_l=1$.

Let us rewrite it in a more illuminating way, for the case that the outcomes are $s_v=1$,
\begin{align}
    H = &H_0 +6\cos\theta H_{t.c.}+\cos^2\theta H_{\text{SPT}}, \nonumber \\
    H_0=&-\sum_l X_l,~~ H_{t.c.}=-\sum_v \prod_{l\ni v}Z_l,~~H_{\text{SPT}}=\sum_l X_l\prod_{v\in l}\prod_{m\ni v}Z_m,
\end{align}
together with gauge contraints, $\prod_{l\in p}X_l=1$. Under a Kramers-Wannier duality from links to vertices, the model becomes
\begin{align}
    H = -\sum_{\langle ij\rangle} \left(X_iX_j +\cos^2\theta Y_iY_j\right)-6\cos\theta \sum_i Z_i.
\end{align}
At $\theta=0$, the model has a $U(1)$ symmetry.

\section{Probability Distribution of Fluxes} \label{app:gauge}
 
In this section, we prove that the correlation functions of all fluxes uniquely specify the probability distribution $P_m(\bmm)$ for the flux configuration $\bmm \equiv \{ m_p \}$. There are total $N-1$ independent plaquettes and $2$ independent cycles. By $\bmm$, we denote fluxes through those independent $N+1$ plaquettes and cycles. 
Consider a closed loop $\gamma$, which is represented as the boundary of (possibly disjoint) areas $A$, i.e., $\gamma = \rd A$. Then, We note that the expectation value of the gauge-invariant object
\begin{equation} \label{eq:correlation_loop}
    \mathbb{E}\big[ \hspace{-4pt} \prod_{e \in \rd A} \hspace{-4pt}  s_e \big] = (\cos \theta)^{\abs{\rd A}} \quad \Rightarrow \quad \mathbb{E}\big[  \prod_{p \in A} m_p    \big] = (\cos \theta)^{\abs{\rd A}}
\end{equation}
for any arbitrary set of plaquettes $A$. Note that along the cycles, we also have $\mathbb{E}\big[ \hspace{-1pt} \prod_{e \in C} \hspace{-1pt}  s_e \big] = (\cos \theta)^{\abs{C}}$. To show that $P_m(\bmm)$ is completely fixed by the correlation in \eqnref{eq:correlation_loop}, we reconstruct the full probability distribution in a inductive manner:
\begin{itemize}
    \item Get $P_m(m_p)$, which is the probability distribution for a certain plaquette $m_p$ where the others are marginalized. Given the flux expectation values, it is straightforward to show that 
    \begin{equation}
        P_m(m_p) = \frac{1}{2} ( 1 + m_p (\cos \theta)^4 )
    \end{equation}
     
    \item We get all single-flux probability distribution.  Based on this information, we can get all two-fluxes probability distributions between any two plaquettes. Basically, \eqnref{eq:correlation_loop} tells that any pair of separated plaquettes are independent, while the neighboring plaquettes are correlated.  We can establish the set of equations for neighboring plaquettes:
    \begin{align}
        (\cos \theta)^6 & = \sum_{m_1, m_2}  (m_1 m_2) P_m(m_1,m_2) \nonumber \\
        P_m (m_1 ) & = \sum_{ m_2}  P_m (m_1, m_2) \nonumber \\
        P_m (m_2) & = \sum_{ m_1 }  P_m (m_1, m_2) \nonumber \\
        1 & = \sum_{m_1, m_2} P (m_1, m_2)
    \end{align}
    Since $P_m(m_1,m_2)$ consists of $4 = 2^2$ variables and we have 4 equations, we can completely determine the two-point probability distribution.
    
    \item The above proof can be simply extended for any three-points probability distribution, and so on. More generally, for a given $1,2,...,k-1$-points probability distributions and the correlation structure \eqnref{eq:correlation_loop}, we can obtain $k$-plaquettes probability distribution. For a given $P_m(\{ m_i \}_{i=1}^k)$ for a set of plaquettes enclosed by the boundary of length $l$, it satisfies the set of equations as the following: 
    \begin{align}
        (\cos \theta)^l & = \sum_{ \{m \}_k }  \qty[\prod_{i=1}^k m_i ]  {P}_m ( \{ m \}_k ) \nonumber \\
        {P}_m ( m_i ) & = \sum_{ \{ m \}_k \setminus \{ m_i \} }  P(\{m \}_k ) \nonumber \\
        {P}_m ( m_i, m_j ) & = \sum_{ \{m \}_k \setminus \{m_i, m_j \} }  P(\{m \}_k ) \nonumber \\
        \vdots & \nonumber \\
        P_m( \{m  \}_k \setminus \{m_i\} ) & = \sum_{ m_i } P_m(\{m \}_k ) \nonumber \\
        1 & = \sum_{ \{m \}_k }P_m(\{m \}_k )
    \end{align}
    Note that the number of equations are simply given by the binomial expansion:
    \begin{equation}
        \textrm{\# of equations} = 1 + \binom{k}{1} + \binom{k}{2} + \cdots + \binom{k}{k-1} + 1 = 2^k
    \end{equation}
    Therefore, since we have $2^k$ independent equations, it should completely determines $P_m(\{ m_i \}_{i=1}^k)$ with $2^k$ variables. 
    \item By induction, since we know how to get all $P_m$ for a single plaquette, we can obtain $P_m(\bmm)$ for all plaquettes. This implies that the loop correlation function is enough to completely specify the probability distribution for the fluxes.
 \end{itemize}




\section{Removing complex phase factor by a shallow quantum circuit} 
Let us show that the difference between the measurement at $\phi\neq 0$, comparing to the case that $\phi=0$. If $\mathcal{P}|\psi\rangle$ is the pure state measured at $\left(\theta, \phi=0\right)$, then $\mathcal{P_\phi}|\psi\rangle$ measured at angle $(\theta, \phi)$ only differs from $\mathcal{P}|\psi\rangle$ by the $U(1)$ phase in their wavefunctions. 

Without loss of generality, let us consider we perform measurements on all links on the cluster state on a $d$-dimensional square lattice. 
\begin{align}
\mathcal P_{\phi}=R_{x,\phi}^{-1}\mathcal{P}R_{x,\phi}, ~~R_{x,\phi}=\prod_{\text{links } l}\left(\cos\frac{\phi}{2}\mathbf{1}_l+\ii \sin \frac{\phi}{2}X_l\right)
\end{align}
\begin{align}
\mathcal{P}|\psi\rangle = \sum_{\{\sigma_{i}=\pm 1\}}\omega_{\{\sigma_i\}}|\{\sigma_i\}\rangle\otimes |s_{l}\rangle 
\end{align}
It follows that 
\begin{align}
\mathcal{P}_\phi|\psi\rangle = R_{x,\phi}^{-1}\mathcal{P}R_{x,\phi}|\psi\rangle
=R_{x,\phi}^{-1}\mathcal{P}\prod_{\text{links } l}\left(\cos\frac{\phi}{2}\mathbf{1}_l+\ii \sin \frac{\phi}{2}\prod_{i\in l}Z_i\right)|\psi\rangle
\end{align}
Since the projectors are operators on links, 
\begin{align}
\mathcal{P}_\phi|\psi\rangle = &R_{x,\phi}^{-1}\prod_{\text{links } l}\left(\cos\frac{\phi}{2}\mathbf{1}_l+\ii \sin \frac{\phi}{2}\prod_{i\in l}Z_i\right)\mathcal{P}|\psi\rangle\nonumber \\
=&R_{x,\phi}^{-1}\prod_{\text{links } l}\left(\cos\frac{\phi}{2}\mathbf{1}_l+\ii \sin \frac{\phi}{2}\prod_{i\in l}Z_i\right)\sum_{\{\sigma_{i}=\pm 1\}}\omega_{\{\sigma_i\}}|\{\sigma_i\}\rangle\otimes |s_{l}\rangle \nonumber \\
=&\sum_{\{\sigma_{i}=\pm 1\}}\omega_{\{\sigma_i\}}e^{\ii\frac{\phi}{2} \sum_{\text{links } l}\prod_{i\in l}\sigma_i}|\{\sigma_i\}\rangle\otimes R_{x,\phi}^{-1}|s_{l}\rangle.
\end{align}

Therefore, 
\begin{align}
    |\mathcal{P}_\phi\psi\rangle = U_Z|\mathcal{P}\psi\rangle, ~~U_Z=\prod_l e^{\ii\frac{\phi}{2}\prod_{i\in l}Z_i}.
\end{align}
The two states are related by a single-depth local unitary. 

Similar results also hold when we measure the qubits on sites. For example, if we start with the $\mathbb{Z}_2^{(1)}\times \mathbb{Z}_2^{(1)}$ SPT state and measure the sites, the post-measurement state are related by 
\begin{align}
    |\mathcal{P}_\phi\psi\rangle = U_Z|\mathcal{P}\psi\rangle, ~~U_Z=\prod_i e^{\ii\frac{\phi}{2}\prod_{l\ni i}Z_l}.
\end{align}

\section{Stochastic Sampling} \label{app:equivalence}

In this appendix, we prove that the \emph{decoded} correlation function calculated by maximum likelihood sampling protocol would give the correlation function of the RBIM. First, note that the conditional probability can be calculated as the following:
\begin{align}
    P^\textrm{RBIM}_{s|m,\varphi}(\tilde{\bs}|\bmm_\bs) = \frac{P^\textrm{RBIM}_{s,\varphi}(\tilde{\bs}) P^\textrm{RBIM}_{m|s,\varphi}(\bmm_\bs | \tilde{\bs}) }{ P^\textrm{RBIM}_{m,\varphi}(\bmm_\bs) }
\end{align}
where $P_\varphi^\textrm{RBIM}$ is given as independent bond distribution where $p(s) = (1 + \cos \varphi)/2$. Note that 
\begin{equation}
    P^\textrm{RBIM}_{m|s,\varphi}(\bmm_\bs | \tilde{\bs}) =  \begin{cases}
        1 &\,\,\,  \textrm{if $\bs \sim \tilde{\bs}$}\\
        0 &\,\,\,  \textrm{otherwise}
        \end{cases} .
\end{equation}
Also, note that
\begin{equation}
   P^\textrm{RBIM}_{s,\varphi}(\bs) = \frac{1}{2^N Z^{\tilde{\beta}}_0} \sum_{\bt} e^{\tilde{\beta} \sum t_i s_{ij} t_j } \quad \Rightarrow \quad  P^\textrm{RBIM}_{m,\varphi}(\bmm_\bs) = \frac{1}{Z^{\tilde{\beta}}_0} \sum_{\bt} e^{\tilde{\beta} \sum t_i s_{ij} t_j }
\end{equation}
where $Z^{\tilde{\beta}}_0 = (2 \cosh \tilde{\beta})^{2N}$ and $\tilde{\beta} = \tanh^{-1} \cos(\varphi)$. It is straightforward that $\sum_{\tilde{\bs}} P^\textrm{RBIM}_{s,\varphi}(\tilde{\bs}|\bmm_\bs) = 1$.

\subsection{Equivalence to the random bond Ising model (RBIM)}

With these results, we can now compare the correlation functions of the RBIM and our decoded correlations. First, the RBIM correlation function is defined as
\begin{align} 
    \overline{ \expval{C_{ij}} }^\textrm{RBIM} = \sum_\bs P^\textrm{RBIM}_{s,\theta}(\bs)  \qty[ \sum_{\bsigma} \frac{1}{Z_\beta[\bs] } \sigma_i \sigma_j  e^{\beta \sum_{\expval{ij}} \sigma_i s_{ij} \sigma_j } ]
\end{align}
while the correlation function calculated by our protocol (where $\varphi = \theta$ is chosen) would be formally expressed as
\begin{align}  
    \overline{ \expval{C^\textrm{decode}_{ij}} }^{\rho_{\theta}} &= \sum_\bs \sum_{\bsigma} P^{\rho_\theta}_s(\bs)  \qty[  \frac{1}{Z_\beta[\bs] } \sum_{\tilde{\bs}} P^\textrm{RBIM}_{s,\varphi}(\tilde{\bs}|\bmm_\bs)  t_i(\bs\mapsto\tilde{\bs}) t_j(\bs\mapsto\tilde{\bs})  \sigma_i  \sigma_j  e^{\beta \sum_{\expval{ij}} \sigma_i s_{ij} \sigma_j } ] \nonumber \\
    &= \sum_\bs  \sum_{\tilde{\bsigma}} P^{\rho_\theta}_s(\bs)  \qty[ \sum_{\tilde{\bs}} \frac{1}{Z_\beta[\tilde{\bs}] }  P^\textrm{RBIM}_{s,\varphi}(\tilde{\bs}|\bmm_\bs)    \tilde{\sigma}_i   \tilde{\sigma}_j  e^{\beta \sum_{\expval{ij}}  \tilde{\sigma}_i s^\textrm{sample}_{ij}  \tilde{\sigma}_j } ] \nonumber \\
    &= \sum_{\tilde{\bs}} \sum_{\tilde{\bsigma}} \qty[ \sum_\bs   P^\textrm{RBIM}_{s,\varphi}(\tilde{\bs}|\bmm_\bs)   P^{\rho_\theta}_s(\bs) ]  \qty[ \frac{1}{Z_\beta[\tilde{\bs}] }    \tilde{\sigma}_i   \tilde{\sigma}_j  e^{\beta \sum_{\expval{ij}}  \tilde{\sigma}_i s^\textrm{sample}_{ij}  \tilde{\sigma}_j } ] \nonumber \\
    &= \sum_{\tilde{\bs}}  \sum_{\tilde{\bsigma}} P^\textrm{RBIM}_{s,\varphi} (\tilde{\bs})  \qty[ \frac{1}{Z_\beta[\tilde{\bs}] }    \tilde{\sigma}_i   \tilde{\sigma}_j  e^{\beta \sum_{\expval{ij}}  \tilde{\sigma}_i s^\textrm{sample}_{ij}  \tilde{\sigma}_j } ] = \overline{ \expval{C_{ij}} }^\textrm{RBIM}
\end{align}
where $\bmm_\bs$ is the flux configuration calculated from the bond configuration $\bs$. Here we used that $Z[\bs] = Z[\tilde{\bs}]$, and the following equality:
\begin{align}
      &\sum_\bs   P^\textrm{RBIM}_{s,\varphi}(\tilde{\bs}|\bmm_\bs)   P^{\rho_\theta}_s(\bs)    = \sum_\bs  \frac{P^\textrm{RBIM}_{s,\varphi}(\tilde{\bs}) P^\textrm{RBIM}_{m,\varphi}(\bmm_\bs | \tilde{\bs}) }{ P^\textrm{RBIM}_{m,\varphi}(\bmm_\bs) }  P^{\rho_\theta}_s(\bs) \nonumber \\
      & \qquad = \sum_{\bs \sim \tilde{\bs}} \frac{1}{2^N} P^\textrm{RBIM}_{s,\varphi}(\tilde{\bs})  \frac{   \sum_{\bt } e^{\beta \sum t_i s_{ij} t_j } / {Z^{{\beta}}_0} }{  \sum_{\bt } e^{\tilde{\beta} \sum t_i s_{ij} t_j } / {Z^{\tilde{\beta}}_0}   }  = \frac{Z_0^{\tilde{\beta}}}{Z_0^\beta} P^\textrm{RBIM}_{s,\varphi}(\tilde{\bs})  \frac{   \sum_{\bt } e^{\beta \sum t_i s^\textrm{sample}_{ij} t_j } }{  \sum_{\bt } e^{\tilde{\beta} \sum t_i s^\textrm{sample}_{ij} t_j }  }\nonumber \\
      &\qquad = P^\textrm{RBIM}_{s,\varphi}(\tilde{\bs})  \frac{ Z_\beta[\tilde{\bs}] }{ Z_{\tilde{\beta}}[\tilde{\bs}] } \cdot \frac{Z_0^{\tilde{\beta}}}{Z_0^\beta} = P_{s,\varphi}^\textrm{RBIM}(\bs^{\textrm{sample}}) \frac{P_{s,\theta}(\bs^{\textrm{sample}}) }{P_{s,\varphi}(\bs^{\textrm{sample}}) }
\end{align}
At $\beta = \tilde{\beta}$ ($\varphi = \theta$),  the above expression simply becomes $P^\textrm{RBIM}_{s,\varphi}(\tilde{\bs}) $. This result rigorously establishes the equivalence between our decoded correlation functions and those of the RBIM upon proposed stochastic sampling.

\subsection{Optimal choice of \texorpdfstring{$\varphi$}{varphi} }

Now one can ask the following question: What if we choose $\varphi \neq \theta$? It means that the decoding scheme does not properly take into account of the internal noise structure. In this case, one can show that
\begin{align}  
     \overline{ \langle C^{\textrm{decode,$\varphi$}}_{ij}   \rangle } ^{\rho_{\theta} } &=  \overline{ \expval{ C_{ij} }_\beta \expval{ C_{ij} }_{\tilde{\beta}}  }^{\rho_\theta}  =  \overline{ \langle  C_{ij} \rangle_{ \tilde{\beta} } }^{\textrm{RBIM,$p(\beta)$}}
\end{align}
which is the correlation function of the RBIM away from the Nishimori condition. The corresponding RBIM would be at the inverse temperature $\tilde{\beta}$ whose disorder distribution is determined by $p_+ = (1+ \cos \theta)/2$, i.e., $\beta$.

Using the interesting observation by Nishimori~\cite{NishimoriDecoding}, the ferromagnetic correlation for a given probability $p(\beta) = (1 + \tanh \beta)/2$ is maximized when the inverse temperature $\tilde{\beta} = \beta$. Therefore, for an optimal estimation of the correlation functions, one should take $\tilde{\beta} = \beta$.

The derivation of \eqnref{eq:corr_max} is a direct generalization of the following equality:
\begin{align} 
    \overline{ \langle {\sigma^\textrm{decode,$\varphi$}_{i}} \rangle  }^{\rho_{\theta}} &= \sum_{\tilde{\bs}}  \sum_{{\bsigma}} P^\textrm{RBIM}_{s,\varphi} (\tilde{\bs}) \frac{P_{s,\theta}(\bs^{\textrm{sample}}) }{P_{s,\varphi}(\bs^{\textrm{sample}}) } \qty[ \frac{1}{Z_{{\beta}} [\tilde{\bs}] }    {\sigma}_i  e^{\beta \sum_{\expval{ij}}  {\sigma}_i s^\textrm{sample}_{ij}  {\sigma}_j } ] \nonumber \\
    &= \frac{1}{2^N Z^{\tilde{\beta}}_0} \sum_{\tilde{\bs} }  \sum_{{\bsigma}} \sum_{\bt}  e^{\tilde{\beta} t_i \tilde{s}_{ij} t_j } \frac{P_{s,\theta}(\bs^{\textrm{sample}}) }{P_{s,\varphi}(\bs^{\textrm{sample}}) } \qty[ \frac{1}{Z_{ {\beta}} [\tilde{\bs}] }    t_i {\sigma}_i  e^{\beta \sum_{\expval{ij}}  {\sigma}_i \tilde{s}_{ij}  {\sigma}_j } ] \nonumber \\
    &= \frac{1}{2^N Z^{\tilde{\beta}}_0} \sum_{\tilde{\bs} } \frac{P_{s,\theta}(\tilde{\bs} ) }{P_{s,\varphi}(\tilde{\bs}) } Z_{\tilde{\beta} } [\tilde{\bs}]    \qty[ \sum_{\bt} \frac{1}{Z_{ \tilde{\beta}} [\tilde{\bs}] }  e^{\tilde{\beta} t_i \tilde{s}_{ij} t_j } ]
    \qty[ \sum_{{\bsigma}} \frac{1}{Z_\beta[\tilde{\bs}]} {\sigma}_i  e^{\beta \sum_{\expval{ij}}  {\sigma}_i \tilde{s}_{ij}  {\sigma}_j } ] \nonumber \\
    &= \sum_{\tilde{\bs} } P_{s,\theta}(\tilde{\bs})    \qty[ \sum_{\bt} \frac{1}{Z_{ \tilde{\beta}} [\tilde{\bs}] } t_i e^{\tilde{\beta} t_i \tilde{s}_{ij} t_j } ]
    \qty[ \sum_{{\bsigma}} \frac{1}{Z_\beta[\tilde{\bs}]} {\sigma}_i  e^{\beta \sum_{\expval{ij}}  {\sigma}_i \tilde{s}_{ij}  {\sigma}_j } ] \nonumber \\
    &= \overline{ \expval{\sigma_i}_\beta \expval{\sigma_i}_{\tilde{\beta}}  }^{\rho_\theta}
\end{align}

\section{Minimum Weight Perfect Matching Problem} \label{app:MWPM}

For a given flux configuration $\bmm = \{ m_p \}$, the problem of finding a bond configuration $\bs$ with the maximum number of ferromagnetic bonds can be mapped to the problem of finding a complete pairing between frustrated plaquettes ($m_p = -1$) where the sum of the distances (in the square grid) between paired plaquettes is minimized~\cite{Bieche1980}.

More precisely, the problem is defined by the weighted graph $G = (V,E, W)$ with the following definitions:
\begin{itemize}
    \item Vertices $V$: the set of negative fluxes $m_p = -1$.
    \item Edges $E$: the set of shortest paths along the square lattice among all vertices $V$.
    \item Weights $W$: the set of (integer) distances for paths path in $E$.
\end{itemize}
Note that the resulting graph has all-to-all connections. Then, the objective is the following:
\begin{align}
    \textrm{Minimize} &\qquad \sum_{v,v' \in V} W(v,v') x(v,v') \nonumber \\
    \textrm{Constraints} &\qquad \sum_{v'} x(v,v') = \sum_{v'} x(v',v) = 1 \nonumber \\
    &\qquad \sum_{v'} x(v,v') \geq 0
\end{align}
where $x(v',v) = x(v,v')$ represents whether two vertices $v$ and $v'$ are matched ($x(v,v') = 1$) or not ($x(v,v') = 0$). The constraint implies that all vertices are perfectly matched.

\end{widetext}

\end{document}